\documentclass{aa}  

\usepackage{graphicx}
\usepackage{float}
\usepackage{txfonts}
\begin{document}

   \title{Confirmation of the outflow in L1451-mm: SiO line and CH$_3$OH maser detections\thanks{The reduced datacubes used for this work are available in electronic form at the CDS via anonymous ftp to cdsarc.u-strasbg.fr (130.79.128.5)
or via http://cdsweb.u-strasbg.fr/cgi-bin/qcat?J/A+A/}
\thanks{This work is based on observations carried out under project number W18AD and ID 214-18 with the IRAM NOEMA Interferometer and 30m telescope. IRAM is supported by INSU/CNRS (France), MPG (Germany) and IGN (Spain).  
}}
   \author{V. Wakelam\inst{1} \and A. Coutens\inst{2} \and P. Gratier\inst{1} \and          
         T.~H~G Vidal\inst{1} \and N. Vaytet \inst{3} }

   \institute{Laboratoire d'astrophysique de Bordeaux, Univ. Bordeaux, CNRS, B18N, all\'ee Geoffroy Saint-Hilaire, 33615 Pessac, France\\
              \email{valentine.wakelam@u-bordeaux.fr}
         \and
        Institut de Recherche en Astrophysique et Plan\'etologie, Universit\'e de Toulouse, UPS-OMP, CNRS, CNES, 9 av. du Colonel Roche, F-31028 Toulouse Cedex 4, France
         \and
         European Spallation Source ERIC, P.O. Box 176, SE-221 00, Lund, Sweden \\
             }

   \date{Received xxx; accepted xxx}

 
  \abstract
   {The observational counterparts of theoretically predicted first hydrostatic cores (FHSC) have been searched for in the interstellar medium for nearly two decades now. Distinguishing them from other types of more evolved but still embedded objects remains a challenge because these objects have a short lifetime, are small, and embedded in a dense cocoon. One possible lead to finding them is the characterization of the outflows that are launched by these objects, which are assumed to have a low velocity and be small extent.}
   {We observed the L1451-mm FHSC candidate with the NOEMA interferometer (and complementary IRAM 30m data) in order to study the emission of several molecules.}
   {Molecular lines were reduced and analyzed with the GILDAS package network, the CASSIS software, and some python packages.  A nonlocal thermodynamic equilibrium analysis of the CH$_3$OH detected lines was performed to retrieve the physical conditions of the emitting region around the central source, together with the CH$_3$OH, SiO, CS, and H$_2$CO column densities. }
   {Of the targeted molecules, we detected lines of c-C$_3$H$_2$, CH$_3$OH, CS, C$^{34}$S, SO, DCN, DCO$^+$, H$_2$CO, HC$_3$N, HDCO, and SiO. One of the methanol lines appears to be a maser line. The detection of this class I maser and the SiO line in  L1451-mm support the presence of a low-velocity and compact outflow. The excitation conditions of the thermal lines of methanol are also compatible with shocks (H$_2$ density of $\sim 3\times 10^6$~cm$^{-3}$ and a temperature higher than 40~K).}
   { Although these low-velocity outflows are theoretically predicted by some models of FHSC, these models also predict the shock temperature to be below 20~K, that is, not evaporating methanol. In addition, the predicted velocities would not erode the grains and release silicon in the gas phase. We therefore conclude that these new observations favor the hypothesis that L1451-mm would be at a very early protostellar stage, launching an outflow nearly on the plane of the sky with a higher velocity than is observed.}

   \keywords{astrochemistry - stars: protostars - stars: low-mass - ISM: molecules - ISM: individual objects: L1451-mm 
               }

   \maketitle
%


\section{Introduction}

The end of the first phase of star formation (isothermal and optically thin dust emission) is characterized by the formation of a small embedded object called the first hydrostatic core (FHSC). These objects are dense and optically thick in the dust. Their temperature rises adiabatically up to $\sim$2000~K until molecular hydrogen starts to dissociate. Then a second phase of collapse starts. First hydrostatic cores are theoretically predicted \citep{1969MNRAS.145..271L,2010ApJ...714L..58T,2012A&A...545A..98C}, and astrophysicists have tried to observe them for nearly two decades now. One difficulty for such a detection is their short lifetimes (a few hundred to a few thousand years). Another difficulty is distinguishing them from more evolved protostellar sources that would still be deeply embedded in their parent cloud and would also have a low luminosity \citep{2006ApJ...651..945D,2014prpl.conf..195D,2017A&A...600A..36V}.
FHSC are supposed to be very small. The predicted size depends on the model and the magnetic field, but its radius is smaller than 10~au, its temperature is higher than 100~K, and its density is higher than $10^{11}$~cm$^{-3}$ \citep{2016ApJ...822...12H}. Depending on the strength and orientation of the magnetic field, it can be surrounded by a disk and launch a low-velocity outflow.  \citet{2012A&A...545A..98C} and \citet{2016ApJ...822...12H} showed that an outflow with a few km s$^{-1}$ is launched only in strongly ($\mu$ = 2) and intermediately ($\mu$ = 10) magnetized cases. Then the predicted extent of the outflow is a few hundred au, with velocities between 1 and 4~km~s$^{-1}$. \citet{2012A&A...545A..98C} proposed to identify FHSCs with respect to more evolved phases through the detection of a low-velocity outflow without its high-velocity counterparts, which are typical of young stellar objects \citep{1996ARA&A..34..111B}. They expected the grains in these outflows to still be intact because the velocity would be too slow to destroy them in the shock at the envelope interface. \\
Alternatively, \citet{2019MNRAS.487.2853Y} predicted that an outflow driven by an FHSC could not be detected in CO, SO, CS, or HCO$^+$ lines even with ALMA because it is so small and the temperatures produced by the outflow are relatively low. \citeauthor{2019MNRAS.487.2853Y} computed synthetic observations of these molecules with predictions made by a simplified chemical model (excluding grain surface chemistry) and a hydrodynamical model. They concluded that rotational structures might be seen as blue and red lobes in the spectra of the CO (4-3) and SO ($8_7 - 7_6$) lines, but not the outflowing motions. They also concluded that CS detections on subarcsecond scales, that is, tracing the FHSC itself, would be unlikely. For these reasons, they questioned the nature of the FHSC candidate L1451-mm because a low-velocity outflow has been seen with CARMA on this source by \citet{2011ApJ...743..201P}. \\
In this paper, we present new NOEMA observations of L1451-mm. We first summarize the current knowledge about L1451-mm (section 2). We then present the new observations and their analysis (sections 3, 4, and 5) and discuss these findings in the context of the nature of the source (sections 6 to 8). 

\section{L1451-mm}

L1451-mm is an infalling core in a very early phase of its evolution \citep{2017ApJ...838...60M}. This source is located in the L1451 cold cloud, in the Perseus molecular cloud. Several references exist for the distance of the source, and we adopted the distance measured by  \citet{2021A&A...645A..55P} of $\sim$282~pc. This value is very close to the value derived by \citet{2018ApJ...869...83Z} by combining stellar photometry, astrometric data (in particular, the Gaia DR2 parallax measurements), and $^{12}$CO spectral line maps (average distance of $279 \pm 17$~pc).
L1451-mm has a dense and compact continuum emission. The ALMA 3mm dust observations of \citet{2020MNRAS.499.4394M} give an effective radius smaller than 429 au, a mass of 0.28 $M_{\odot}$, and a density higher than $1.3\times 10^8$~cm$^{-3}$. Infall within the source was proposed by \citet{2017ApJ...838...60M} based on a kinematic analysis of an inverse P-Cygni profile in the HCN observed line and a blue asymmetry in the N$_2$H$^+$ double-peak profile. They inferred an infall velocity of ~0.17~km~s$^{-1}$. \citet{2011ApJ...743..201P} found evidence of a very compact low-velocity outflow seen in $^{12}$CO (2-1) with the SMA. The outflow would be in the south-north direction almost perpendicular to the RA axis, with a characteristic velocity of 1.3~km~s$^{-1}$. \citet{2019MNRAS.487.2853Y} argued that the CO channels observed by \citet{2011ApJ...743..201P} could trace rotation rather than an outflow, which is unlikely because the rotation was found with other molecules in the inner envelope of the object to be  perpendicular to the outflow direction \citep{2017ApJ...838...60M,2019ApJ...882..103P,2020MNRAS.499.4394M}.
 Later ALMA observations from \citet{2020MNRAS.499.4394M} confirmed its presence in CH$_3$OH and detected it weakly in HCN. SMA observations from the Mass Assembly of Stellar Systems and their Evolution with the SMA (MASSES) survey \citep{2019ApJS..245...21S} indicate that there is no C$^{18}$O (2-1) emission in the source, and therefore no obvious heating source within the object to release large quantities of CO from the grains. The summary paper by \citet{2019ApJS..245...21S} mentioned a detection of C$^{18}$O in their Table 7,  but the line cannot be seen in  the public data \citep[see also][]{2020MNRAS.499.4394M}. The very compact dust emission, the very low deduced mass, the absence of central heating source, the absence of C$^{18}$O emission, and the very slow compact outflow make this source one of the most promising detected FHSCs in the literature \citep{2011ApJ...743..201P,2017ApJ...838...60M,2019ApJS..245...21S,2020MNRAS.499.4394M}, although, according to \citet{2019MNRAS.487.2853Y}, theoretical simulations of an outflow launched by an FHSC could not be detected even with ALMA.  Last, a bolometric luminosity lower than $<$0.05~L$_{\odot}$ was determined by \citet{2011ApJ...743..201P} from {\it Spitzer} observations. This places this object in the very low luminosity class of objects (VELLOs), while the {\it Herschel} (PACS) 70~$\mu$m observations yielded an upper limit of 44~mJy \citep{2012A&A...547A..54P}. According to the numerical simulations by \citet{2017A&A...600A..36V}, such a low luminosity would more likely be produced by an FHSC than by protostellar sources.

\section{Observations and analysis tools}

The source was observed with the NOEMA interferometer (project number W18AD) on 2019 March 26 and April 7. The D configuration was used with ten antennas. The phase and amplitude calibrators were 0333+321 (both dates), 0234+285 (March 26), and J0329+351 (April 7), and the bandpass calibration was carried out on 3C84. MWC349 and LkH$\alpha$ 101 were used for the flux calibration on April 7 and March 26, respectively. 
A high number of 64 MHz wide spectral windows were observed between 128 and 152 GHz with a spectral resolution of 62.5 kHz using the Polyfix correlator.  Short spacing observations with the 30m telescope (project ID 214-18) were obtained during June 2019 with the FTS50 backend (d$\nu$ = 50 kHz). The NOEMA and 30m data were merged after continuum subtraction using the GILDAS software\footnote{http://www.iram.fr/IRAMFR/GILDAS/} \citep[CLIC and MAPPING;][]{2005sf2a.conf..721P}. The beam size of the merged data is $\sim$ 2.6$\arcsec$ $\times$ 2.2 $\arcsec$. Imaging was performed with natural weighting. No mask was used during cleaning, and absolute residuals showed no sign of spatially structured emission The data were resampled with a resolution of 86 kHz (0.2 km~s$^{-1}$). Primary beam (39$\arcsec$) correction was applied on the merged data. The rms of the data is about 8~mJy~beam$^{-1}$.

The line identification and analysis were made with the GILDAS/LINEDB package as well as the CASSIS software\footnote{http://cassis.irap.omp.eu} (developed by IRAP-UPS/CNRS). 
 The spectroscopic catalogs used for the line identification are the JPL database\footnote{https://spec.jpl.nasa.gov/} \citep{1998JQSRT..60..883P} and the CDMS database\footnote{https://cdms.astro.uni- koeln.de/classic/} \citep{2005JMoSt.742..215M}.
When discussed in the paper, the line emission sizes were obtained by fitting an elliptical Gaussian in the (u,v) plane using the GILDAS package. The resulting parameters are the minor and major axes of the full width at half maximum (FWHM) of the elliptical Gaussian. 
Velocity channel maps were created with GILDAS, and the moment maps were made with the SpectralCube python package \citep[within the Astropy project;][]{astropy:2013,astropy:2018} and plotted with APLpy \citep[an open-source plotting package for Python;][]{2012ascl.soft08017R}. The conversion of NOEMA data from Jy~beam$^{-1}$ into K was made with SpectralCube.

\section{Continuum}

The 3mm ALMA observations from \citet{2020MNRAS.499.4394M} were best fit with a point-like source model, and the integrated flux density was 4.17 mJy (103~GHz). 
We fit our continuum observations obtained with the LSB receiver (at 132.25~GHz) with a point-like source and a circular or elliptical Gaussian using GILDAS in the (u,v) plan. The three fits give a faint but not negligible residual. For all three fits, the central position is ($\alpha_{2000}$, $\delta_{2000}$)=(03$^{\rm h}$25$^{\rm m}$10.25$^{\rm s}$, +30$^{\rm o}$23$^{\arcmin}$55.09$\arcsec$). Fluxes are (6.7$\pm$0.02) mJy for a point source and (8.8$\pm$0.04) mJy for the two other models. The elliptical fitting gives an almost circular shape of (1.5$\pm$0.02)$\arcsec$$\times$(1.4$\pm$0.01)$\arcsec$ and a position angle of (80$\pm$6)$^{\rm o}$ , and the circular fitting gives an FWHM of (1.4$\pm$0.01)$\arcsec$.  We also fit our continuum obtained with the USB receiver (at 147.75~GHz), that is, at higher frequency, with a circular Gaussian source and obtained a higher flux of  (11.25$\pm$0.04)~mJy with a same FWHM as with the LSB receiver. Our difference in integrated flux density with \citet{2020MNRAS.499.4394M} can then be explained by the fact that they used observations at lower frequency (103~GHz). We discuss the dust emissivity index computed from our continuum fluxes at two different frequencies in Appendix~\ref{dust_emissivity}.

\section{Detected molecular lines}

\subsection{Integrated intensity maps}
In the data cube, we detected lines of the following molecules: c-C$_3$H$_2$, CH$_3$OH, CS, C$^{34}$S, SO, DCN, DCO$^+$, H$_2$CO, HC$_3$N, HDCO, and SiO. These lines together with their spectroscopic parameters are listed in Table~\ref{detect_lines}. Their integrated intensity maps are shown in Appendix~\ref{integrated_maps}. 
HC$_3$N, CH$_3$OH, and SiO molecules peak on the source itself only. SO, DCN, DCO$^+$, CS, H$_2$CO, and HDCO have a strong emission on the source but also an extended emission. Last, c-C$_3$H$_2$ does not peak on the source, but northeast of it.


\begin{table*}
\caption{List of detected transitions}
\begin{center}
\begin{tabular}{lccccc}
\hline
\hline
Molecule & Frequency (MHz) & E$_{\rm up}$ (K) & g$_{\rm up}$ & A$_{\rm ij}$ (s$^{-1}$) & Transition  \\
\hline
c-C$_3$H$_2$    &    145089.595  &  16.0 &  21  & $7.44\times 10^{-5}$  &      3 1 2 0 -- 2 2 1 0    \\
c-C$_3$H$_2$   &     150436.545 &  9.7  &  5 & $5.89\times 10^{-5}$    &    2 2 0 0 -- 1 1 1 0   \\
c-C$_3$H$_2$    &    150820.665 &  19.3  &  9 & $1.80\times 10^{-4}$   &     4 0 4 0 -- 3 1 3 0     \\
c-C$_3$H$_2$   &     150851.908 &  19.3  & 27 & $1.80\times 10^{-4}$   &     4 1 4 0 -- 3 0 3 0  \\
CH$_3$OH  &        145093.754   &    27.1 &   7 & $1.23\times 10^{-5}$     &   3 0   0 -- 2 0   0      \\
CH$_3$OH    &     145097.435 &  19.5  &  7 & $1.10\times 10^{-5}$   &     3-1   0 -- 2-1   0       \\
CH$_3$OH    &     145103.185 &   13.9  &  7 & $1.23\times 10^{-5}$  &      3 0 + 0 -- 2 0 + 0   \\    
CH$_3$OH$^a$  &        145126.191 &  36.2   & 7 & $6.77\times 10^{-6}$     &   3 2   0 -- 2 2   0      \\
CH$_3$OH$^a$  &        145126.386 &  39.8   & 7 & $6.87\times 10^{-6}$     &   3-2   0 -- 2-2   0      \\
CH$_3$OH  &      145131.864 &  35.0  &  7 & $1.12\times 10^{-5}$     &   3 1   0 -- 2 1   0    \\
CH$_3$OH  &        146618.697  &  104.4 & 19 & $8.04\times 10^{-6}$    &    9 0 + 0 -- 8 1 + 0  \\
SiO   &         130268.610 &    12.5 &   7 & $1.06\times 10^{-4}$   &           3 -- 2        \\
CS  &          146969.033  &  14.1 &   7  & $6.11\times 10^{-5}$    &          3 - 2 \\
C$^{34}$S &          144617.109 & 13.9 &   7 & $5.82\times 10^{-5}$ &             3 - 2 \\
SO     &       129138.923 &   25.5  &  7&  $2.29\times 10^{-5}$      &      3 3 -- 2 2      \\
H$_2$CO &         145602.949 & 10.5  &  5  &$7.81\times 10^{-5}$     &     2 0 2 -- 1 0 1  \\
H$_2$CO   &       150498.334 &  22.6   &15 & $6.47\times 10^{-5}$   &       2 1 1 -- 1 1 0  \\
HC$_3$N  &       145560.946 &   59.4 &  99&  $2.41\times 10^{-4}$    &         16 -- 15    \\
HDCO    &      128812.860 &   9.3  &  5 & $5.40\times 10^{-5}$   &       2 0 2 -- 1 0 1  \\
DCN$^b$   &        144826.573  &  10.4 &   5&  $3.15\times 10^{-5}$  &          2 2 -- 1 2         \\
DCN $^b$   &       144826.825 &  10.4  &  3 & $7.00\times 10^{-5 }$  &         2 1 -- 1 0   \\
DCN$^c$    &       144827.991 &   10.4 &   3 & $3.50\times 10^{-6}$  &          2 1 -- 1 2  \\
DCN$^c$    &       144827.991 &  10.4  &  3 & $5.25\times 10^{-5}$    &        2 1 -- 1 1 \\
DCN$^c$    &       144827.991 &  10.4 &   5 & $9.45\times 10^{-5}$    &        2 2 -- 1 1  \\
DCN$^c$    &       144827.991 & 10.4  &  7 & $1.26\times 10^{-4}$      &      2 3 -- 1 2  \\
DCO$^+$$^d$  & 144077.214 &    10.4  &  3 & $8.77\times 10^{-5}$  &      2 0 0 1 -- 1 0 0 1  \\
DCO$^+$$^d$  & 144077.262  & 10.4  &  3 & $5.85\times 10^{-6}$  &      2 0 0 1 -- 1 0 0 2    \\
DCO$^+$$^d$ & 144077.280 &  10.4  &  7 & $2.10\times 10^{-4}$  &      2 0 0 3 -- 1 0 0 2     \\
DCO$^+$$^d$  & 144077.285  &10.4  &  5 & $1.58\times 10^{-4}$    &    2 0 0 2 -- 1 0 0 1    \\
DCO$^+$$^d$ & 144077.324  & 10.4   & 3 & $1.17\times 10^{-4}$     &   2 0 0 1 -- 1 0 0 0    \\
DCO$^+$$^d$ & 144077.333 &  10.4  &  5 & $5.26\times 10^{-5}$  &     2 0 0 2 -- 1 0 0 2    \\
\hline
\end{tabular}
\end{center}
$^a$ These two lines are blended. When the integrated intensity maps are computed, both lines are summed together and shown in Fig.~\ref{moment_maps2} with the frequency name 145126.191 MHz.  \\
$^b$ to $^d$ are blended hyperfine components. 
\label{detect_lines}
\end{table*}%

\subsection{Channel maps and line profiles}

Based on the DCN and DCO$^+$ observed lines, which present a Gaussian narrow emission, the $\rm v_{LSR}$ of the cloud is 3.9 km~s$^{-1}$ \citep[in agreement with ][]{2011ApJ...743..201P}. 
The DCN 144827~MHz line on the source itself is slightly shifted toward lower velocities ($\sim 3.5$~km~s$^{-1}$) with respect to the $\rm v_{LSR}$ (see the DCN spectrum on the source in Fig.~\ref{spectra_onsource6}). 
To study the velocity structure of the source, we constructed velocity channel maps (obtained with the MAPPING package of the GILDAS software), which are shown in Appendix~\ref{channel_maps} (Figs~\ref{channel_maps_1}-\ref{channel_maps_9}) for one representative line of each molecule. Molecular spectra are also shown in Figs.~\ref{spectra_onsource1} to \ref{spectra_onsource6} at the continuum peak position. The line widths discussed here were obtained by a Gaussian fit of the spectra at the continuum source position (except for SO, which was too weak; we fit it at an offset of +1.5$\arcsec$, -3.4$\arcsec$ from the center), and represent the FWHM of the lines.  \\
The velocity of the four detected lines of c-C$_3$H$_2$ is clearly restrained to the cloud velocity. The lines have a Gaussian shape with a typical line width of 0.8~km~s$^{-1}$. Both DCO$^+$ and DCN have a narrow emission (with typical line widths of  0.7-0.8~km~s$^{-1}$). 
The two line emissions do not peak at the same location around the source, and if both emissions are present at the source position, their maximum emission is not on the source. Similarly to the two other deuterated molecules, HDCO has a narrow emission  (0.9~km~s$^{-1}$) but also red wings that extend at higher velocities. The lines peak at 3.9 km~s$^{-1}$, which is the velocity of the extended emission, while the high-velocity emission is only present at the source location. The 129.138~GHz line of SO presents weak and narrow (0.3~km~s$^{-1}$) Gaussian shapes. The maximum emission is located at a similar position as DCO$^+$ (southwest of the central dust peak). The HC$_3$N line presents a larger width ($\sim$ 2.4~km~s$^{-1}$) with a more complex shape. The weak emission peaks only on the source. The velocity channel maps appear to show two velocity peaks (see Fig.~\ref{channel_maps_6}). Although the two peaks cannot be seen in the spectrum on the source in Fig.~\ref{spectra_onsource5}, shifting to an offset of (1$\arcsec$, $-1\arcsec$) reveals a double-peak profile. \\
These two velocity peaks are more clearly seen in the last four molecules CS, SiO, H$_2$CO, and CH$_3$OH. Two of these four molecules present strong extended emission at the $\rm v_{LSR}$ (CS and H$_2$CO), and two are only associated with the source position (SiO and CH$_3$OH). All four molecules have two emission peaks at slightly different velocities. Taking the 146.618 GHz line of CH$_3$OH, whose transition upper energy is very high (104.4~K), the first peak is at 3.34~km~s$^{-1}$ and the second is at about 4.74~km~s$^{-1}$. 
All the other detected methanol lines do not show this double-peak profile but several of them show redshifted wings. We discuss the methanol case in section~\ref{methanol}.
 The extended emission of H$_2$CO and CS is removed in the NOEMA data alone (without the 30m data), and we then find similar trends as for the two other molecules. 
 
\section{Tracing the outflow}

\begin{figure*}
\includegraphics[width=0.49\linewidth]{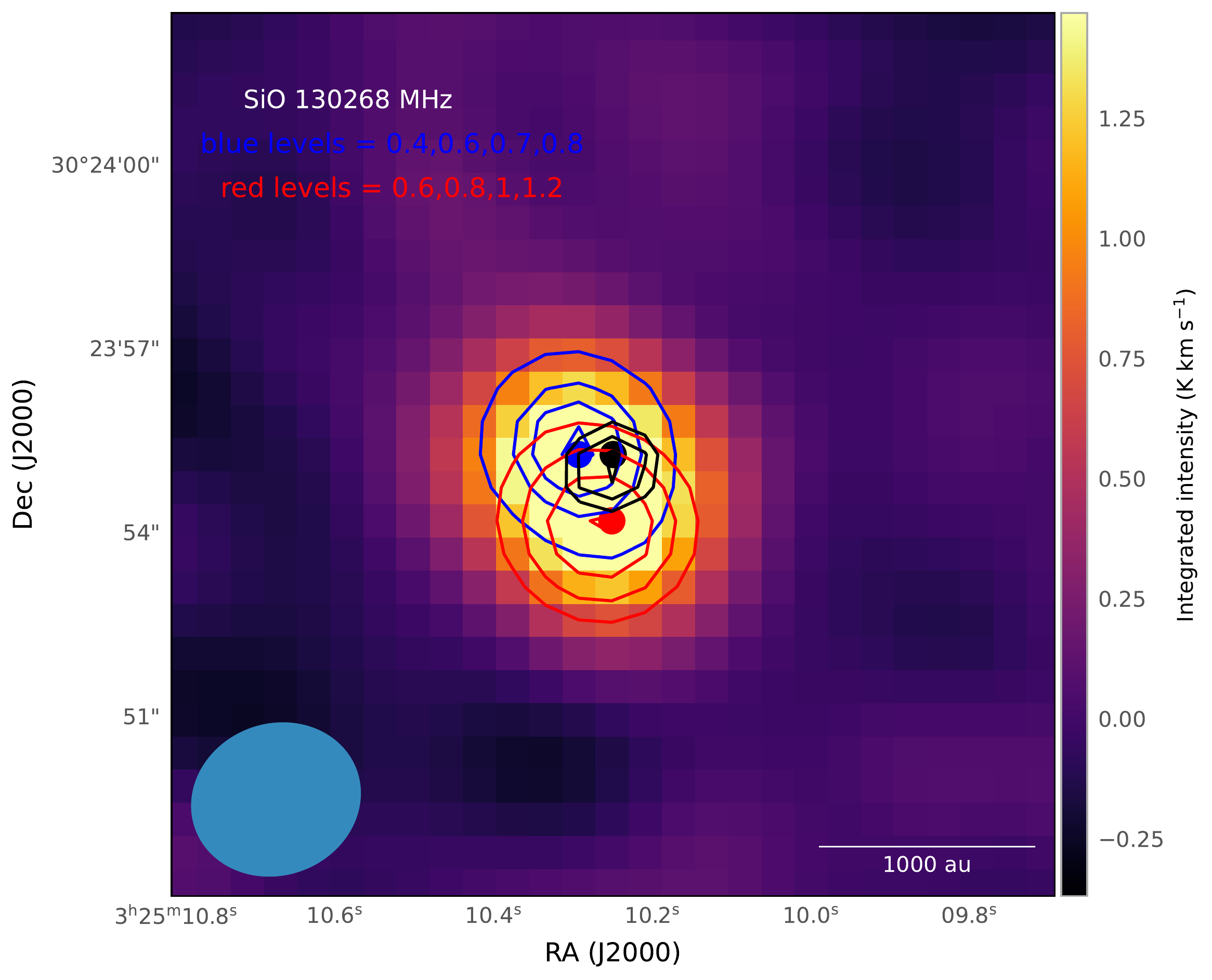}
\includegraphics[width=0.49\linewidth]{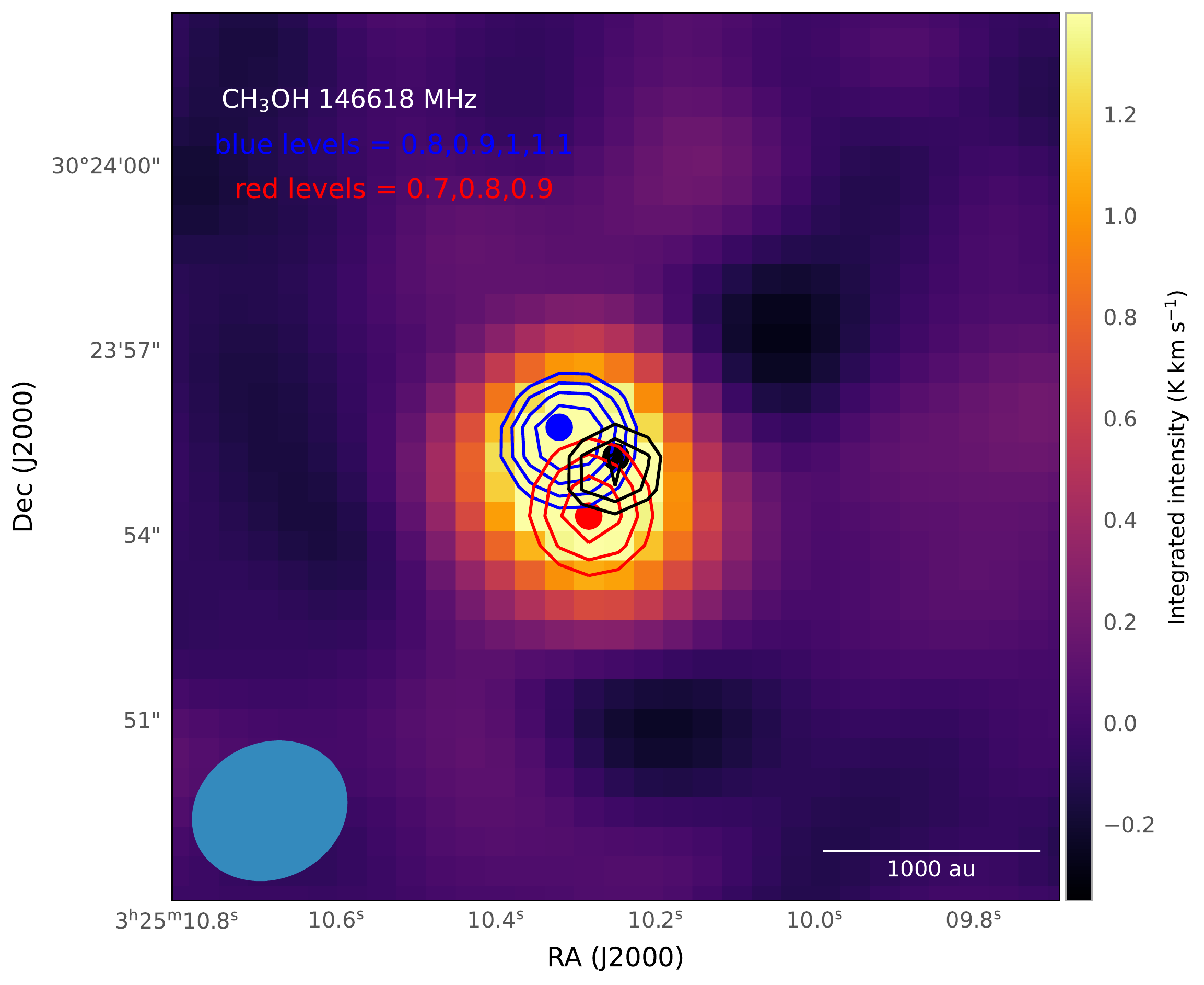}
\includegraphics[width=0.49\linewidth]{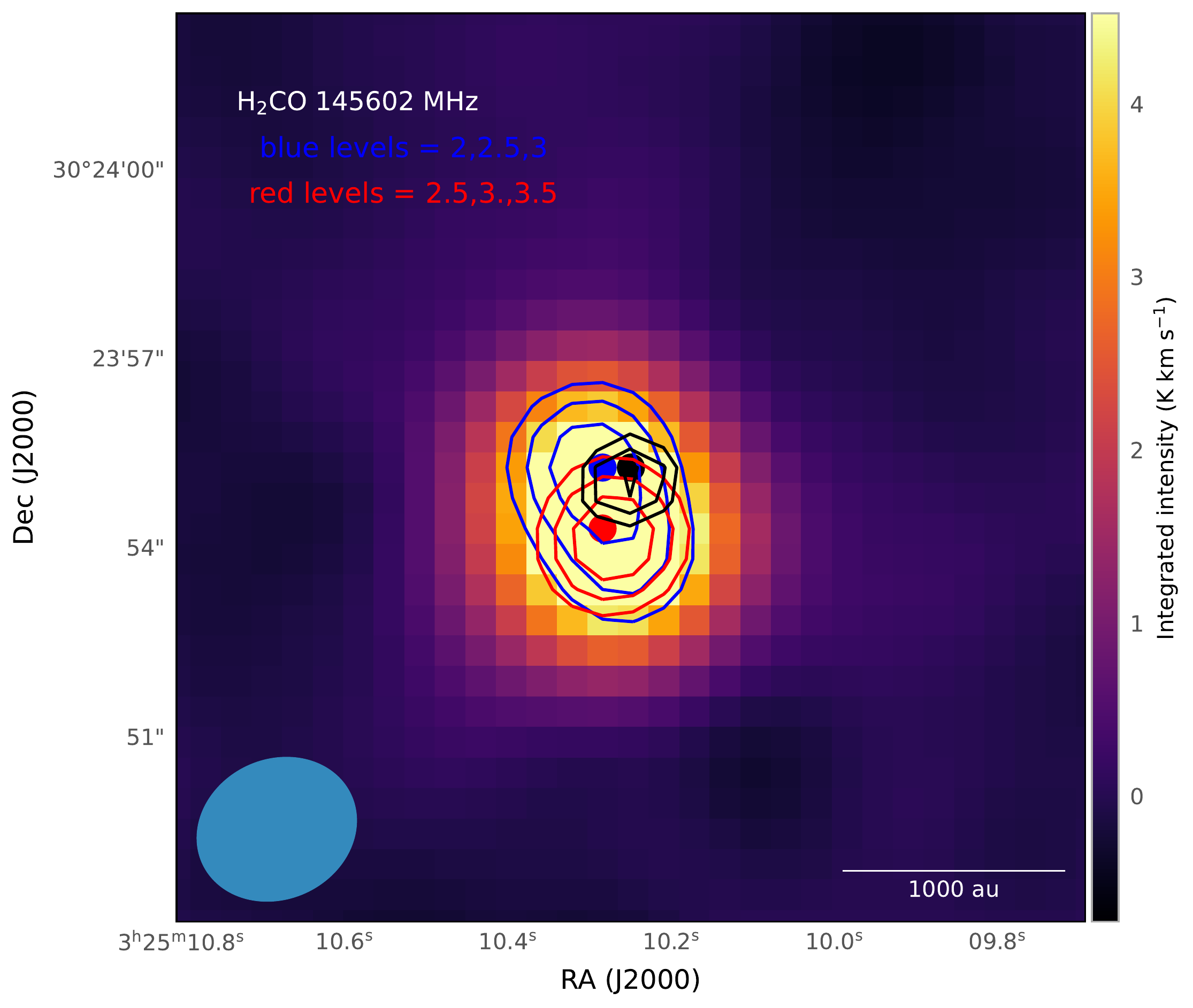}
\includegraphics[width=0.49\linewidth]{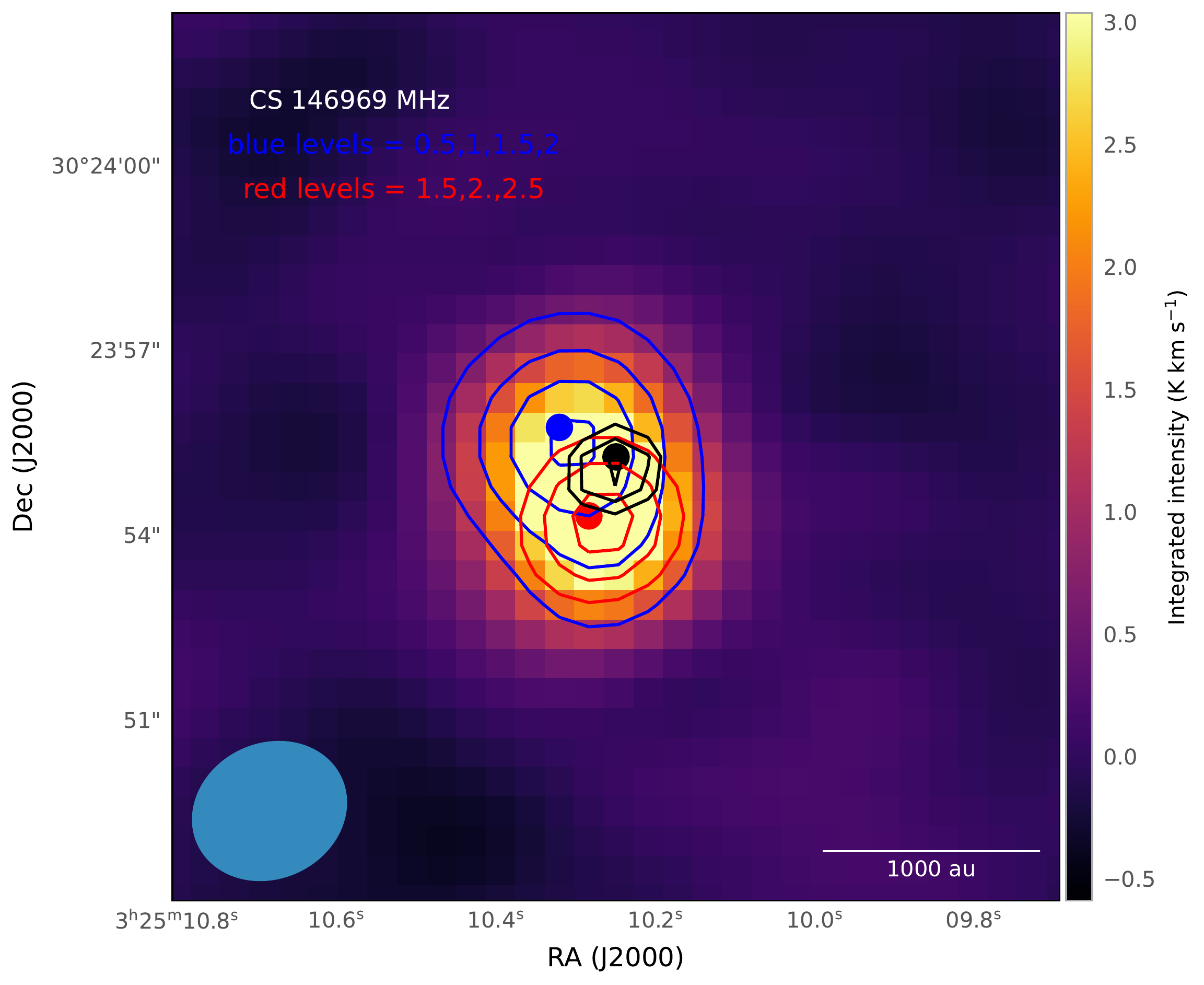}
\caption{Integrated intensity maps over all velocities for SiO, CH$_3$OH, H$_2$CO, and CS (color coding). Superimposed are the blue (2.09-3.94~km~s$^{-1}$) and red (3.94-7.32~km~s$^{-1}$) emissions for each molecule. Blue and red dots indicate the position of the maximum emissions. The level of each contour (in K~km~s$^{-1}$) is indicated in each figure. The color map is the integrated intensity map of each line integrated over all velocities. The black contours show the continuum emission, and the black dot shows its maximum. The levels for the continuum are $5.5\times 10^{-3}$, $6.0\times 10^{-3}$, and $6.5\times 10^{-3}$ mJy. \label{red_blue_maps}}
\end{figure*}

\begin{figure*}
\includegraphics[width=0.32\linewidth]{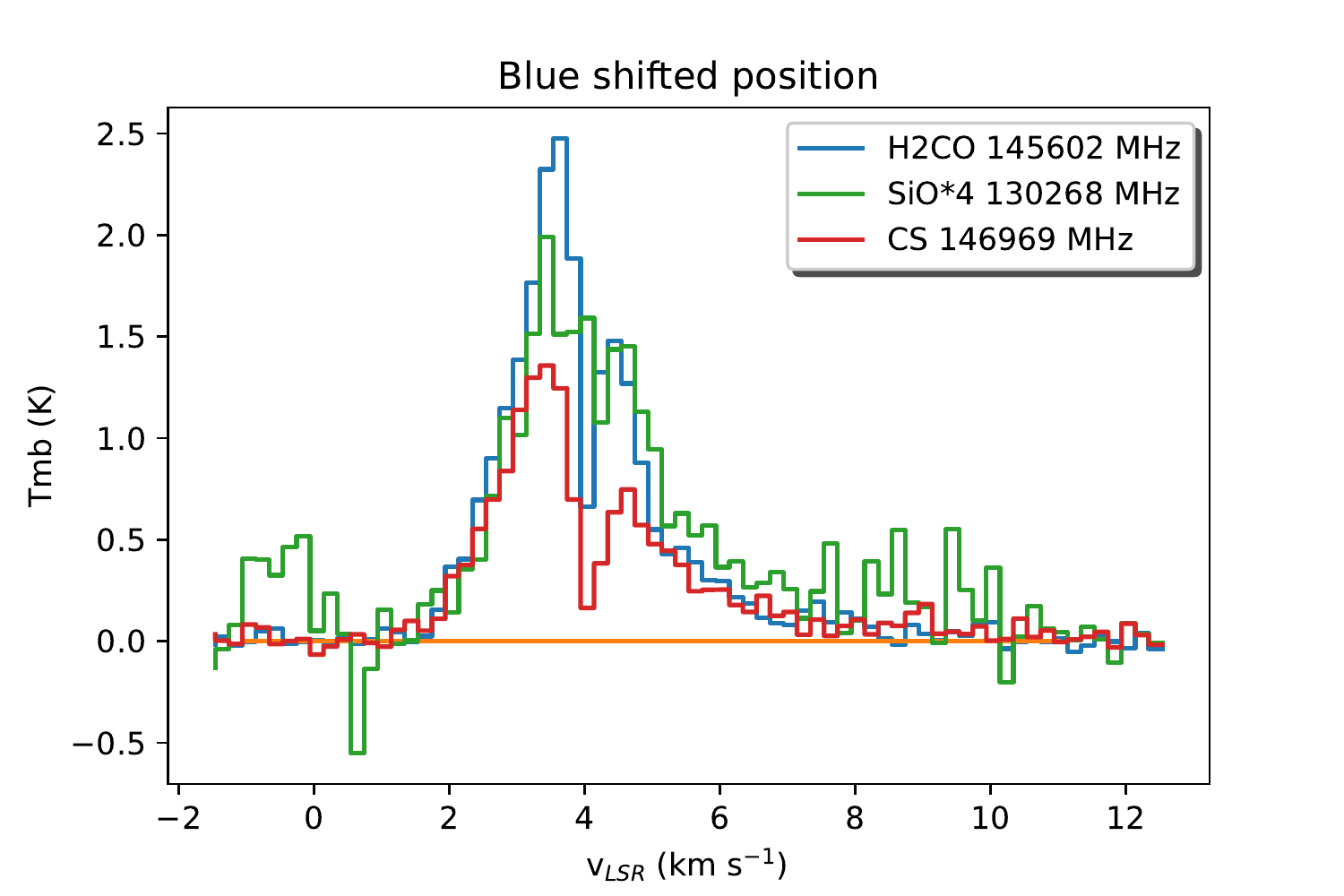}
\includegraphics[width=0.32\linewidth]{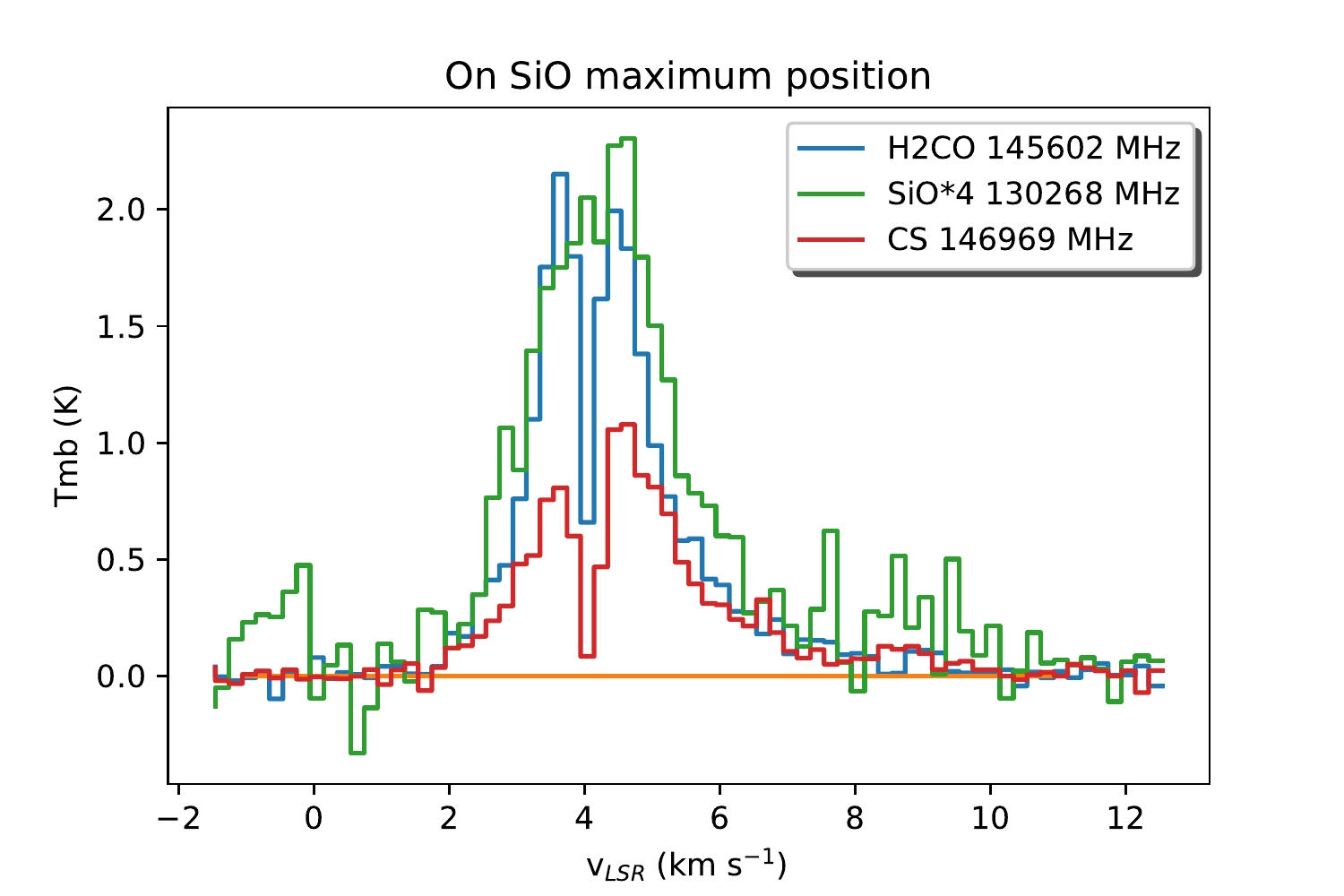}
\includegraphics[width=0.32\linewidth]{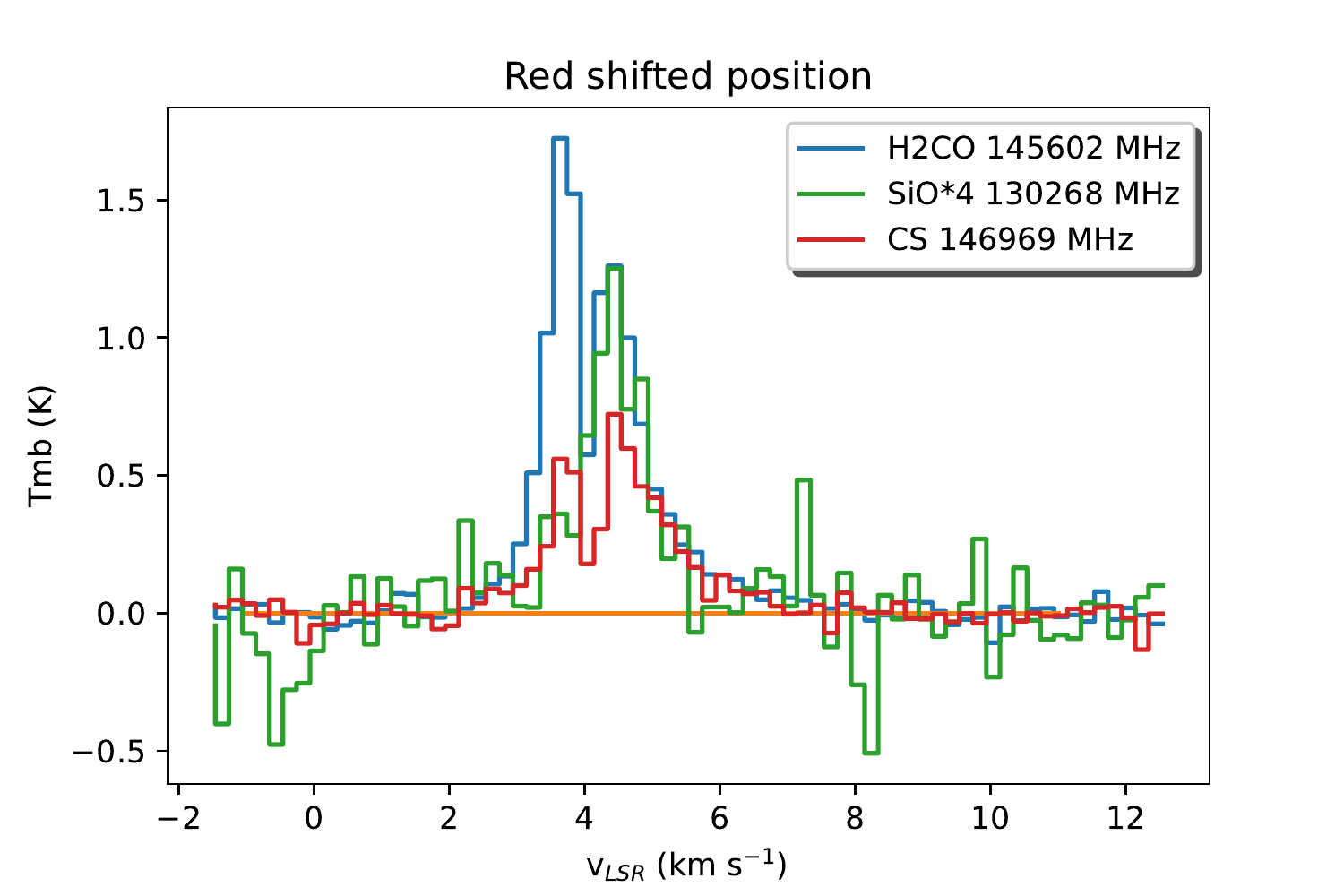}
\includegraphics[width=0.32\linewidth]{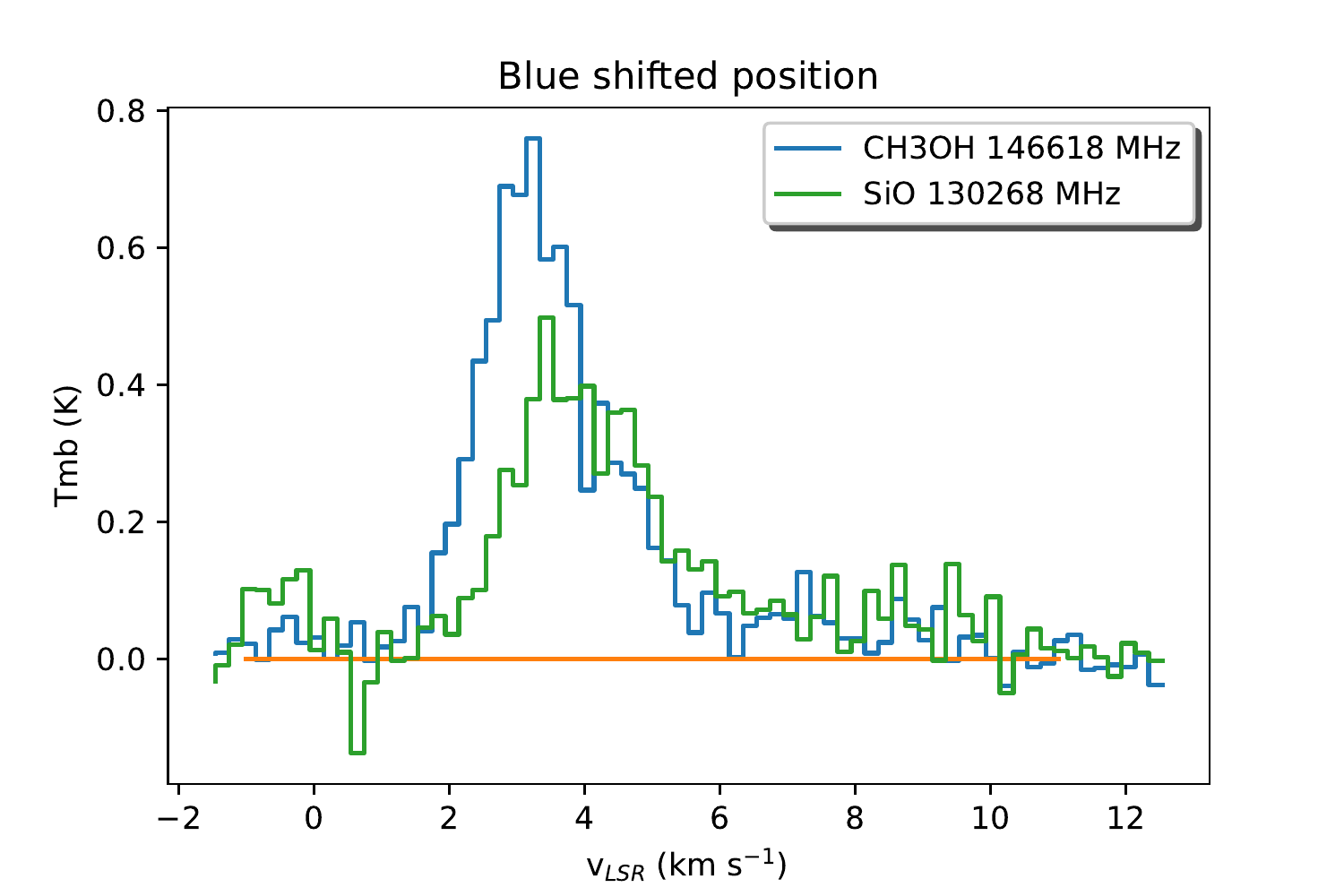}
\includegraphics[width=0.32\linewidth]{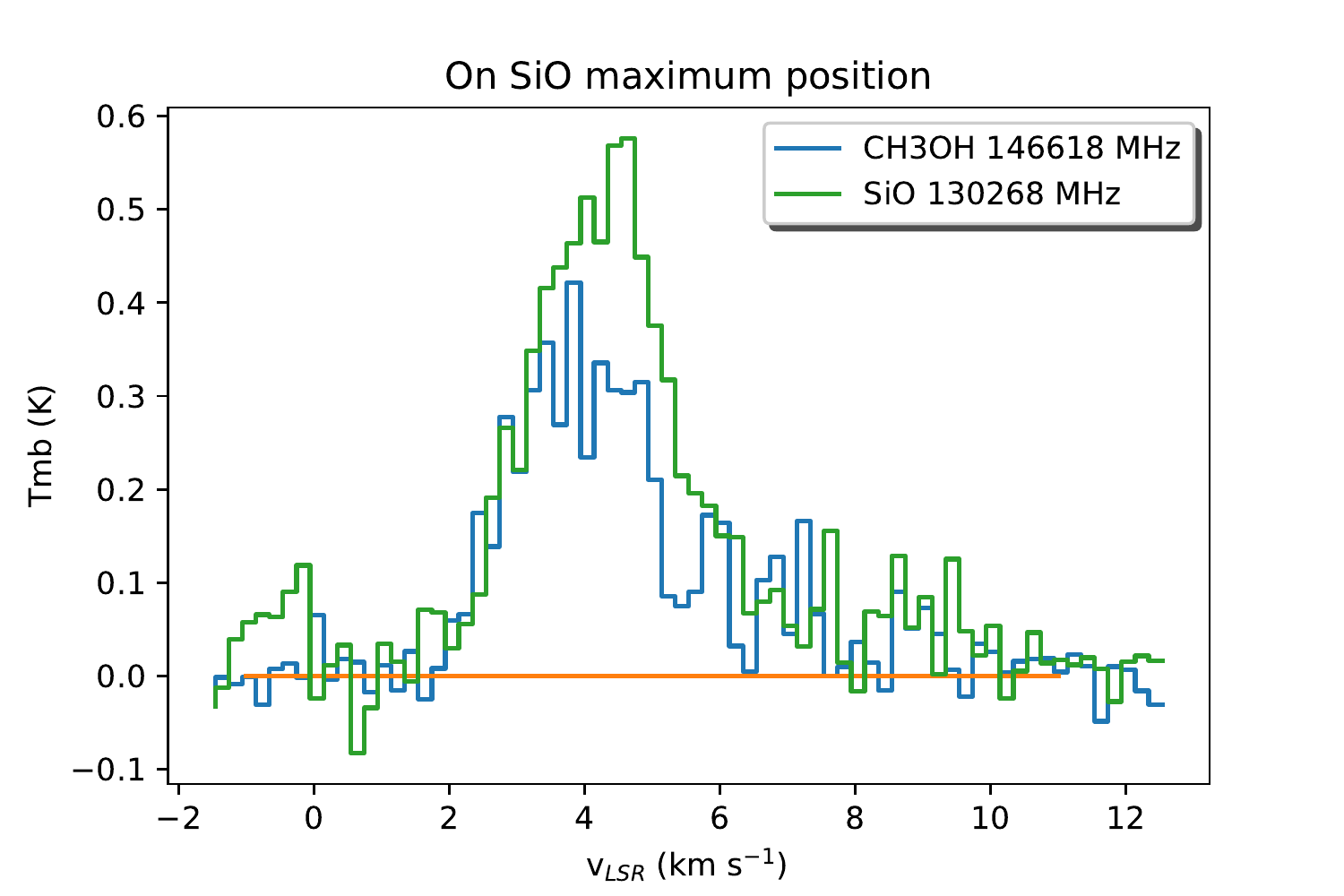}
\includegraphics[width=0.32\linewidth]{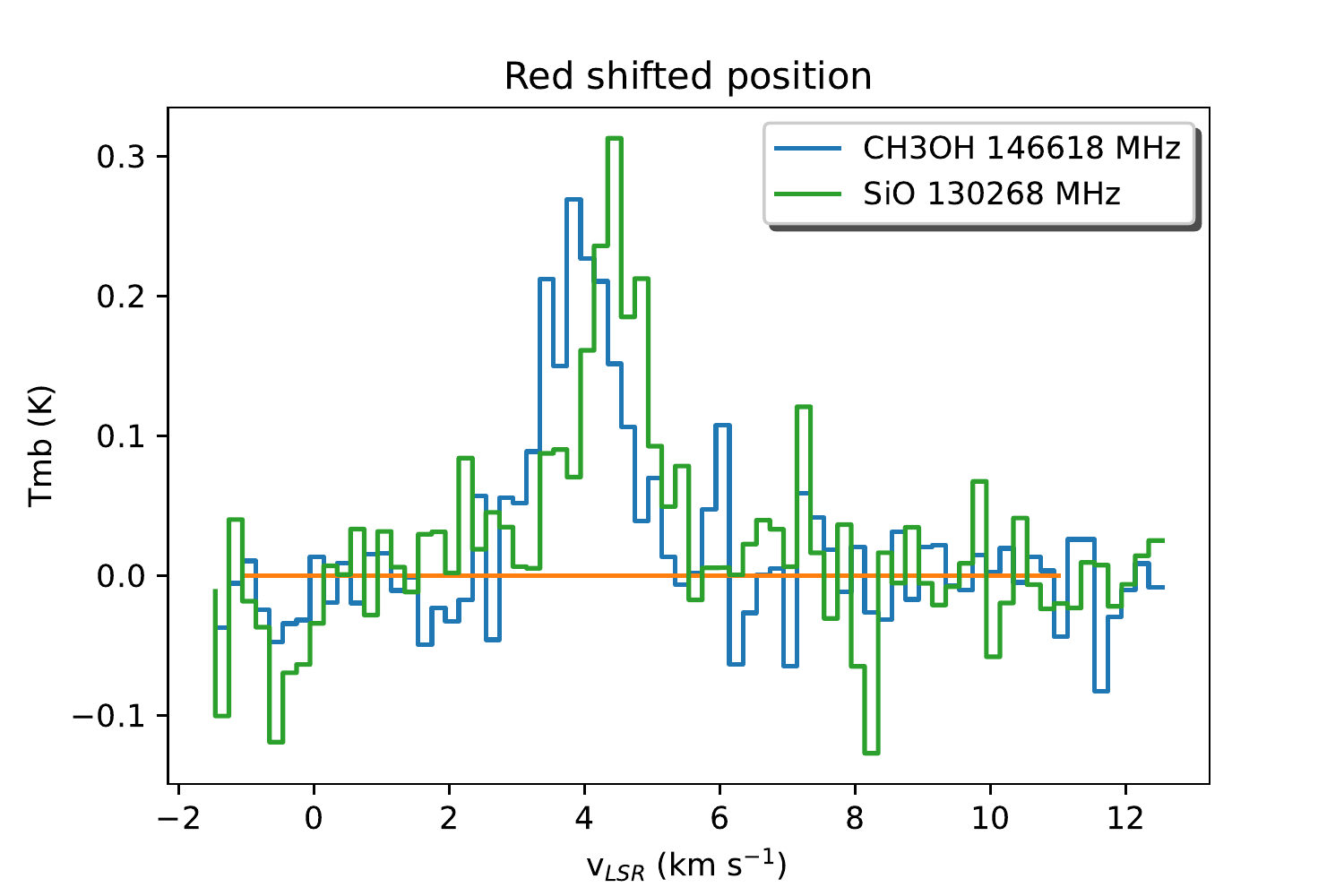}
\caption{ Comparison of the spectral line profiles of the transitions of H$_2$CO, CS, SiO, and CH$_3$OH at three different positions: maximum SiO total integrated intensity position (middle panel), at the blueshifted position ((+0.60$\arcsec$,+0.73$\arcsec$),  left panel), and at the redshifted position ((0$\arcsec$, -0.87$\arcsec$), right panel).
\label{red_blue_spectra}}
\end{figure*}

The two velocity components previously discussed are very likely associated with the outflow identified by \citet{2011ApJ...743..201P} and \citet{2020MNRAS.499.4394M}. In this section, we use the data obtained from NOEMA data alone because we are only interested in the emission at small scales. For SiO and CH$_3$OH, the fluxes are the same, but for CS and H$_2$CO, the extended emission is filtered out. 
Based on the line profiles, we integrated the line intensities over two different ranges of velocities, between 2.1 and 3.9~km~s$^{-1}$ (blueshifted emission) and between 3.9 and 7.3~km~s$^{-1}$ (redshifted emission). These two emissions are shown in Fig.~\ref{red_blue_maps} for the four molecules, superimposed on the total integrated intensity for each transition. The positions of the maximum for each emission are shifted with respect to the continuum maximum position and are slightly different for the four molecules. For SiO, as an example, the maximum of the total SiO emission is shifted by (0.52$\arcsec$, -0.33$\arcsec$) from the continuum. The position of the maximum of the redshifted emission is shifted by (0$\arcsec$, -0.87$\arcsec$) and the position of the blueshifted emission is  (0.52$\arcsec$, 0.21$\arcsec$).   
The difference between the red and blue emissions is then 1.2$\arcsec$. At the distance of 282~pc, the separation represents 339~au. All these shifts are smaller than the beam size, however.\\
To better appreciate the morphology of the lines, we show the molecular lines (of H$_2$CO, SiO, CS, and CH$_3$OH) in Fig.~\ref{red_blue_spectra} at three positions in the maps: 1) the maximum intensity of SiO integrated over the entire velocity range, 2) the maximum intensity of SiO integrated over the blueshifted emission (2.1-3.9~km~s$^{-1}$, blue point in the upper left panel of Fig.~\ref{red_blue_maps}, above position 1), and 3) the position maximum of the intensity of SiO integrated over the redshifted emission (3.9-7.3~km~s$^{-1}$, red point in the upper left panel of Fig.~\ref{red_blue_maps}, below position 1). 
The line width of these four molecules and the positions of the red and blue emissions agree well. 


\section{Discussions}

\subsection{Methanol and the possible detection of a methanol maser}\label{methanol}

In the spectral bands of our observations, two other lines of methanol (145.124 and 145.133~GHz) are covered but undetected.  Both lines have an upper energy at 51~K ($A_{ij}$ = $6.89\times 10^{-6}$~s$^{-1}$) within the range of energies we detected. These nondetections raise some questions concerning our detection of the 146.618~GHz CH$_3$OH line with an upper energy of 104~K. A radiative transfer modeling, at local thermodynamical equilibrium (LTE), of the methanol emission with CASSIS cannot reproduce the observed intensity of all the detected lines (including the 146.618~GHz line) without overpredicting the 145.124 and 145.133~GHz lines. In addition, the double-peak profile observed at 146.618~GHz is not seen in other methanol lines. The 146.618~GHz methanol transition has been proposed to be a possible class I methanol maser by  \citet{1991ASPC...16..119M}, \citet{2004A&A...428.1019M}, and \citet{2021arXiv210908557N}, but as far as we know, it has not yet been reported in the literature \citep[although it was mentioned in][with a reference to a paper in preparation]{1991ASPC...16..119M}. Class I masers are pumped by collisional excitation and are then associated with shocked gas within star-forming regions. In contrast, class II methanol masers are pumped by radiative excitation and are usually associated with accretion disks. Based on multitransition studies, indicative physical conditions in which methanol class I masers are observed are a density of $10^5$~cm$^{-3}$ and temperature of 100~K \citep{1997IAUS..178..163M}. \citet{2016A&A...592A..31L} reported a theoretical study of the physical conditions under which maser transitions can be populated. They found that methanol masers trace high densities (n(H$_2$) $\sim$ $10^7 - 10^8$~cm$^{-3}$) and high temperatures ($>$100~K). However, they did not study the line we detected. Using a shock model with a velocity shock of 17.5~km~s$^{-1}$, \citet{2021arXiv210908557N} predicted this maser to be associated with a moderate preshock density of $\sim10^5$~cm$^{-3}$ , and a postshock density would be two to ten times higher. These conditions may not reflect our case, however, because the shock velocity would be lower in our case. In low-mass star-forming regions, methanol masers are associated with low-velocity outflows \citep{2010MNRAS.405..613K}, as is the case here. The emission size of our 146.618~GHz CH$_3$OH line is approximately (3.4$\pm$0.3)$\arcsec$$\times$(1.8$\pm$0.3)$\arcsec$ (fit in the (u,v) plan using MAPPING from GILDAS package). In our case, together with the SiO detection, this maser could be the signature of a shock produced by the very young low-velocity outflow.  

When the other detected lines of methanol (excluding the maser line) are taken into account, we have detected four lines of form E (two of them are blended) and one line of form A. Using the CLASS software, we fit the observed lines on the source position with a Gaussian profile. Only the NOEMA data were used because the line fluxes are the same as the combined data NOEMA + 30m and the noise is smaller. The observed profiles are not all Gaussians. In particular, the lines at 145097 and 145103 MHz have red wings. The integrated intensity computed over the overall velocity range is less than 20\% higher than the velocity obtained through a Gaussian fit of the lines. 
 The blended lines at 145126 MHz were fit with two different Gaussians so that the determined integrated intensity of each line is uncertain. Comparing the observed integrated intensities with a grid of modeled intensities computed with the RADEX non-LTE radiative transfer code \citep{2007A&A...468..627V}, we derived the physical conditions traced by the methanol emission. The full method is explained in Appendix~\ref{ndradex}. We found an H$_2$ density of about $3\times 10^6$~cm$^{-3}$. As expected, this value (representative of the outflow conditions) is lower than the value estimated from the 3mm dust observations by \citet{2020MNRAS.499.4394M} that traced the inner object. They found a lower limit on the H$_2$ density of $1.3\times 10^8$~cm$^{-3}$ within a source smaller than 417~
au\footnote{The effective radius was recomputed with respect to the value given in the paper to take the change in the distance of the source into account.}. The temperature found from the methanol emission is about 50~K. The methanol column density is about  $10^{14}$~cm$^{-2}$. Assuming a source emission of about 3$\arcsec$ (approximately the emission size of the methanol lines), an H$_2$ density of $3\times 10^6$~cm$^{-3}$, and a CH$_3$OH column density of $10^{14}$~cm$^{-2}$, the abundance of methanol would be about $3\times 10^{-9}$ with respect to H$_2$. If the methanol enhancement in the source is due to shocks, the emitting zone might be smaller and the abundance beam diluted. Using the physical conditions derived from CH$_3$OH, we derived approximate column densities for H$_2$CO, CS, and SiO of $(8.0\pm 0.3)\times 10^{13}$~cm$^{-2}$, $(1.2\pm 0.1)\times 10^{13}$~cm$^{-2}$, and 
$(1.6 \pm 0.3) \times 10^{12}$~cm$^{-2}$ respectively (see Appendix~\ref{ndradex} for details).

\subsection{Observation of SiO}

We computed an approximate SiO column density in the outflow that is two orders of magnitude lower than that of methanol. Two main questions are raised by the SiO observations. The first question is the processes needed to release silicon in the gas phase, and the second question is the form in which it is released. Gaseous SiO is not observed in quiescent cold cores. In the diffuse interstellar medium, most of the silicon is stored as silicates in the refractory part of the grains. The remaining part in the gas phase, as shown by UV observations, is approximately 1\% \citep[abundance of about $1.8\times 10^{-6}$ with respect to total proton density,][]{2009ApJ...700.1299J}. If Si-bearing molecules are observed in energetic regions \citep[diffuse and translucent clouds, and shocked regions,][]{1998ApJ...495..804T,2008A&A...482..809G}, they are not detected in cold dense cores with upper limits on the abundances of about $10^{-12}$ for SiO \citep{1989ApJ...343..201Z}. If the 1\% cosmic abundance of silicon were used in astrochemical models, it would produce SiO gas-phase abundances under cold core conditions that would be much higher than the observed upper limit \citep{2021ApJ...920...37M}, even if the reservoir of silicon were SiH$_4$ in the grains, as suggested by \citet{1996MNRAS.278...62M}. When the elemental abundance of silicon were decreased to fit the nondetection of SiO in cold cores, the SiO predicted by the models in the ices would then be very low.\\
 When we assume that the observed silicon has been released by a shock, its velocity should be higher than 20~km~s$^{-1}$ \citep{2000MNRAS.318..809M,2008A&A...482..809G}, that is, much higher than the SiO velocities that we observed \citep[see also Table 12 of][]{2011ApJ...743..201P}. For such a high velocity to produce an apparent velocity of 1~km~s$^{-1}$ , an angle inclination of 87$^{\rm o}$ would be required (i.e., close to the plan of the sky). The apparent outflow extinction would then be similar to the real length:  1.5$\arcsec$ (half of the emission size of methanol), which represents 423~au at the distance of our source. In this case, its dynamical age would be about 95~yr. If L1451-mm is a protostellar source, this means that it is in a very early stage, maybe the youngest protostar ever observed. 
SiO is observed in low velocity shocks as well, however, which are not strong enough to destroy the refractory part of the grains \citep{2010MNRAS.406..187J,2013ApJ...775...88N,2013ApJ...773..123S,2016A&A...595A.122L}. \citet{2013ApJ...775...88N} and \citet{2016A&A...595A.122L} were able to reproduce their observations in low-velocity shocks with shock models only if a substantial fraction (10\%) of the silicon was already in the form of free Si, either in the gas phase or on the grains, or directly in the form of SiO in the grain mantles. SiO was also observed in the low-mass protostar IRAS16293B at low velocities \citep{2019A&A...626A..93V}. Again, the authors suggested that this SiO was produced by some silicon that was still available in the gas phase.  In the direction of the low-mass star forming region NGC 1333 IRAS4, \citet{1998ApJ...504L.109L} detected a very narrow emission of SiO. Comparing the spectral line width of SiO with their narrow H$_2$O emission, \citet{2013A&A...560A..39C} concluded that both H$_2$O and SiO should come from the photodesorption from the ice mantles in this source or sputtering from grains induced by shocks further decelerated and diluted. In these very dense regions, it is difficult to keep Si in atomic form, however. \\
Another source of SiO that could be discussed, although more speculatively, is the possible presence of amorphous SiO polymer grains. \citet{2014ApJ...782...15K} showed experimentally that the polymerization of SiO compounds is barrierless. This process might occur before the formation of the clouds, locking some silicates, and this material might break more easily than silicates in low-velocity shocks and directly produce the observed SiO. Work is still needed to reconcile the nondetection of SiO in cold cores and its observation in these regions, where refractory grains are assumed to be intact.

\subsection{Nature of the source}

With the SiO and class I methanol maser line detections, a shock induced by outflowing gas within $\sim 3\arcsec$ (i.e., a radius of 430 au, the emission size of the methanol and SiO lines) appears to be established. Shocked regions in FHSCs were predicted by \citet{2012A&A...545A..98C} in the case of intermediately and strongly magnetized models. The apparent velocity of the outflow in L1451-mm is closer to the case of the intermediately (their $\mu = 10$ case) magnetized models (outflow with an extent of $\sim$ 300 au and a velocity $\sim$ 1~km~s$^{-1}$). The question now is whether a low-velocity outflow like this can release silicon from the grains into the gas phase. With a shock model, \citet{2016A&A...595A.122L} showed that this is not the case. They needed a shock velocity higher than 7~km~s$^{-1}$ so they had to consider some free silicon in the gas phase to reproduce the SiO observations. 
 \citet{2012A&A...545A..98C} suggested that one diagnostic of FHSCs would be that the grains in the outflow would not be destroyed (and then would probably not produce SiO). \citet{2016ApJ...822...12H} found that the temperatures in the outflow produced by an FHSC would be lower than $\sim 25$~K, except in the very inner (few au) regions. These temperatures would not be high enough to release methanol from the ices, and some potential silicon if any is stored in the ices. 
As discussed in section~\ref{methanol}, the observation of the methanol maser cannot indicate the physical conditions as it can be produced under a large range of physical conditions. High concentrations of gas-phase methanol released from the grains are required, however. It is unlikely that the outflows and shocked regions predicted by hydrodynamical models for FHSCs can explain the methanol maser and the SiO lines. We therefore argue that L1451-mm would more probably be a very young low-mass protostar and that the apparent low velocity of the outflow is due to a strong inclination of its axis.

\section{Summary and conclusions}

We have presented some NOEMA + 30m data of a few molecules on the L1451-mm source. In particular, we have detected and mapped H$_2$CO, CS, and SiO lines. These three molecules present a double-peak profile that agrees well with an outflow that has been identified by previous publications. We also detected several transitions of methanol. One of these lines appears to be the class I maser line at 146.618~GHz. This specific transition also seems to present a double-peak profile, while some of the methanol transitions have redshifted wings. Non-LTE radiative transfer analysis of the thermal lines of methanol indicates an H$_2$ density of about $3\times 10^6$~cm$^{-3}$ and a temperature of about 50~K. We detected DCO$^+$, DCN, HDCO, and SO lines, but their emission is rather extended or not on the source. We also detected weak HC$_3$N emission on the source. 
The SiO and CH$_3$OH maser emission tends to confirm the launch of a compact (expanding to 430 au) and apparently low-velocity (with a characteristic velocity around 1~km~s$^{-1}$) outflow from the central object. Based on current predictions by hydrodynamical models, it is unlikely that an outflow launched by an FHSC could explain our observations of SiO and methanol maser.  We therefore argue that L1451-mm more probably is at a very early protostellar stage, launching an outflow with a higher velocity than is observed, and lying nearly in the plan of the sky.

\begin{acknowledgements}

The authors acknowledge the CNRS program "Physique et Chimie du Milieu Interstellaire" (PCMI) co-funded by the Centre National d'Etudes Spatiales (CNES).  AC is supported by the European Research Council (ERC) under the European Union's Horizon 2020 research and innovation programme through ERC Starting Grant "Chemtrip" (grant agreement no 949278). The authors thank the IRAM staff, especially Jan Martin Winters, for their help with the observations and calibration of the data. VW thanks Fabrice Herpin from the Laboratoire d'astrophysique de Bordeaux for interesting discussions on masers. The authors are grateful to the anonymous referee for the very helpful suggestions that improved the quality of the paper.

\end{acknowledgements}





\bibliographystyle{aa}
\bibliography{bib}



\begin{appendix}
 
\section{Computation of the dust emissivity index}\label{dust_emissivity}

We used our continuum data to compute the dust emissivity index $\beta$ with equation 2 of \citet{2019A&A...632A...5G}. We obtain  $\beta$ = 0.2. This low index in this wavelength range is usually attributed to millimeter grain sizes \citep{2019A&A...631A..88Y}. Typical values of $\beta$ in the diffuse interstellar medium are about 1.6 \citep{2014A&A...571A..11P,2015A&A...584A..94J}. 
In the starless core TMC-1C \citep{2010ApJ...708..127S} and the prestellar core L1544 \citep{2019A&A...623A.118C}, $\beta$ was found to be similar to the diffuse interstellar medium or even higher ($\sim$2). In class 0 protostars, a gradient in the values of $\beta$ was found from interstellar medium-like values in the outer parts of the protostellar envelopes to values lower than one in the inner parts \citep{2019A&A...632A...5G}. In the class 0 protostar IRAS 2A1, for instance, $\beta$ was found to be 0.38 at 500 au \citep[see Table 2 of][]{2019A&A...632A...5G}. Considering the early stage of our object, such low values of $\beta$ are quite surprising. The dust emissivity index in this source should be investigated with additional observations. 

\section{Integrated intensity maps}\label{integrated_maps}

\begin{figure*}
\includegraphics[width=0.5\linewidth]{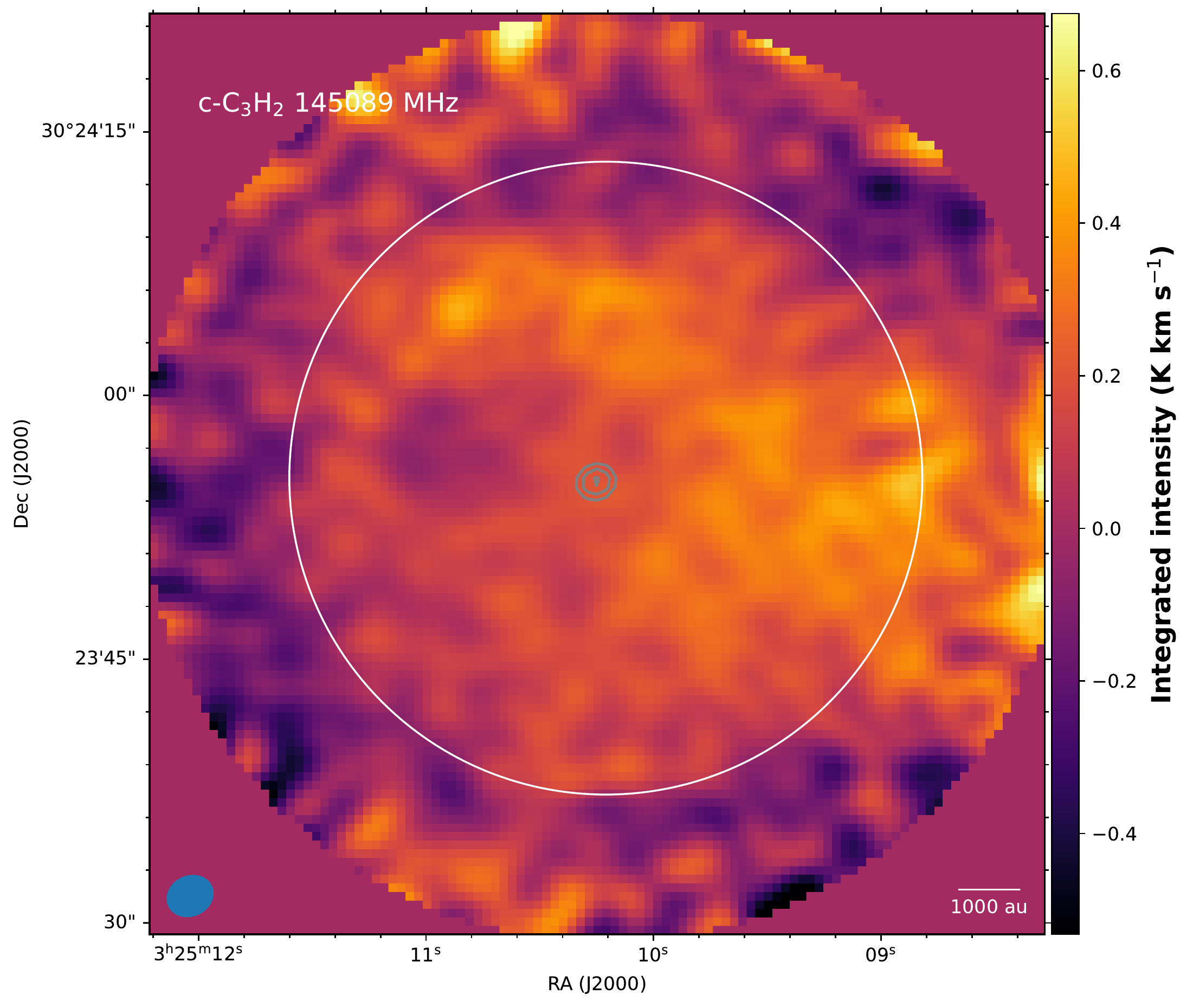}
\includegraphics[width=0.5\linewidth]{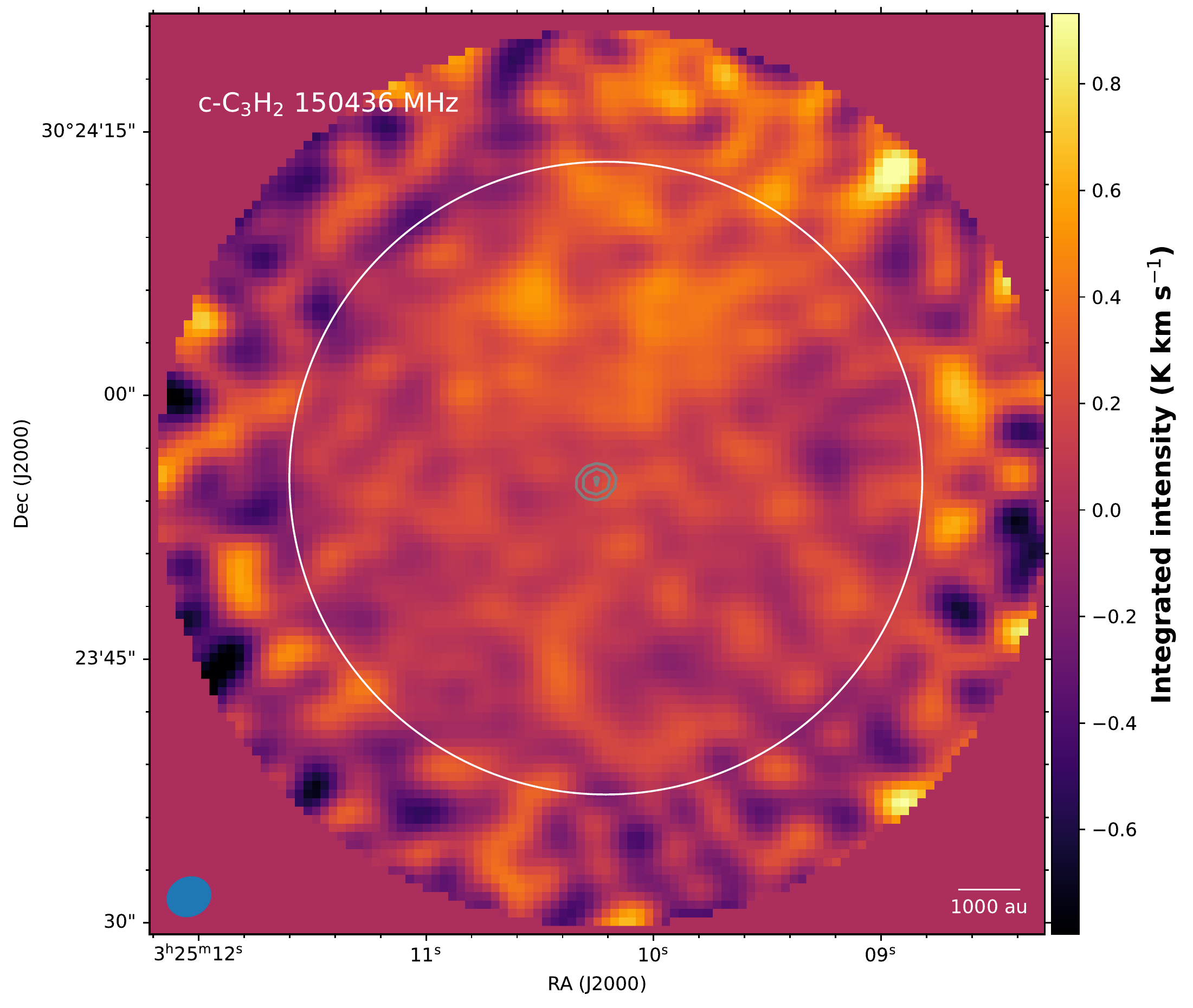}
\includegraphics[width=0.5\linewidth]{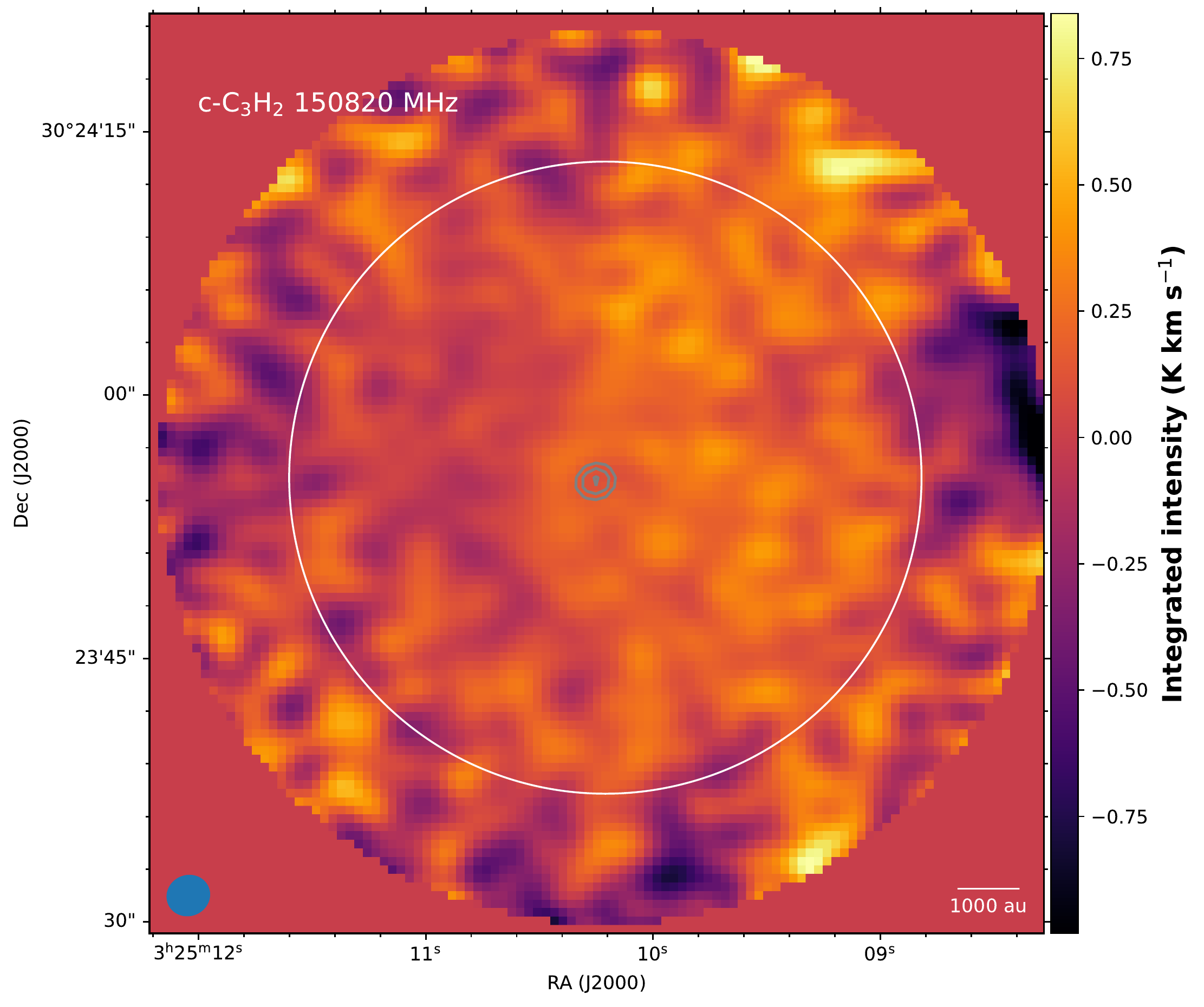}
\includegraphics[width=0.5\linewidth]{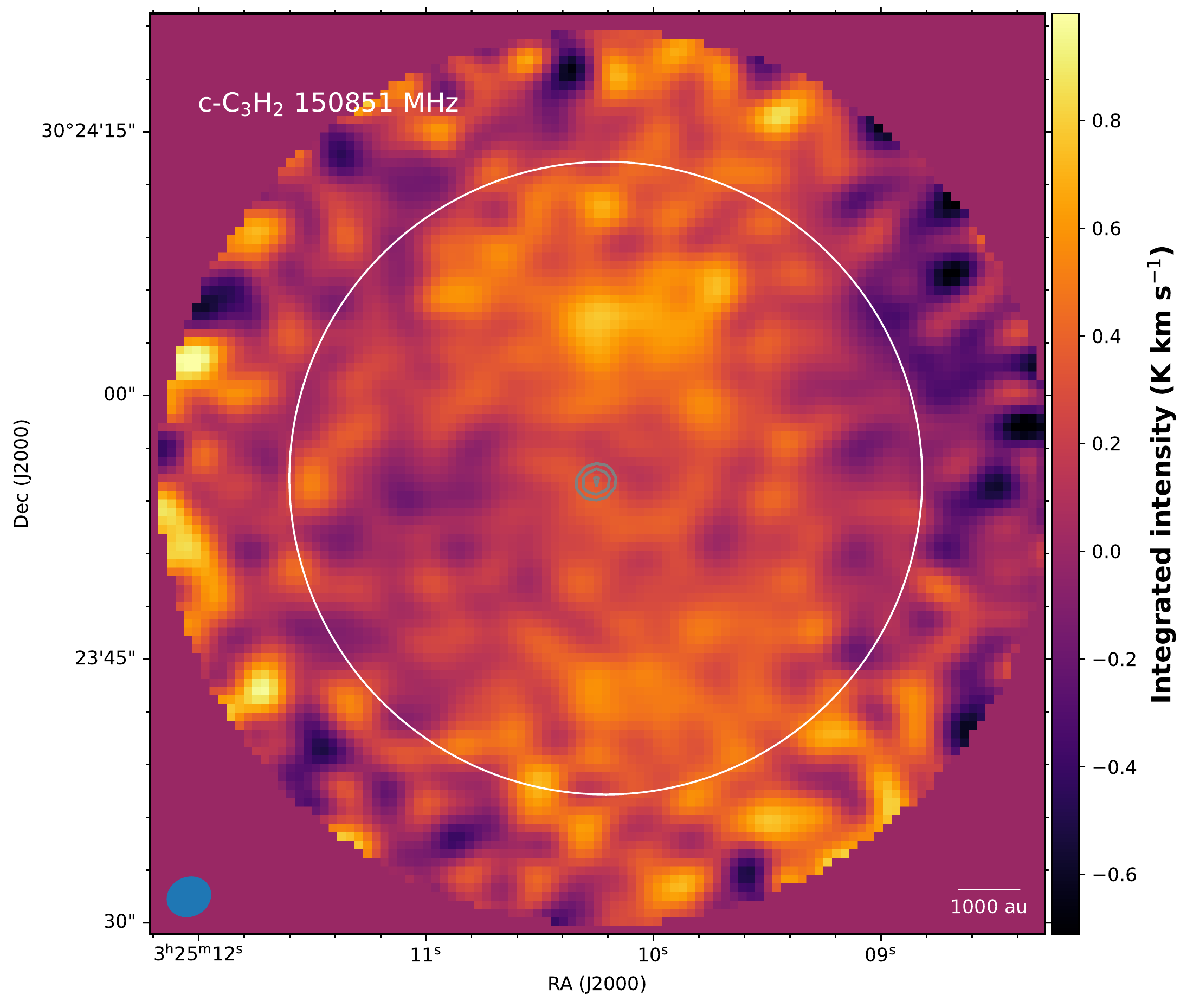}
\includegraphics[width=0.5\linewidth]{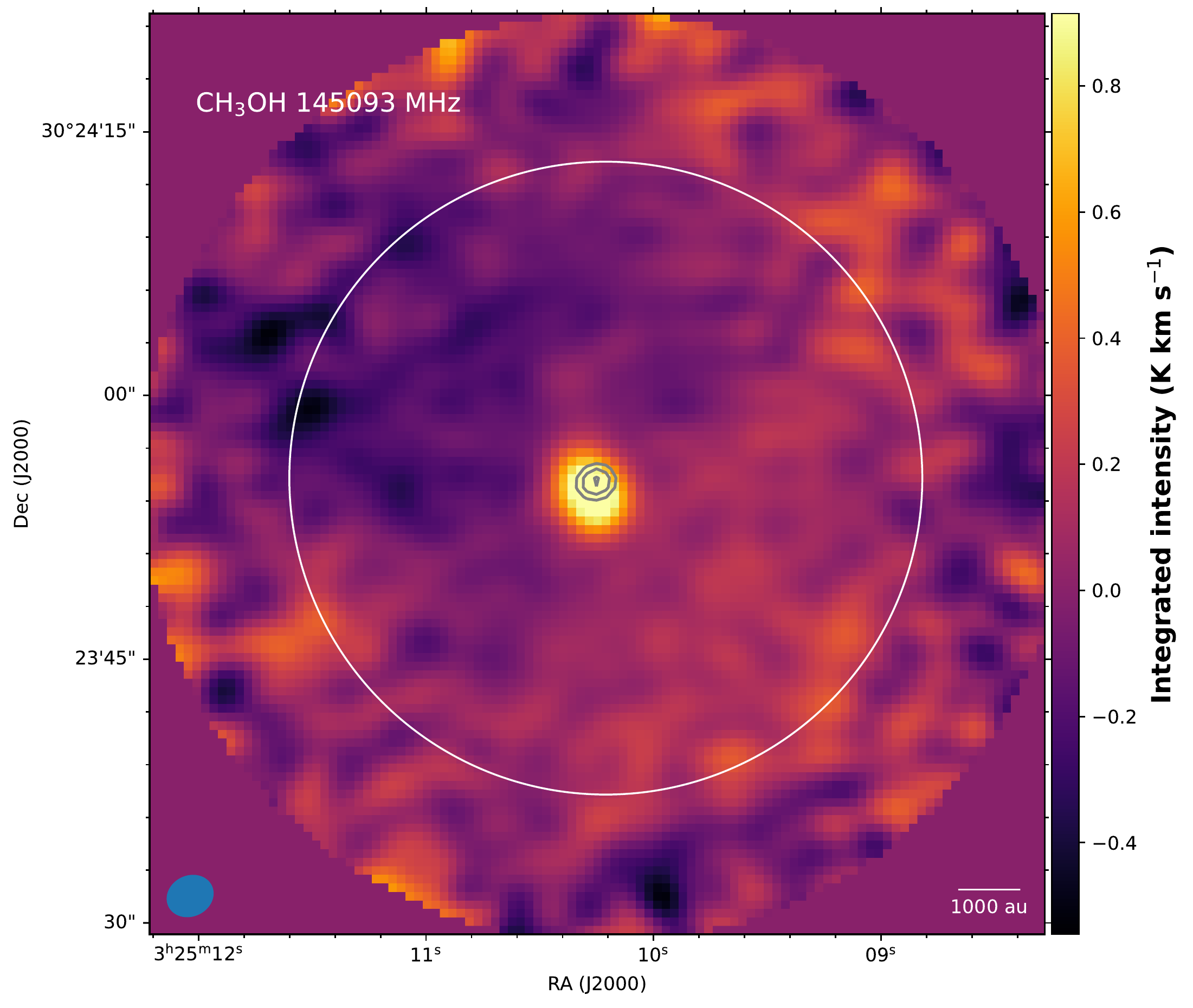}
\includegraphics[width=0.5\linewidth]{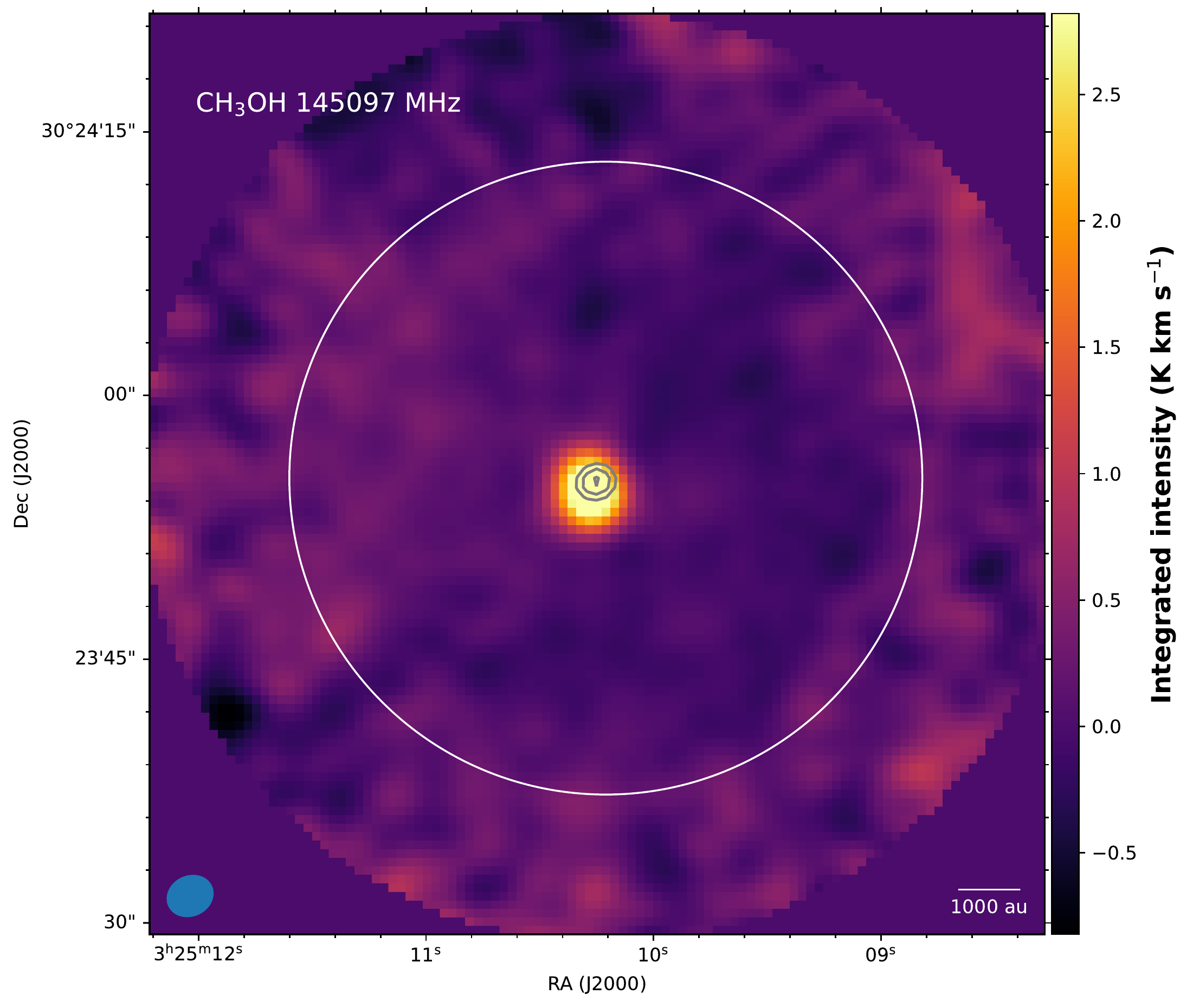}
\caption{Integrated intensity maps of detected lines (color scale) with the continuum flux (gray contours at 4.5, 5.5, and 6.5 mJy~beam$^{-1}$). The synthetic beam is indicated by the blue ellipses, and the primary beam is indicated by the large white circle. \label{moment_maps1}}
\end{figure*}

\begin{figure*}
\includegraphics[width=0.5\linewidth]{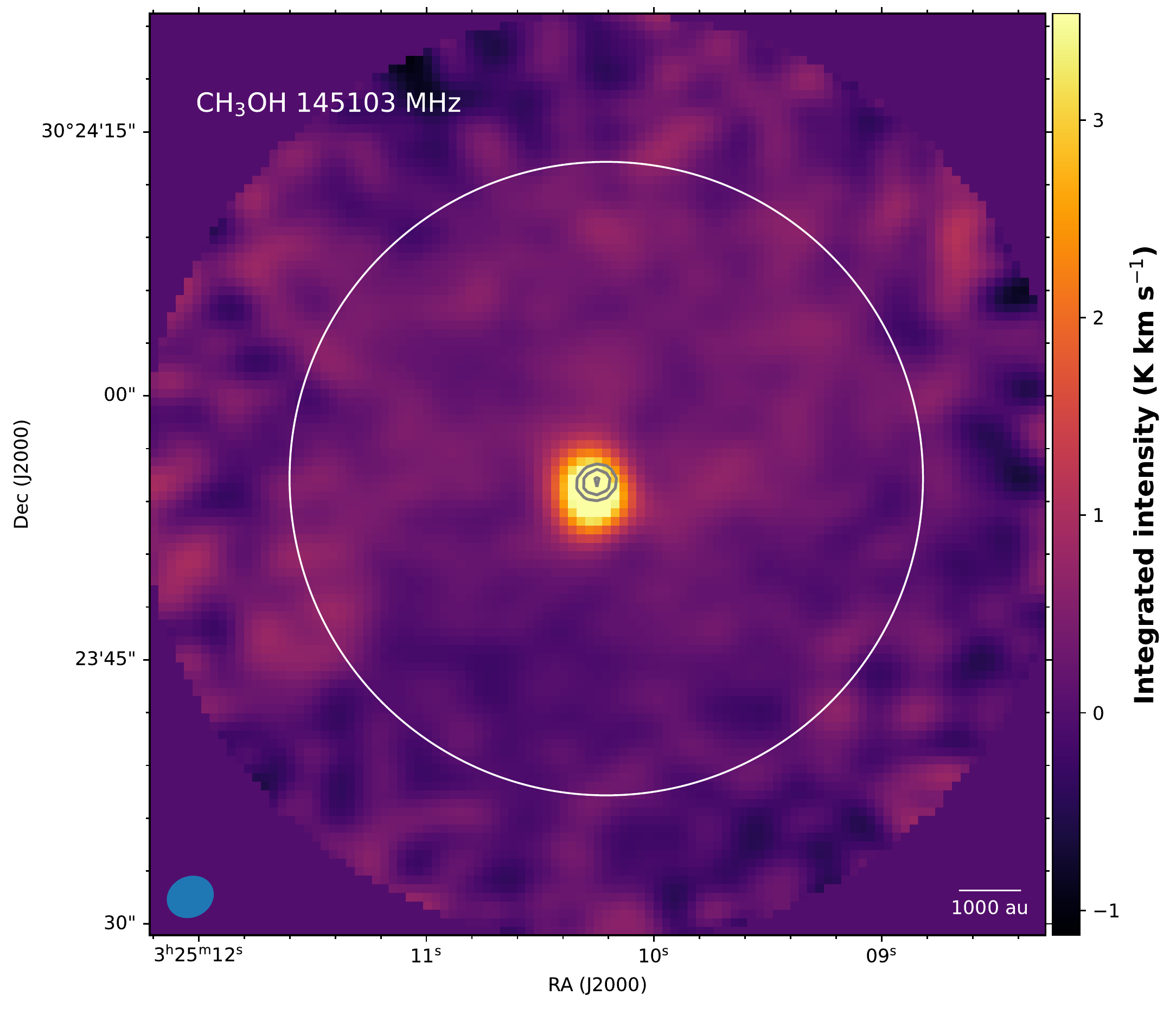}
\includegraphics[width=0.5\linewidth]{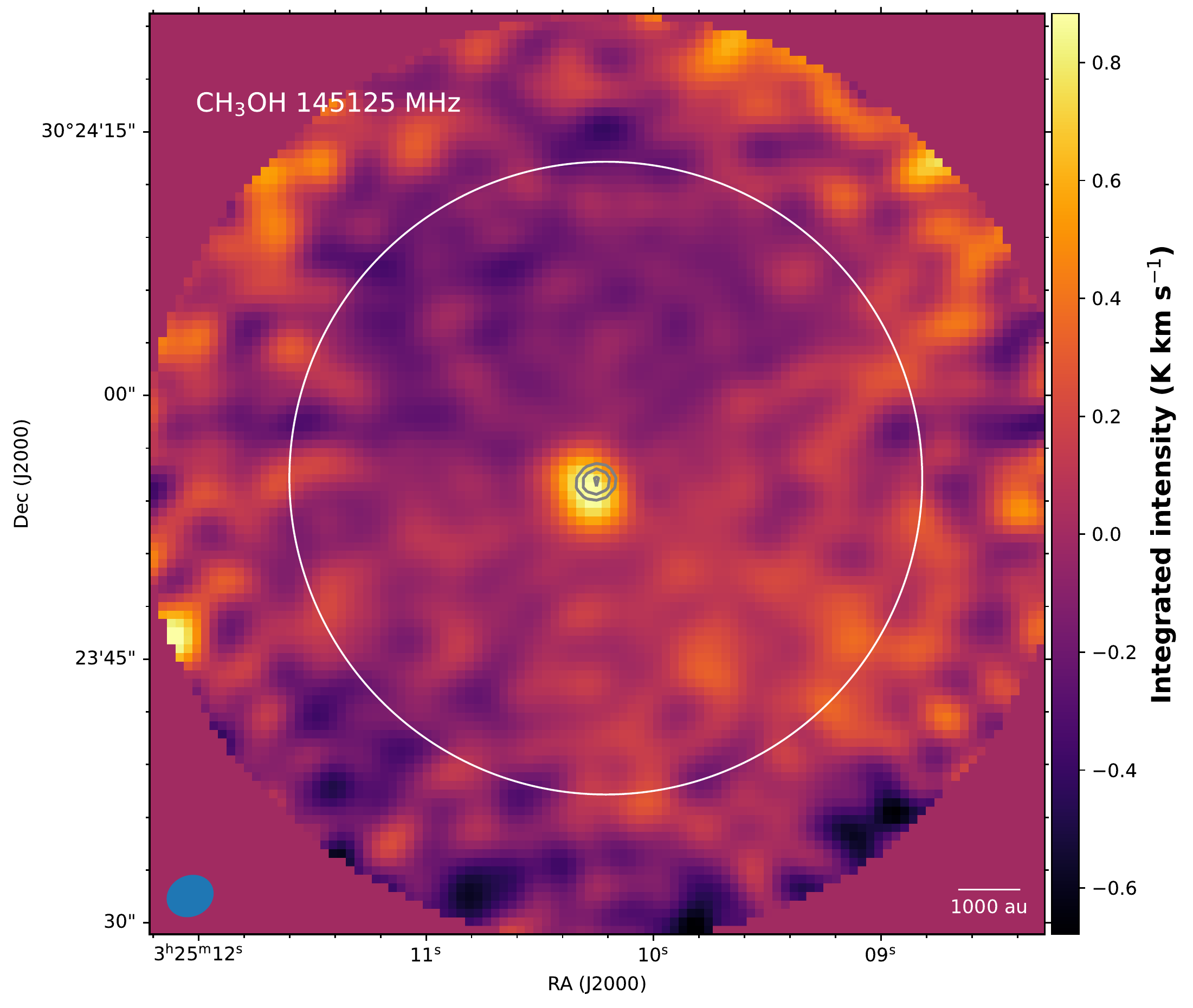}
\includegraphics[width=0.5\linewidth]{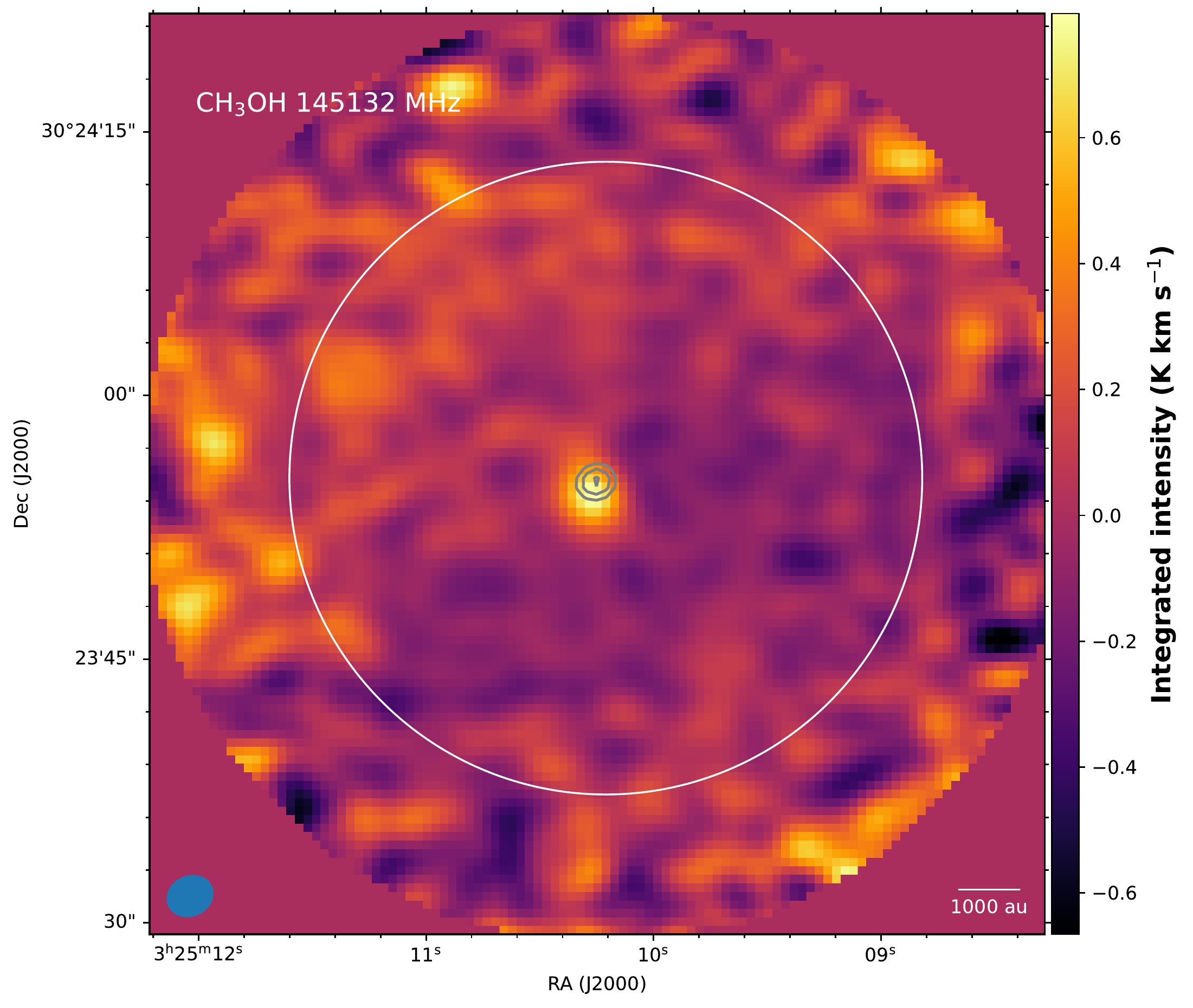}
\includegraphics[width=0.5\linewidth]{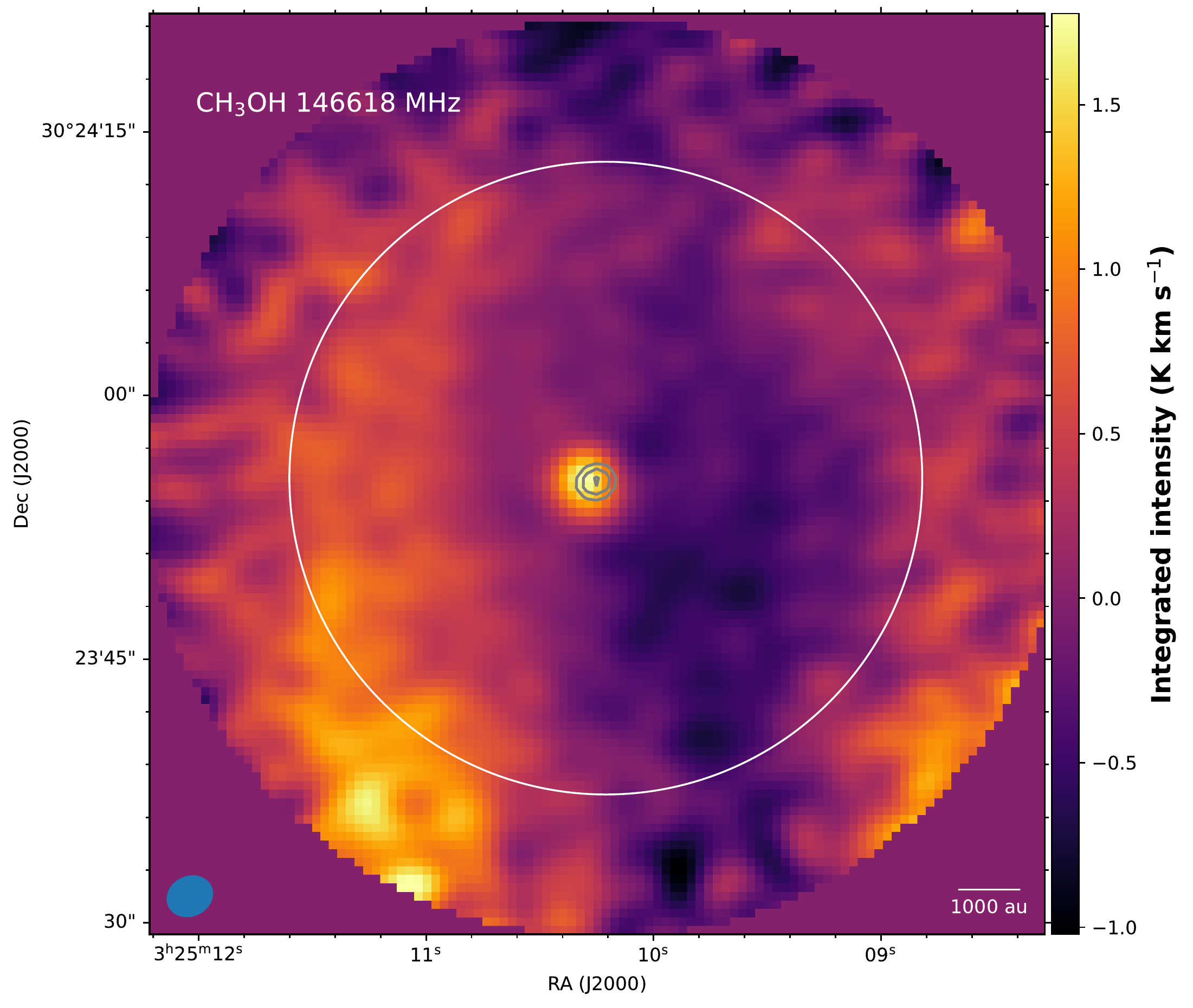}
\includegraphics[width=0.5\linewidth]{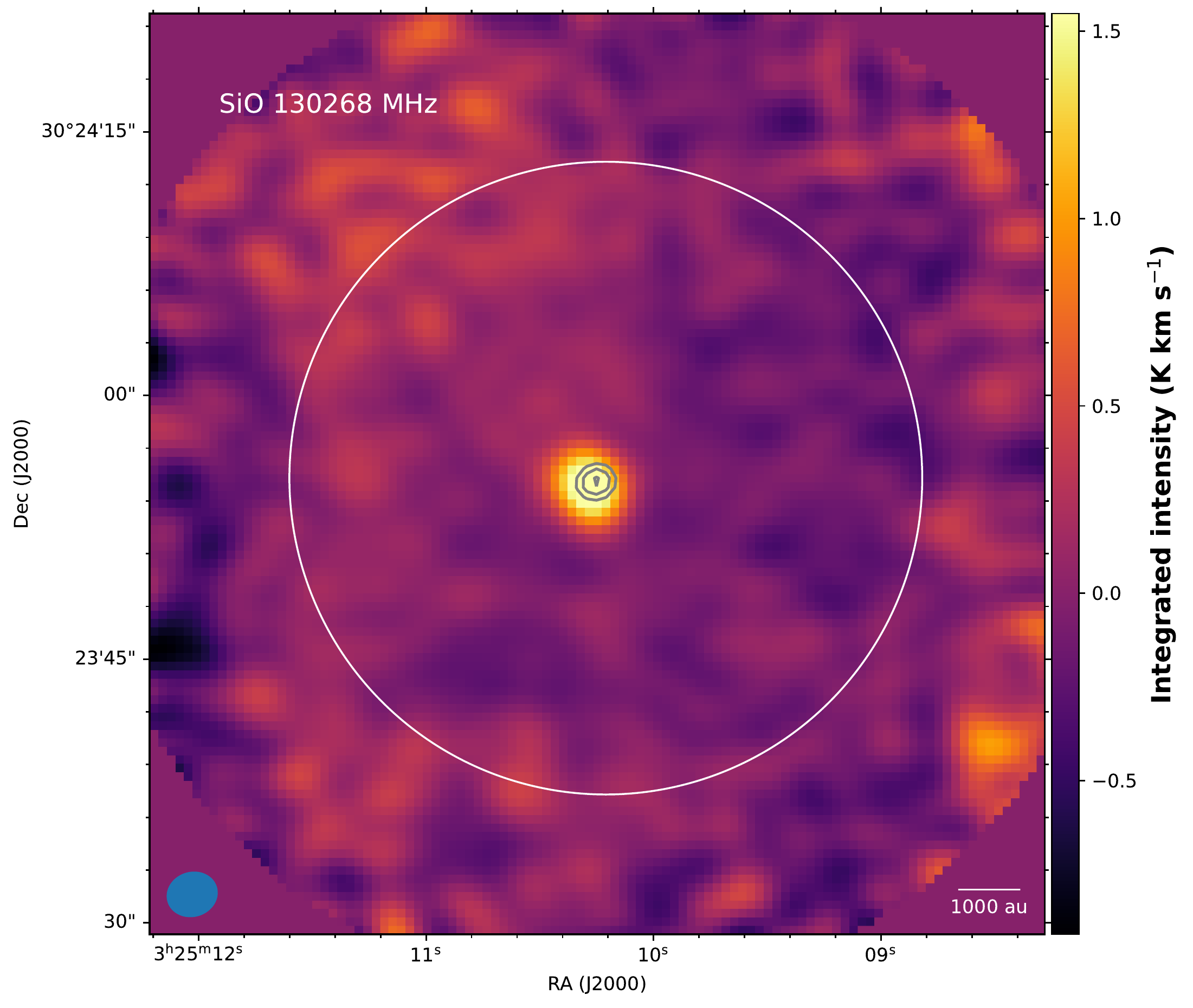}
\includegraphics[width=0.5\linewidth]{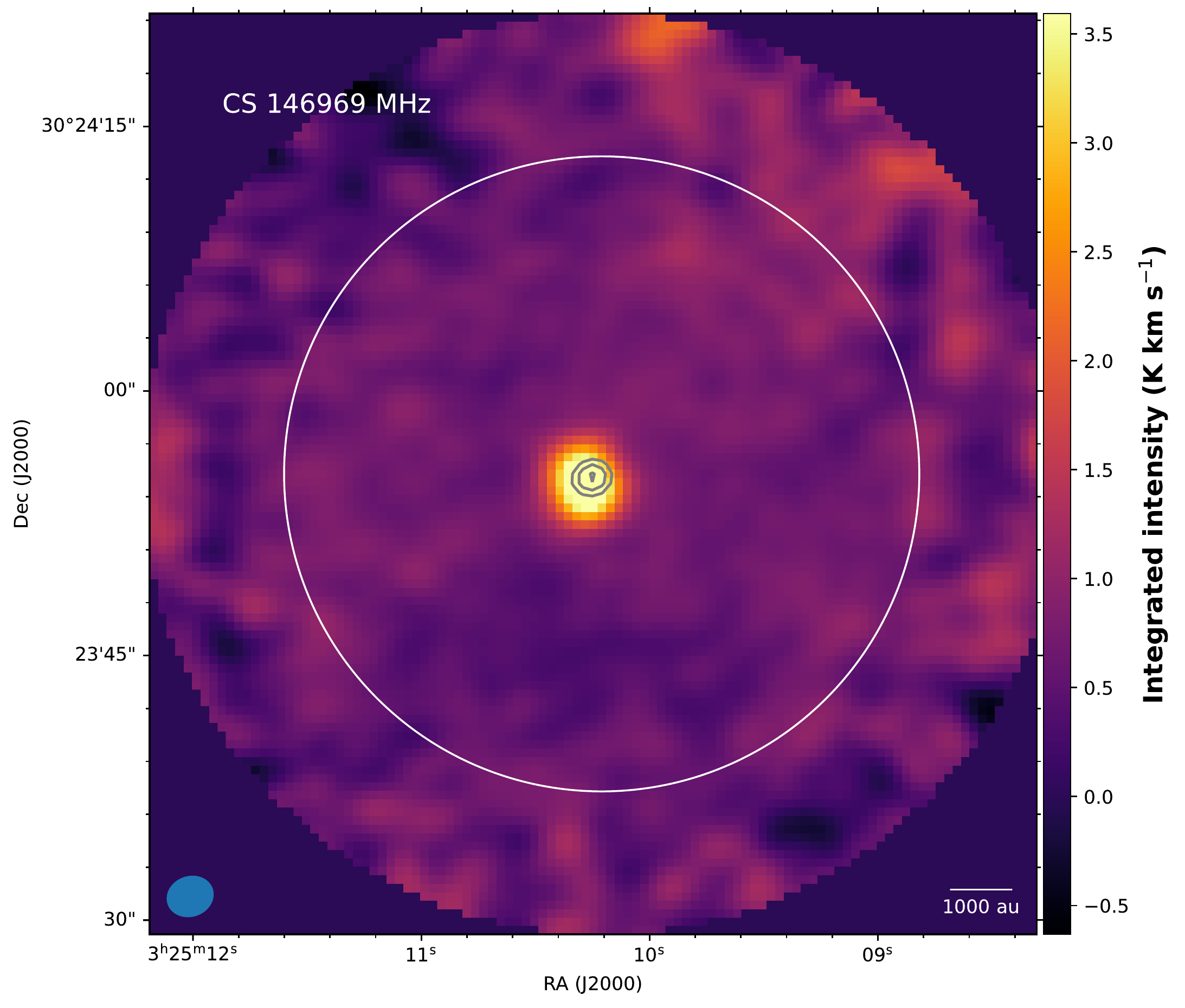}
\caption{Integrated intensity maps of detected lines (color scale) with the continuum flux (gray contours at 4.5, 5.5, and 6.5 mJy~beam$^{-1}$). The synthetic beam is indicated by the blue ellipses, and the primary beam is indicated by the large white circle. \label{moment_maps2}}
\end{figure*}

\begin{figure*}
\includegraphics[width=0.5\linewidth]{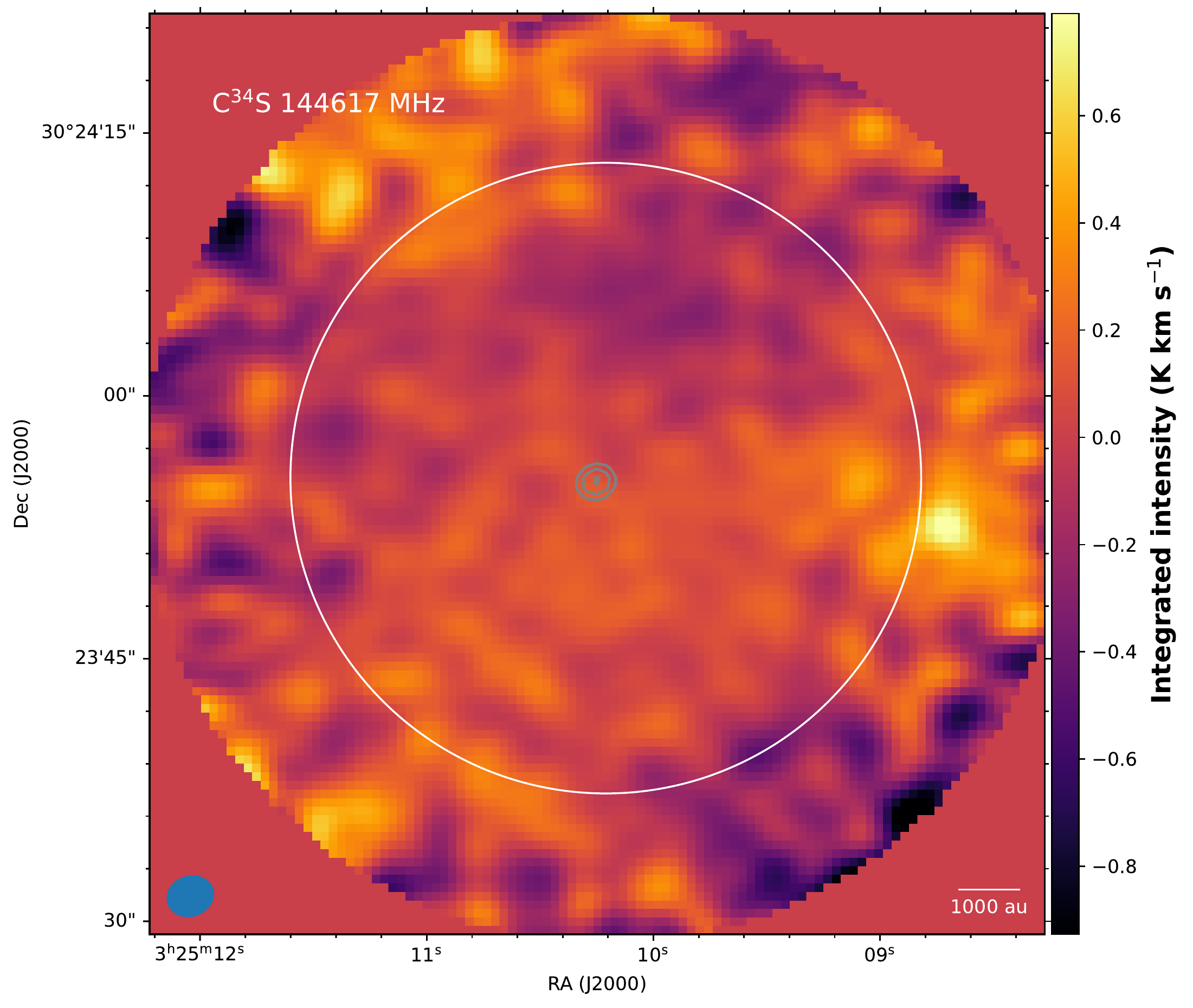}
\includegraphics[width=0.5\linewidth]{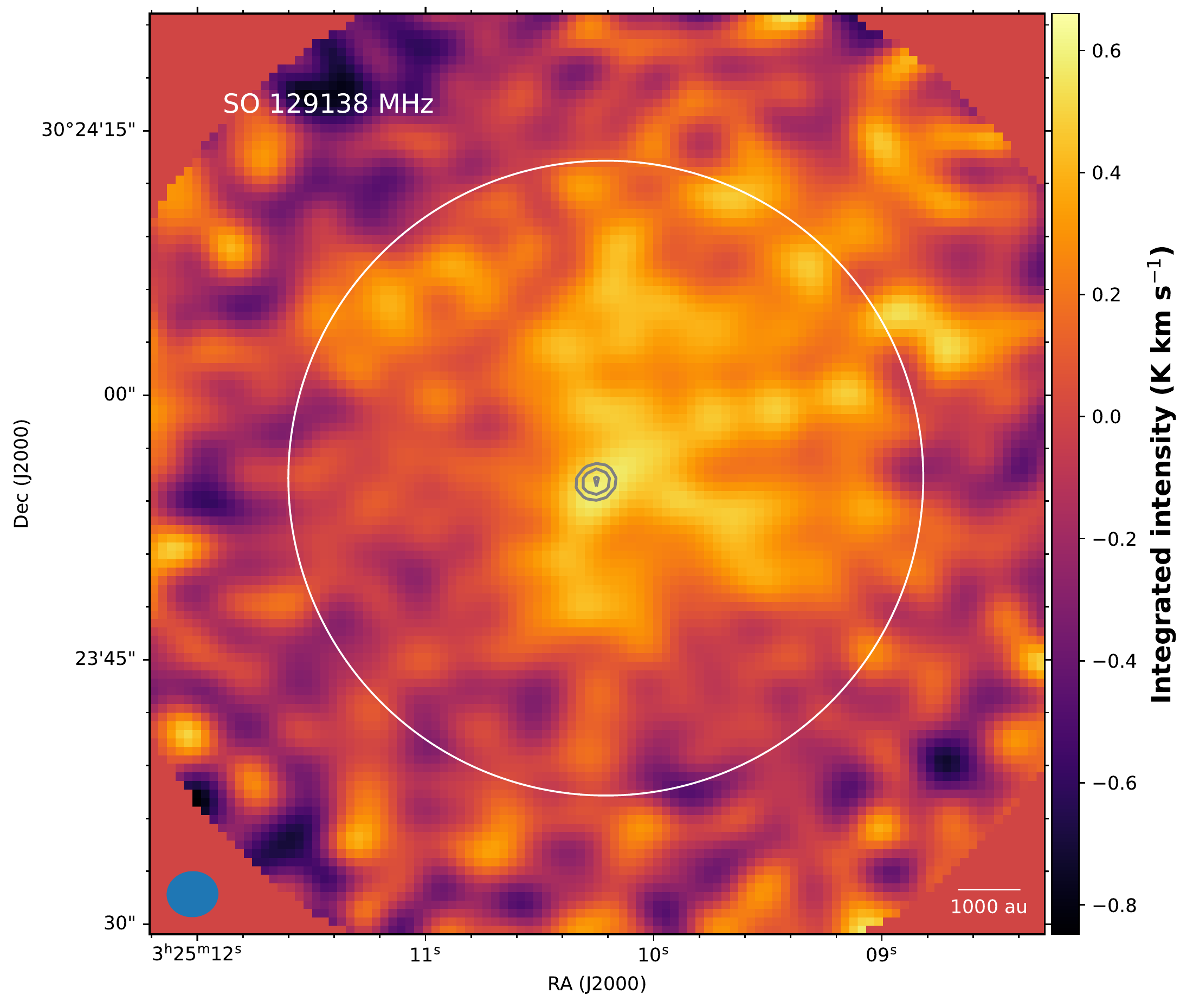}
\includegraphics[width=0.5\linewidth]{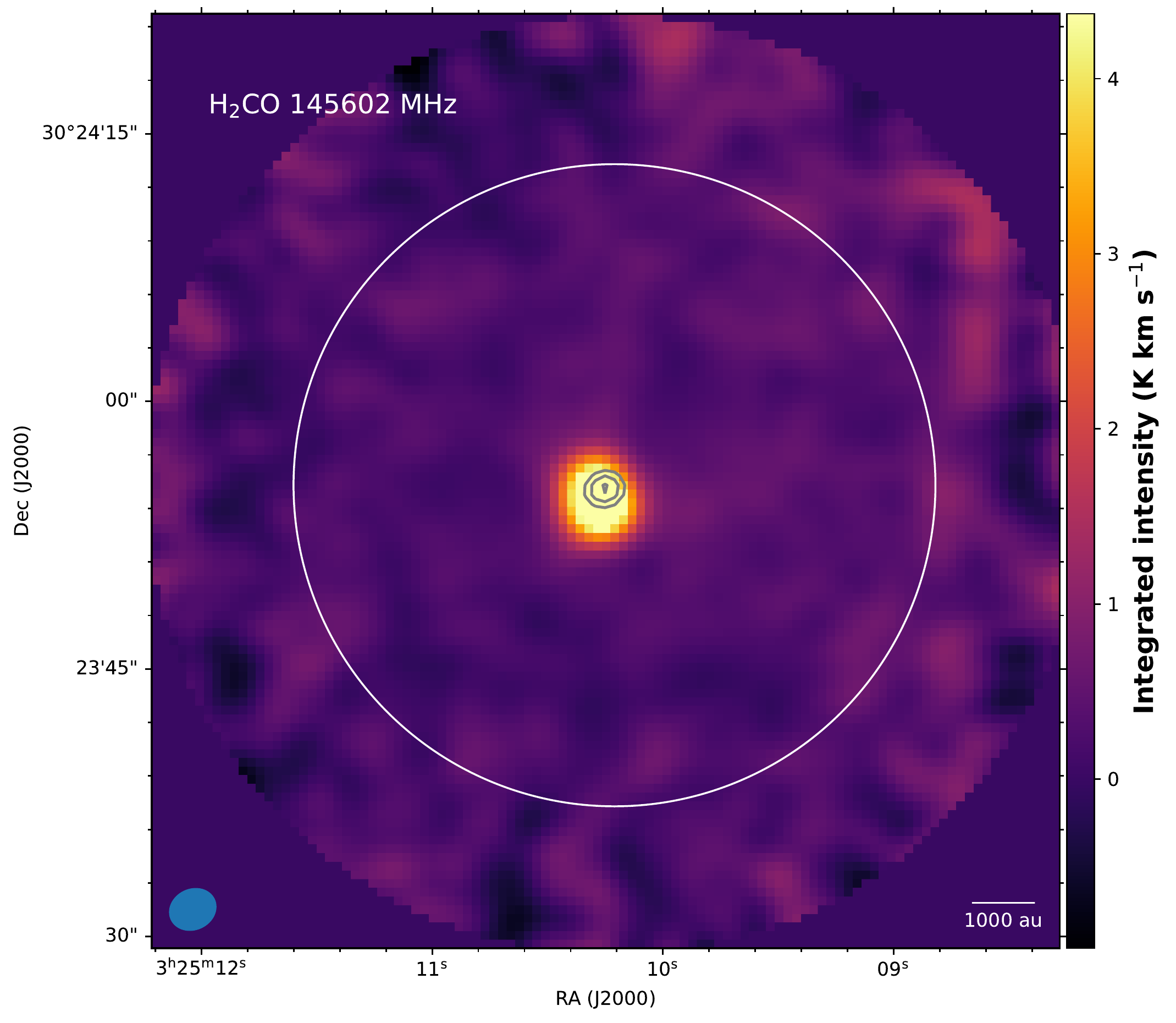}
\includegraphics[width=0.5\linewidth]{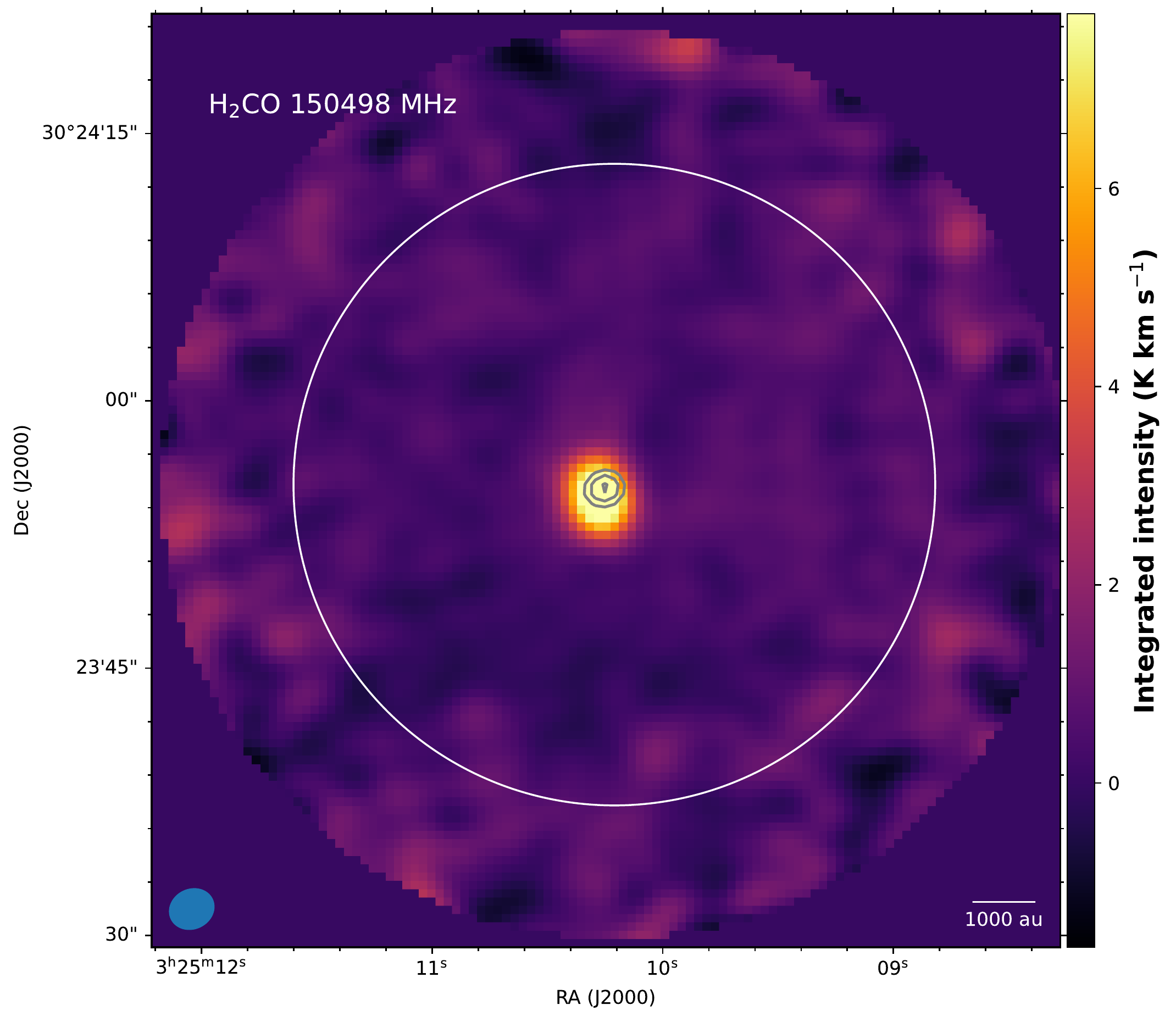}
\includegraphics[width=0.5\linewidth]{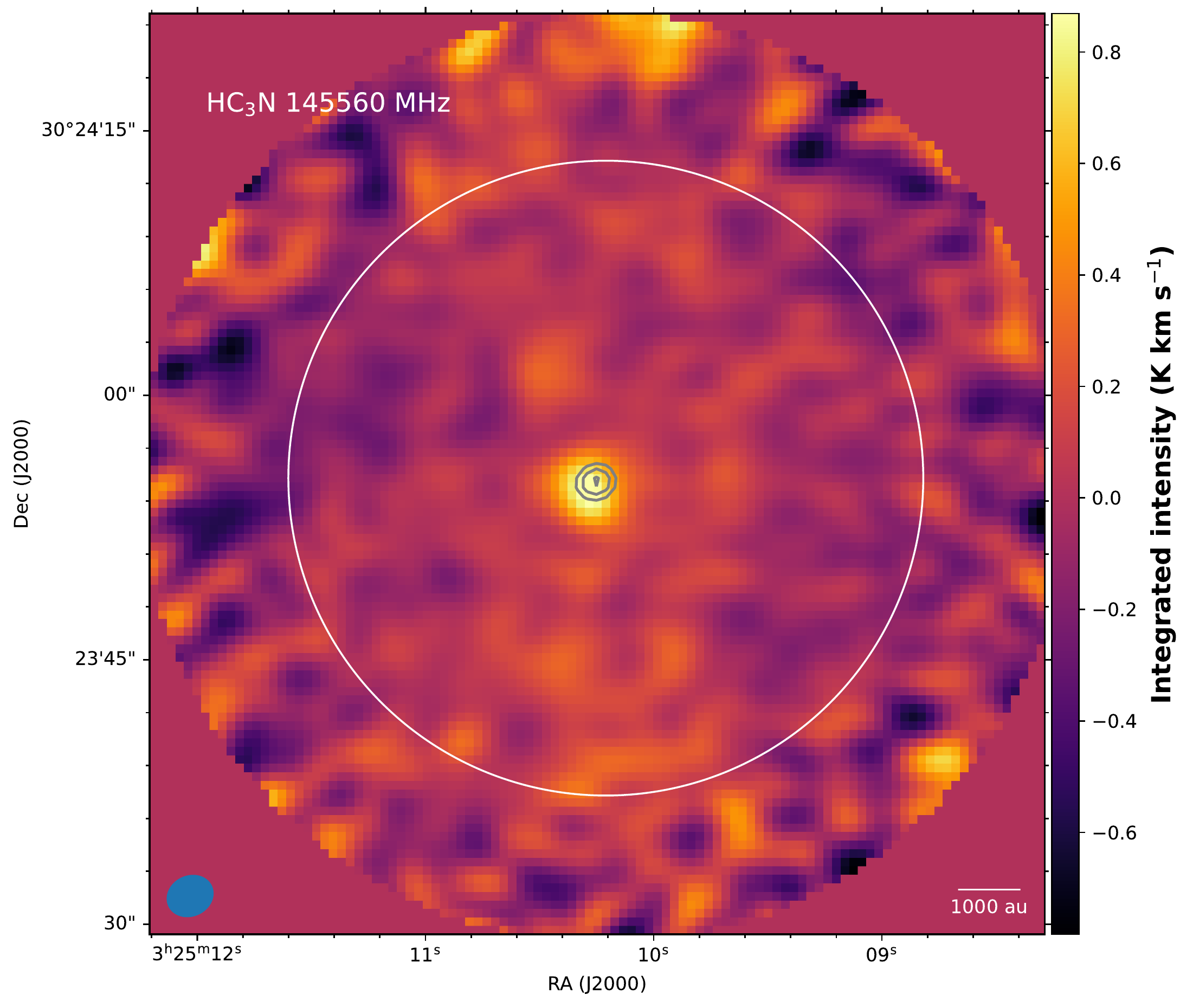}
\includegraphics[width=0.5\linewidth]{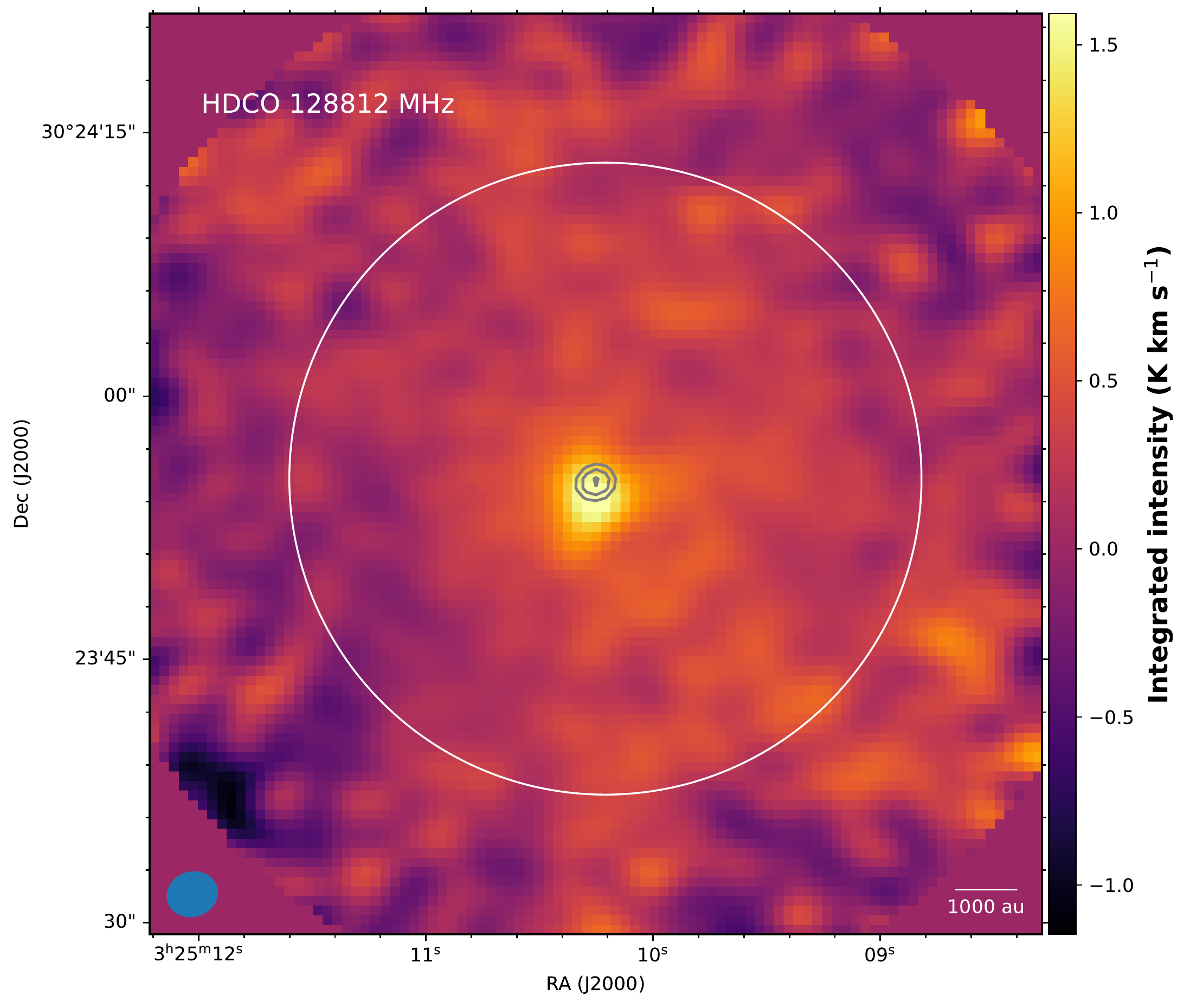}
\caption{Integrated intensity maps of detected lines (color scale) with the continuum flux (gray contours at 4.5, 5.5, and 6.5 mJy~beam$^{-1}$). The synthetic beam is indicated by the blue ellipses, and the primary beam is indicated by the large white circle. \label{moment_maps3}}
\end{figure*}

\begin{figure*}
\includegraphics[width=0.5\linewidth]{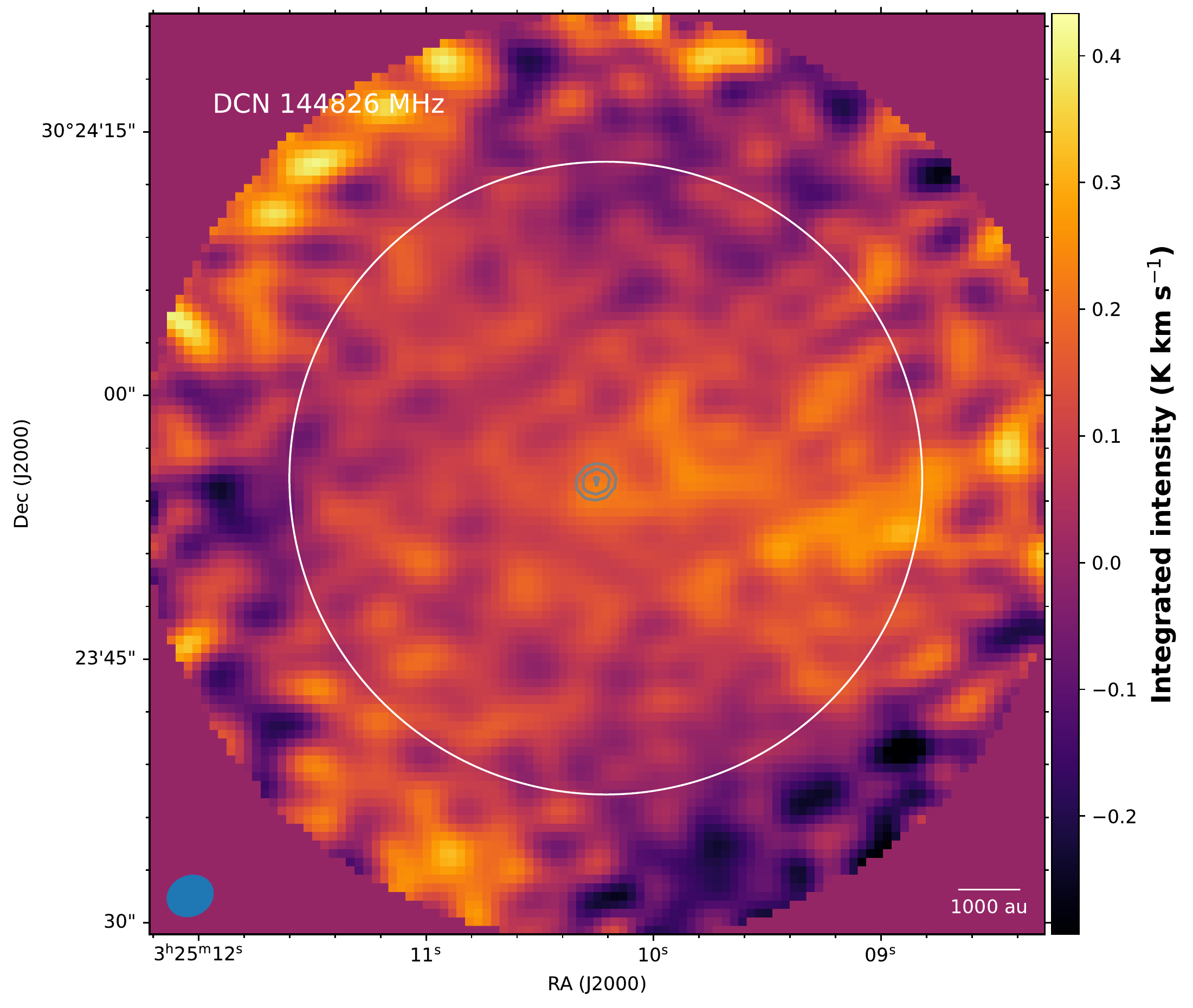}
\includegraphics[width=0.5\linewidth]{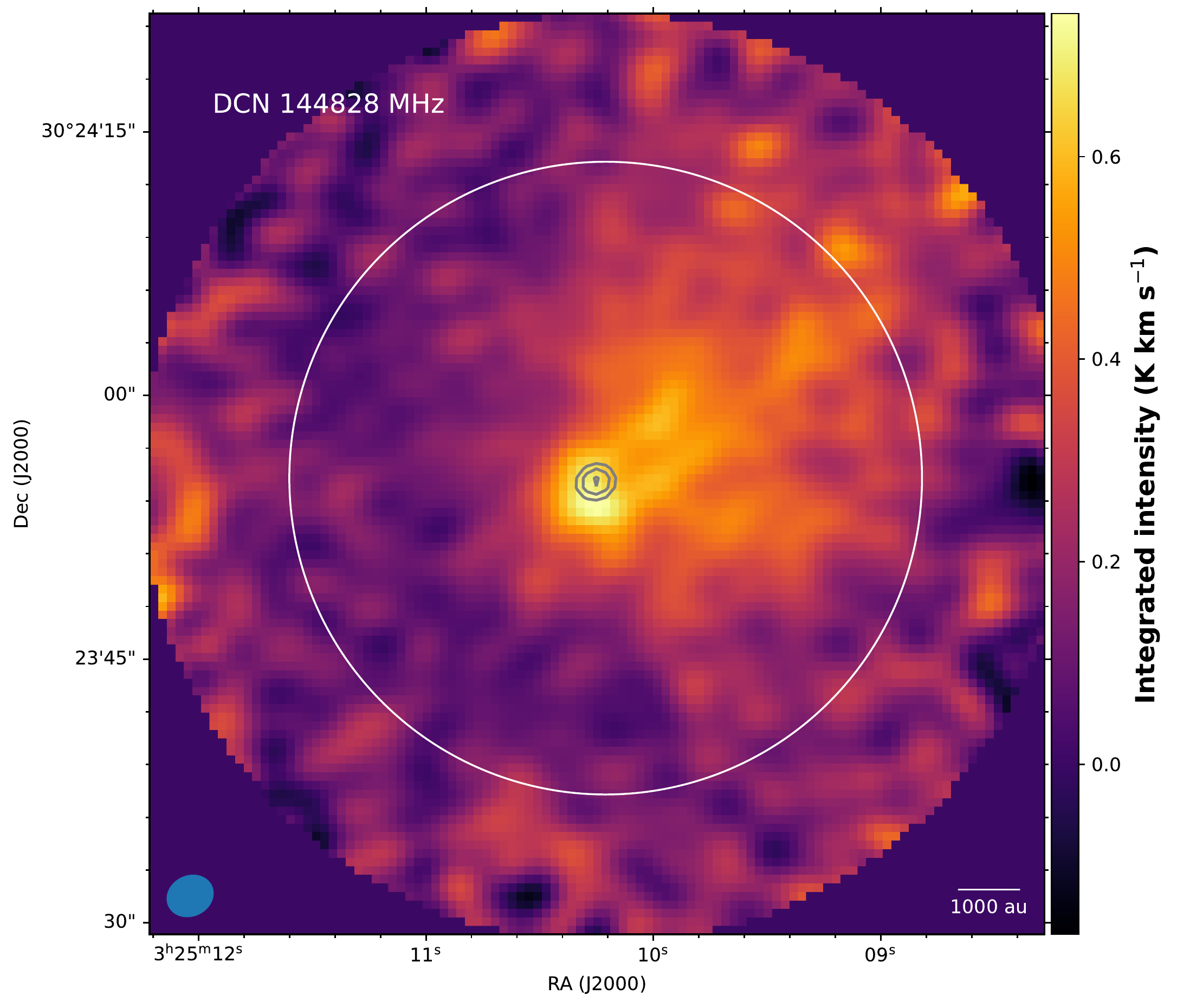}
\includegraphics[width=0.5\linewidth]{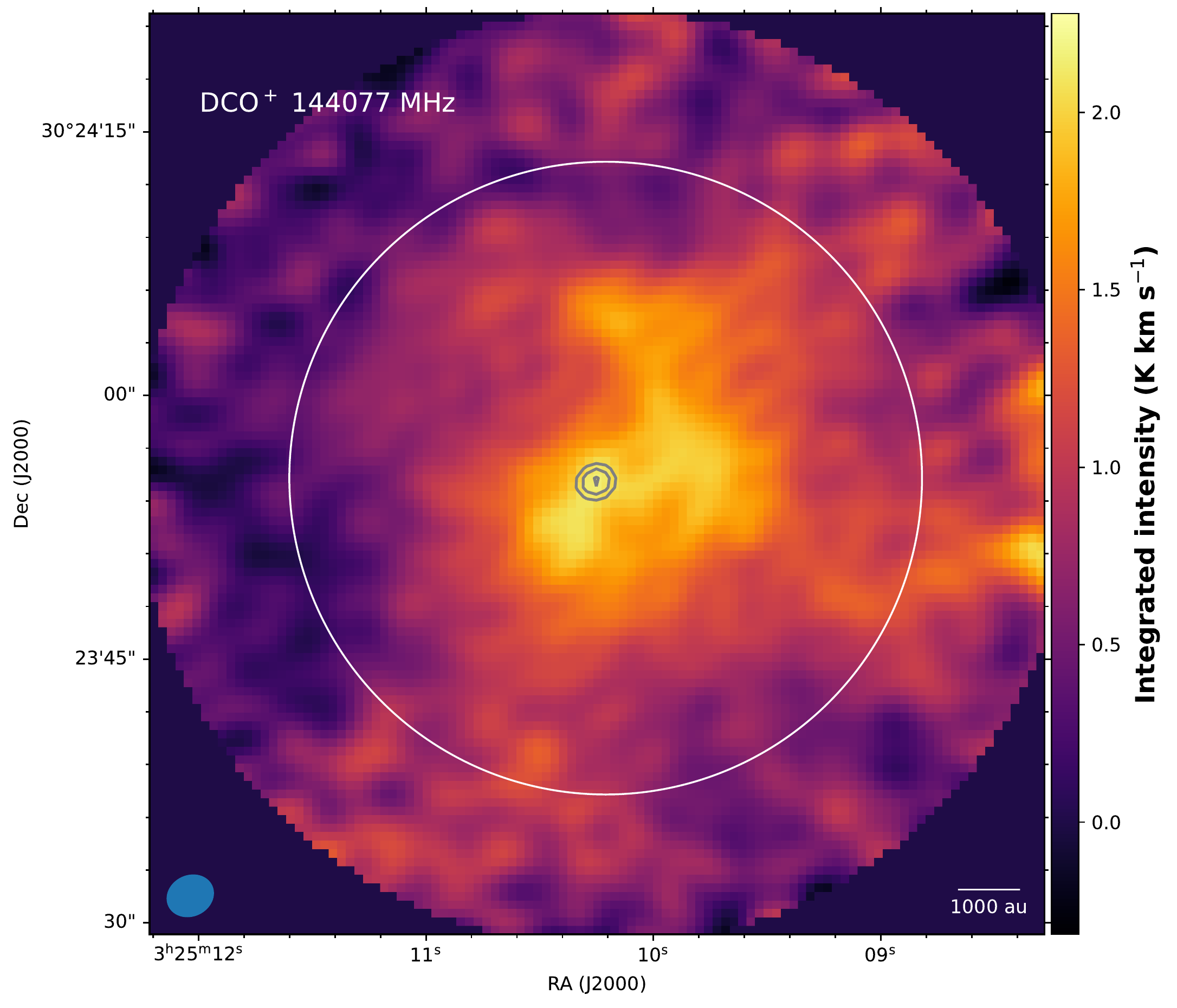}
\caption{Integrated intensity maps of detected lines (color scale) with the continuum flux (gray contours at 4.5, 5.5, and 6.5 mJy~beam$^{-1}$). The synthetic beam is indicated by the blue ellipses, and the primary beam is indicated by the large white circle. \label{moment_maps4} }
\end{figure*}

The integrated intensity maps of the detected lines were computed with SpectralCube python package \citep[within the Astropy project;][]{astropy:2013,astropy:2018}. A mask was applied on the data to remove signal below the 3 rms limit on each point. The resulting maps are shown in Figs.~\ref{moment_maps1} to \ref{moment_maps4} (color scale) with the continuum emission in gray contours. 

\clearpage
\section{Velocity channel maps}\label{channel_maps}

 This appendix presents the velocity channel maps of the NOEMA+30m data obtained with the GILDAS software (MAPPING). The central plus indicates the position of the continuum maximum. When several lines of a specific molecule have been detected, we only show the brightest line.


\begin{figure}[h]
\includegraphics[width=1\linewidth]{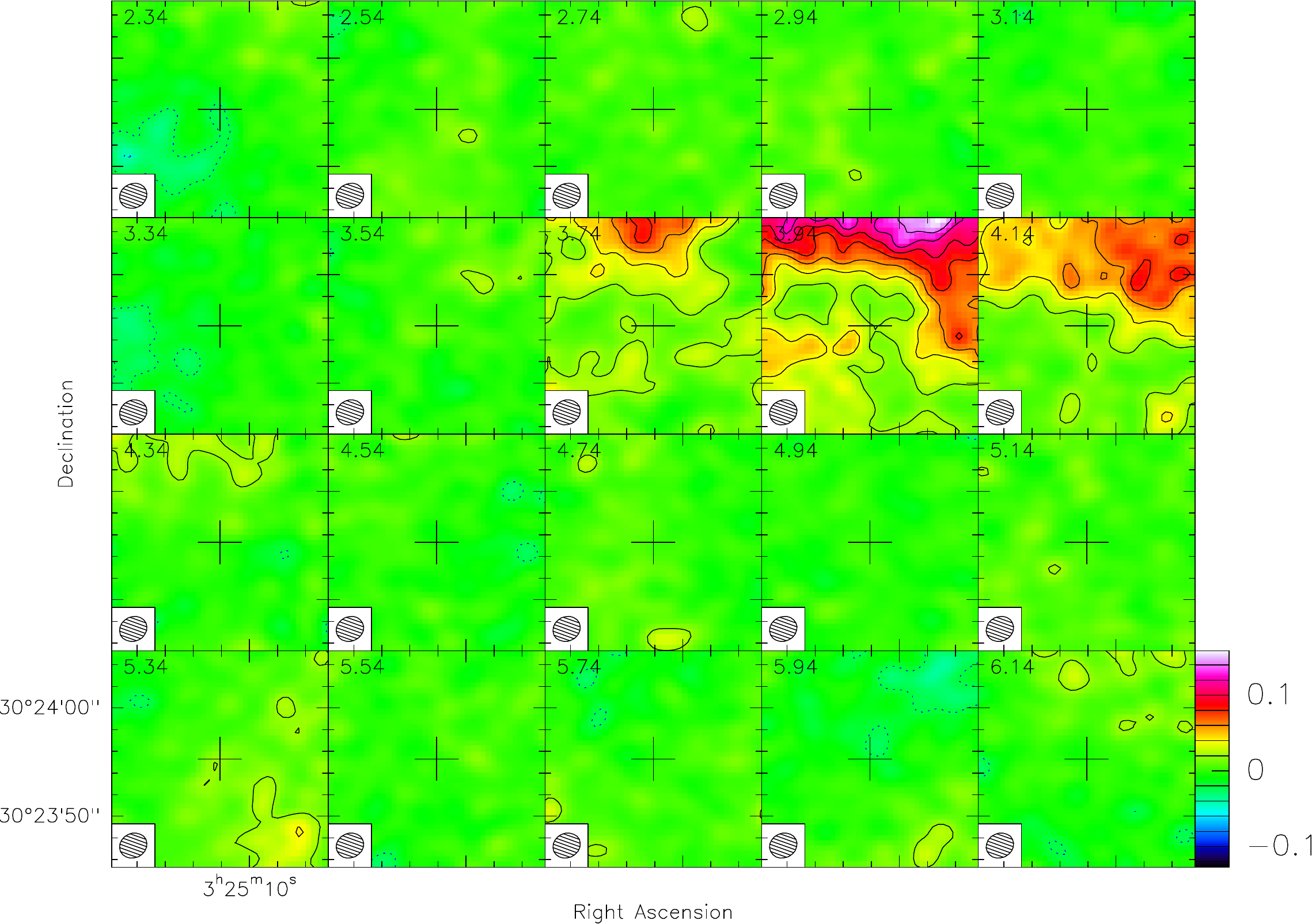}
\caption{Channel map for the c-C$_3$H$_2$ line at 150.851~GHz. Color levels are in mJy~beam$^{-1}$. \label{channel_maps_1}}
\end{figure}

\begin{figure}[h]
\includegraphics[width=1\linewidth]{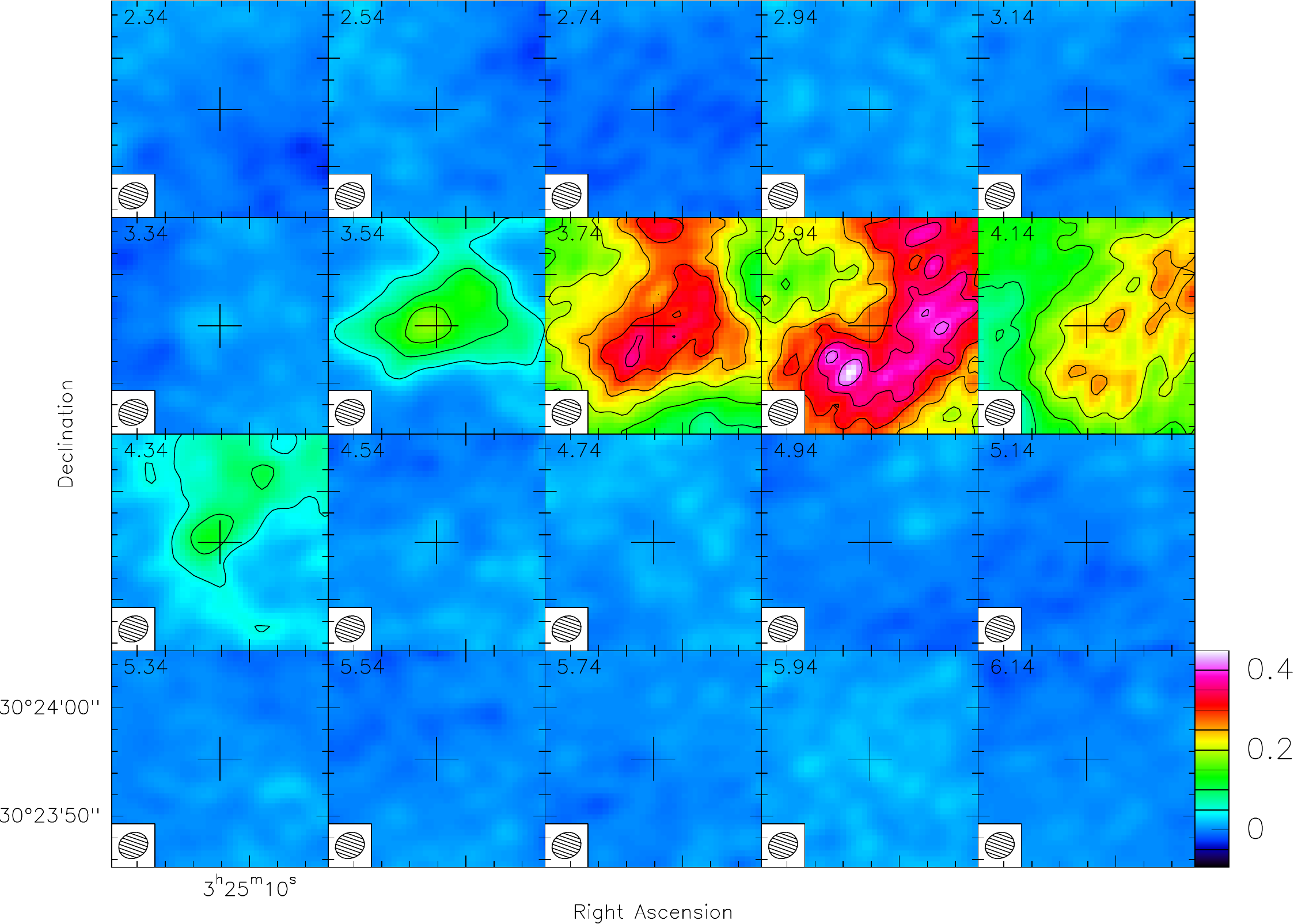}
\caption{Channel map for the DCO$^+$ line at 144.077~GHz. Color levels are in Jy~beam$^{-1}$.\label{channel_maps_2}}
\end{figure}

\begin{figure}[h]
\includegraphics[width=1\linewidth]{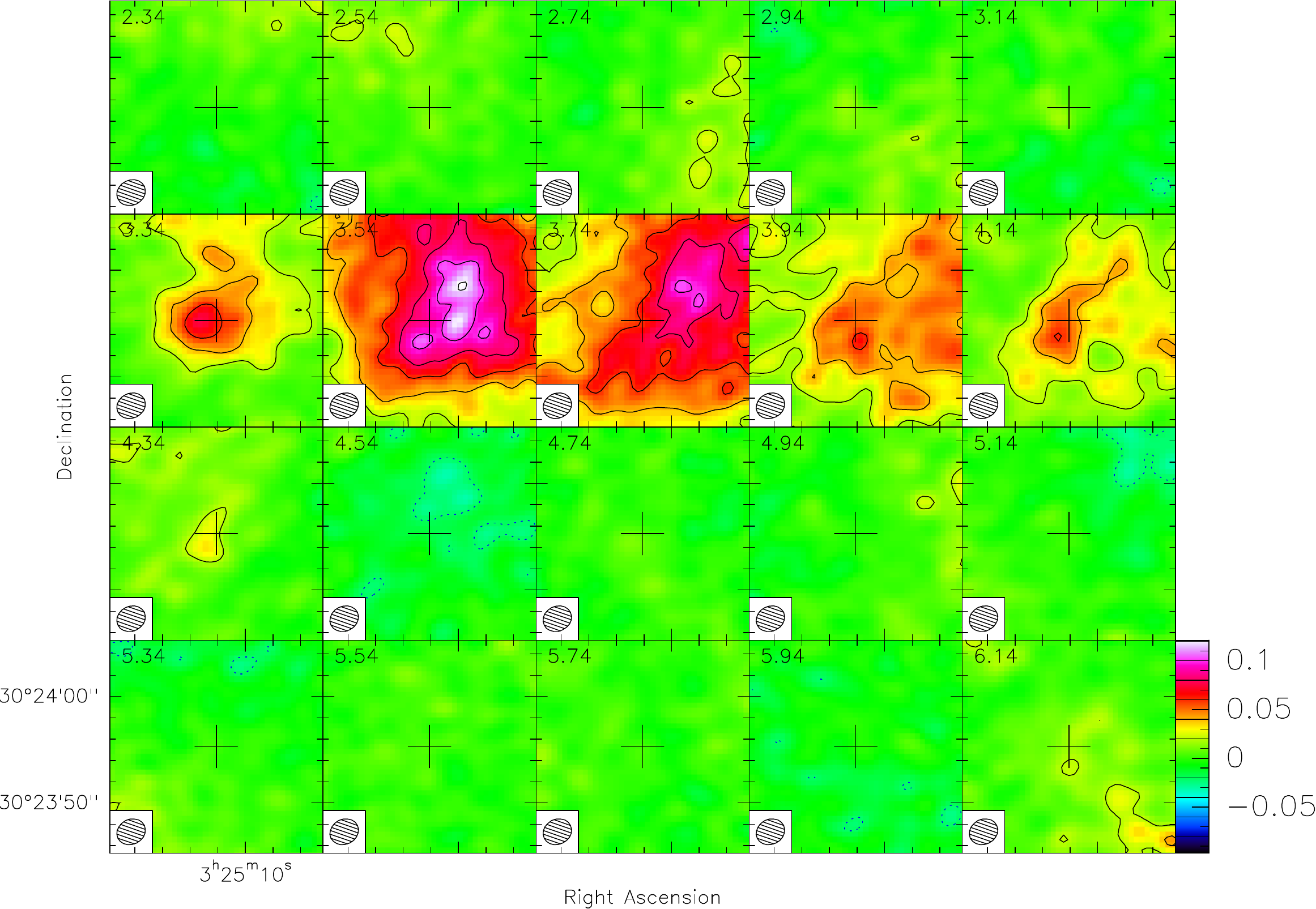}
\caption{Channel map for the DCN line at 144.828~GHz. Color levels are in Jy~beam$^{-1}$.\label{channel_maps_3}}
\end{figure}

\begin{figure}[h]
\includegraphics[width=1\linewidth]{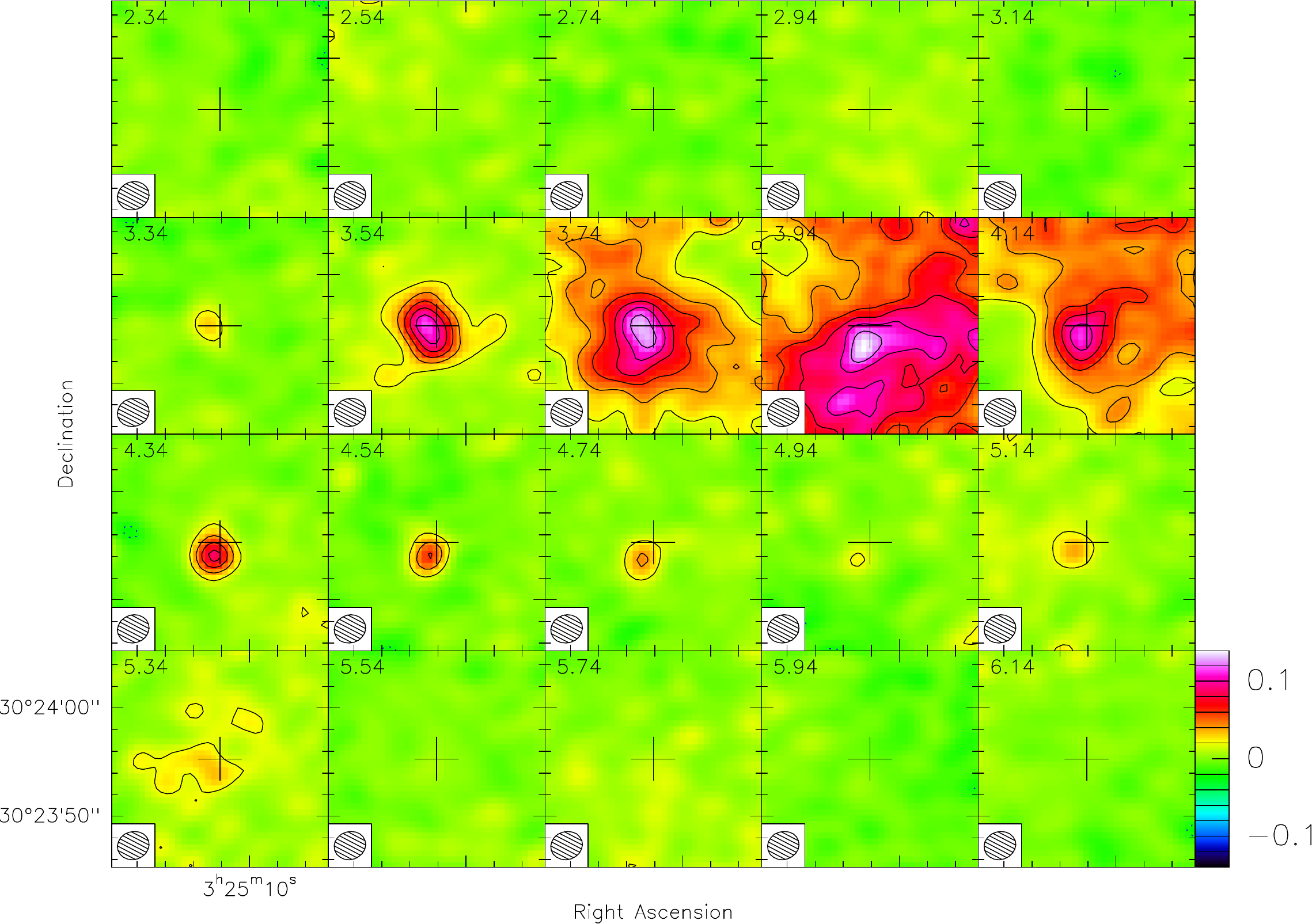}
\caption{Channel map for the HDCO line at 128.812~GHz. Color levels are in Jy~beam$^{-1}$.\label{channel_maps_4}}
\end{figure}

\begin{figure}[h]
\includegraphics[width=1\linewidth]{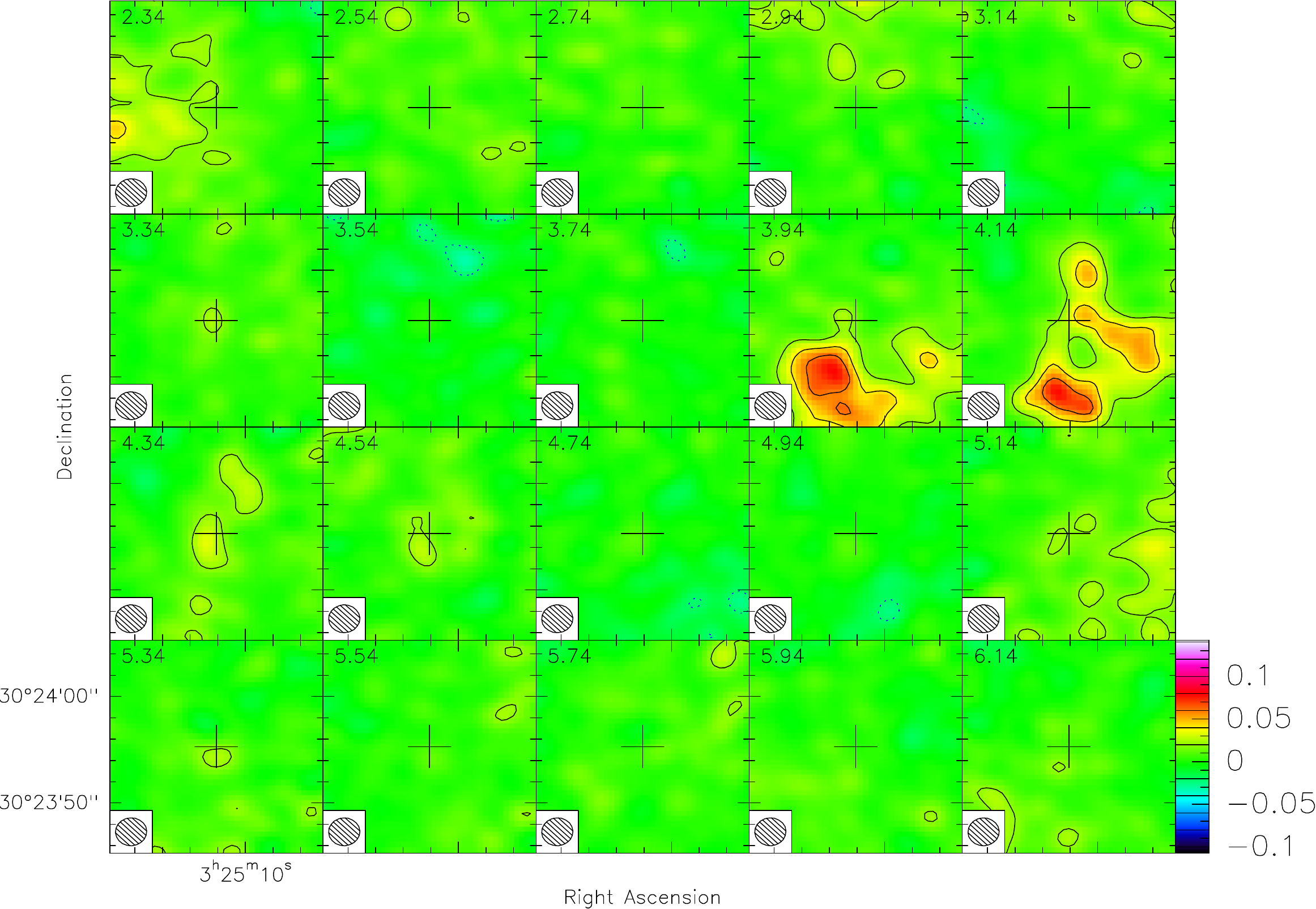}
\caption{Channel map for the SO line at 129.138~GHz. Color levels are in Jy~beam$^{-1}$.\label{channel_maps_5}}
\end{figure}

\begin{figure}[h]
\includegraphics[width=1\linewidth]{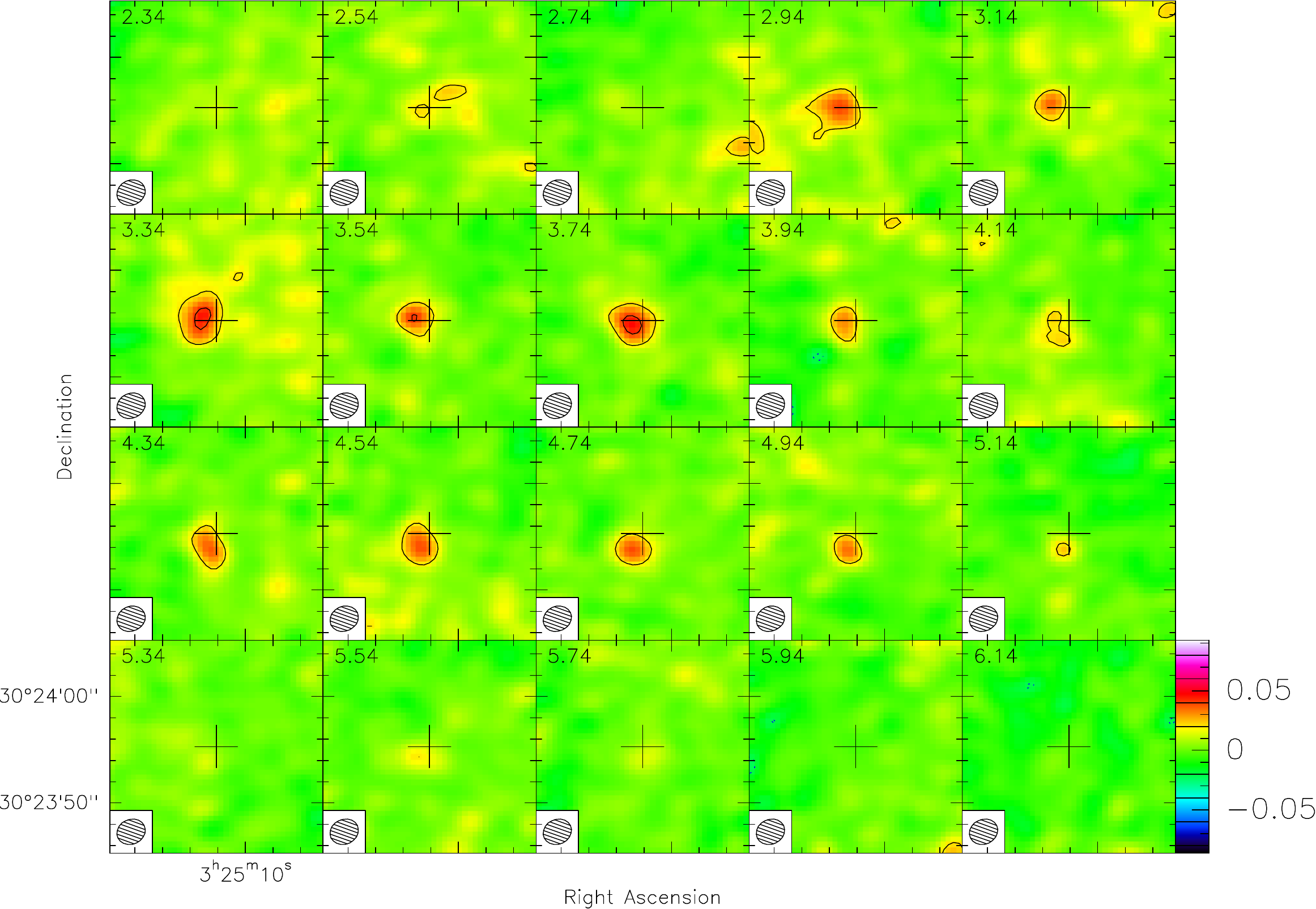}
\caption{Channel map for the HC$_3$N line at 145.560~GHz. Color levels are in Jy~beam$^{-1}$.\label{channel_maps_6}}
\end{figure}

\begin{figure}[h]
\includegraphics[width=1\linewidth]{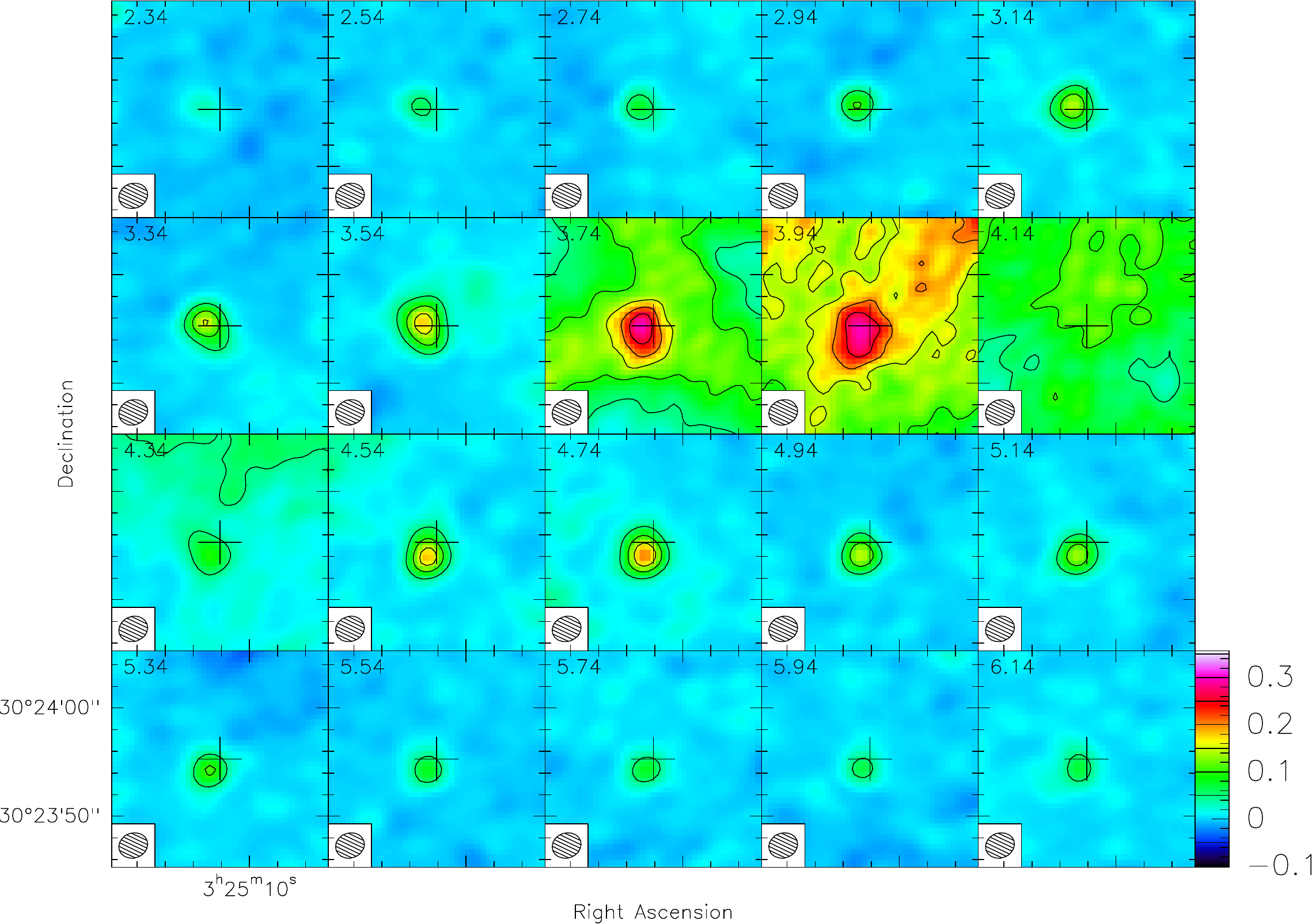}
\caption{Channel map for the CS line at 146.969~GHz. Color levels are in Jy~beam$^{-1}$.\label{channel_maps_7}}
\end{figure}

\begin{figure}[h]
\includegraphics[width=1\linewidth]{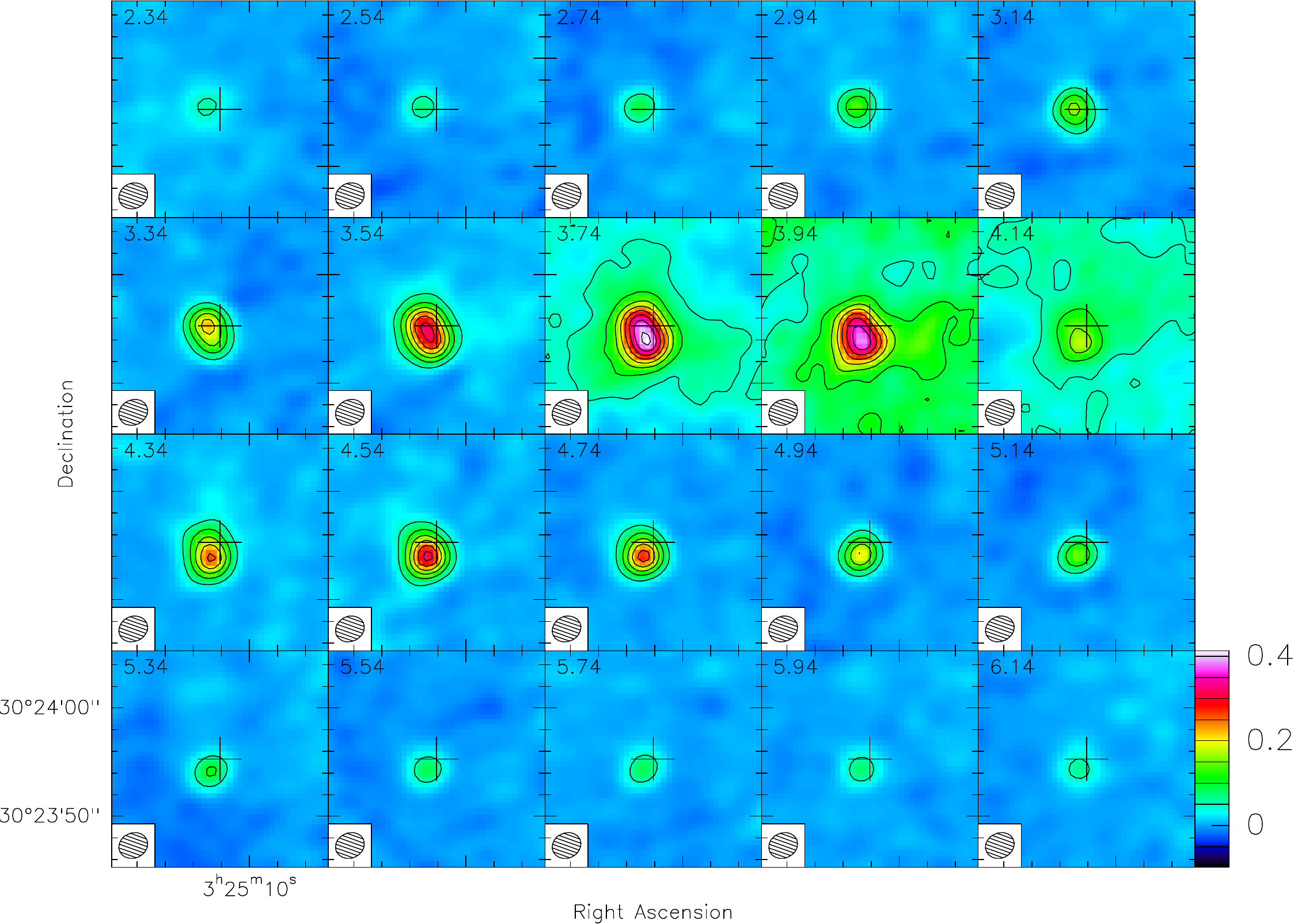}
\caption{Channel map for the H$_2$CO line at 145.602~GHz. Color levels are in Jy~beam$^{-1}$.\label{channel_maps_8}}
\end{figure}

\begin{figure}[h]
\includegraphics[width=1\linewidth]{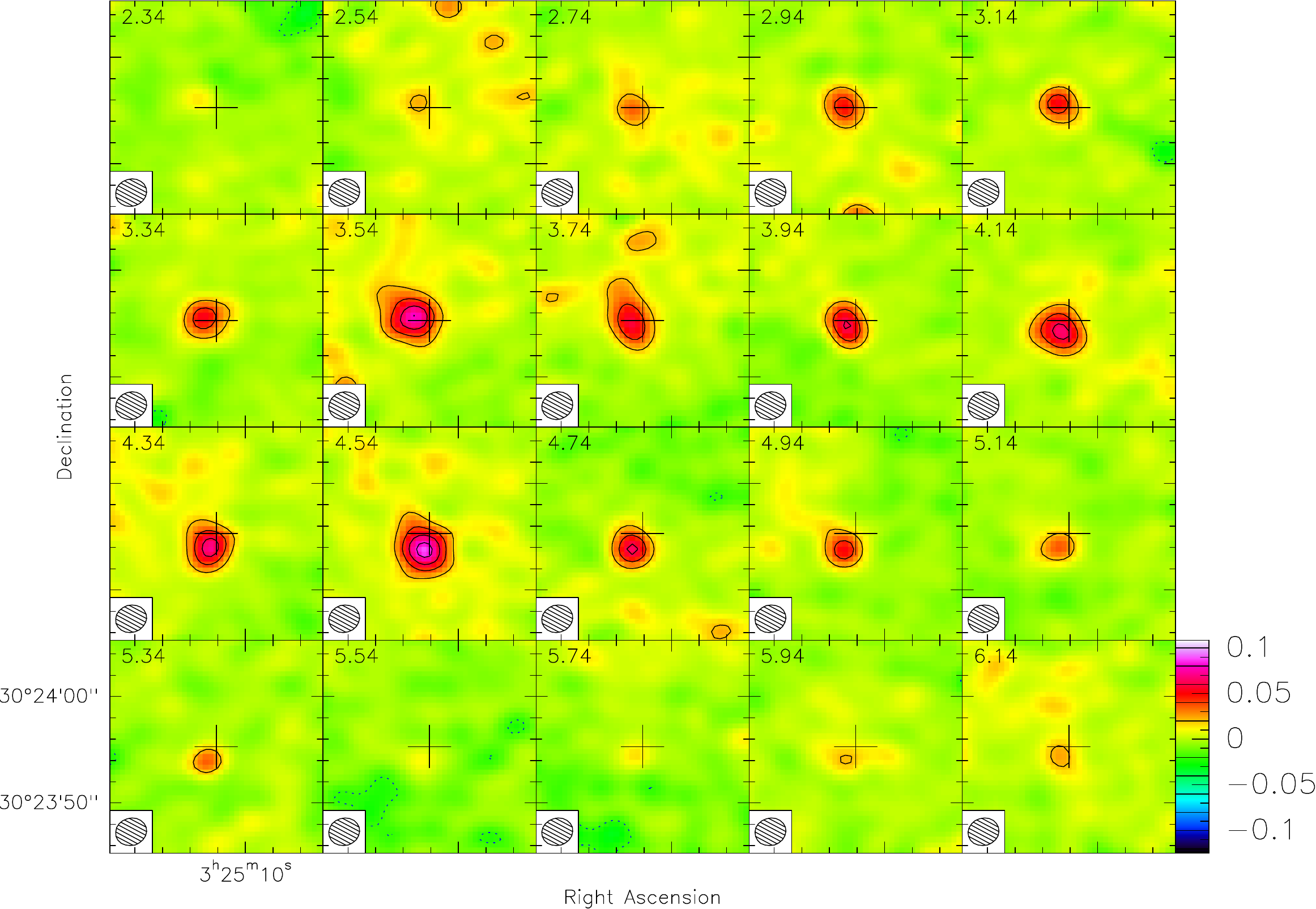}
\caption{Channel map for the SiO line at 130.268~GHz. Color levels are in Jy~beam$^{-1}$.\label{channel_maps_9}}
\end{figure}

\begin{figure}[h]
\includegraphics[width=1\linewidth]{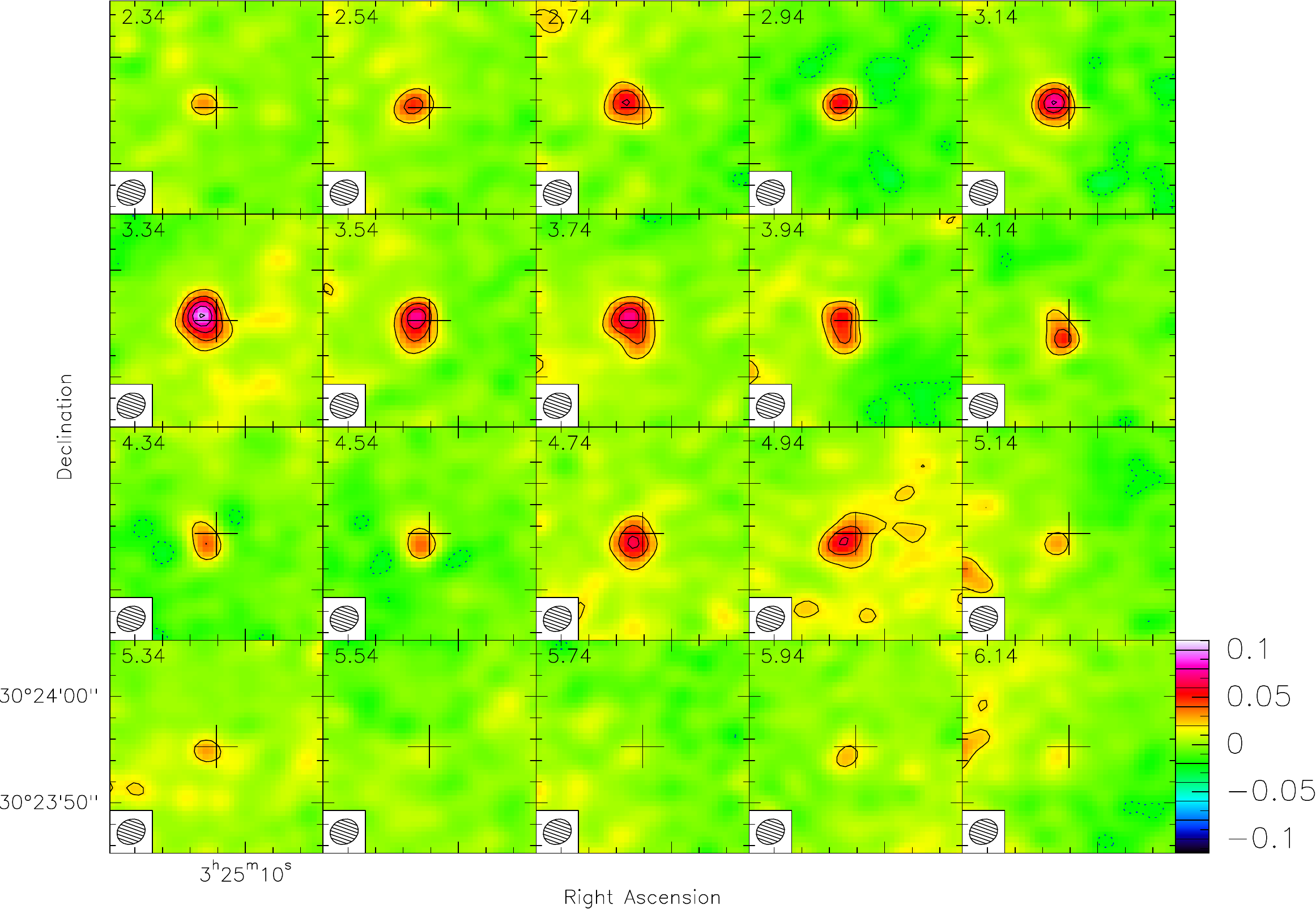}
\caption{Channel map for the CH$_3$OH line at 146.618~GHz. Color levels are in Jy~beam$^{-1}$.\label{channel_maps_9}}
\end{figure}

\clearpage
\section{Molecular spectra on the source position}\label{spectra_source}

 This appendix presents all molecular lines detected at the maximum continuum position for the NOEMA+30m data. The vertical dashed lines on the spectra show the local rest velocity of the cloud.

\begin{figure}[h!]
\includegraphics[width=0.95\linewidth]{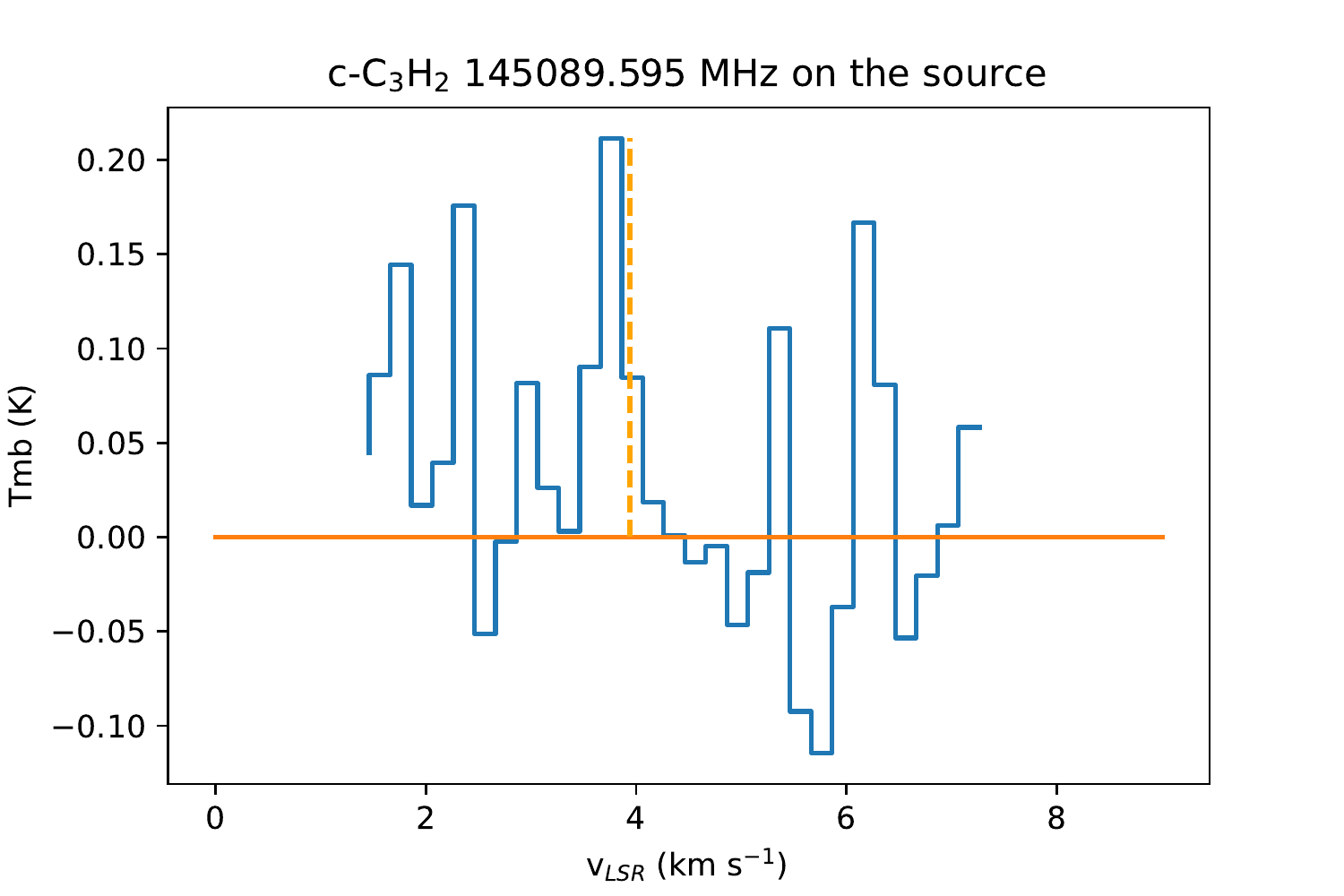}
\includegraphics[width=0.95\linewidth]{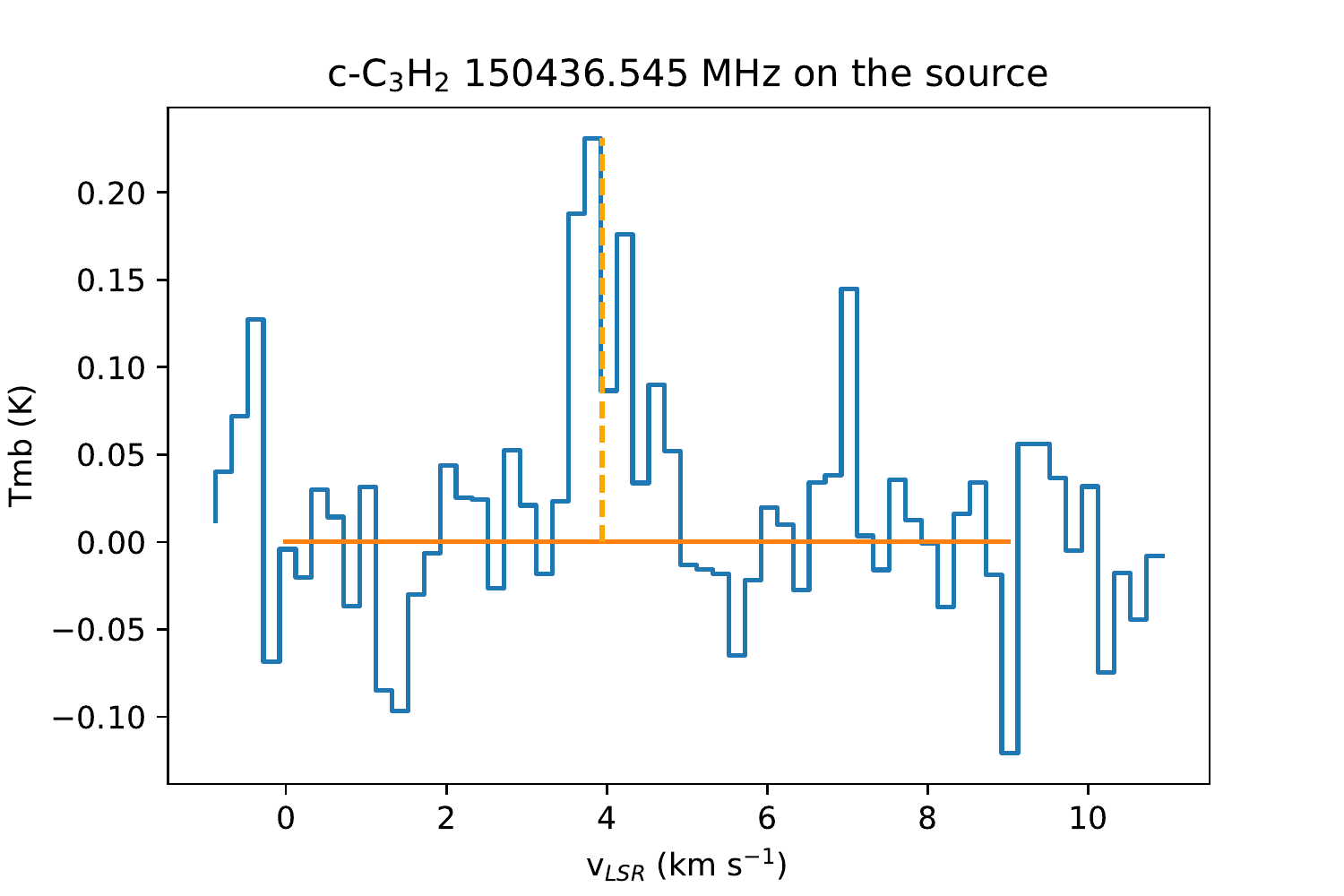}
\includegraphics[width=0.95\linewidth]{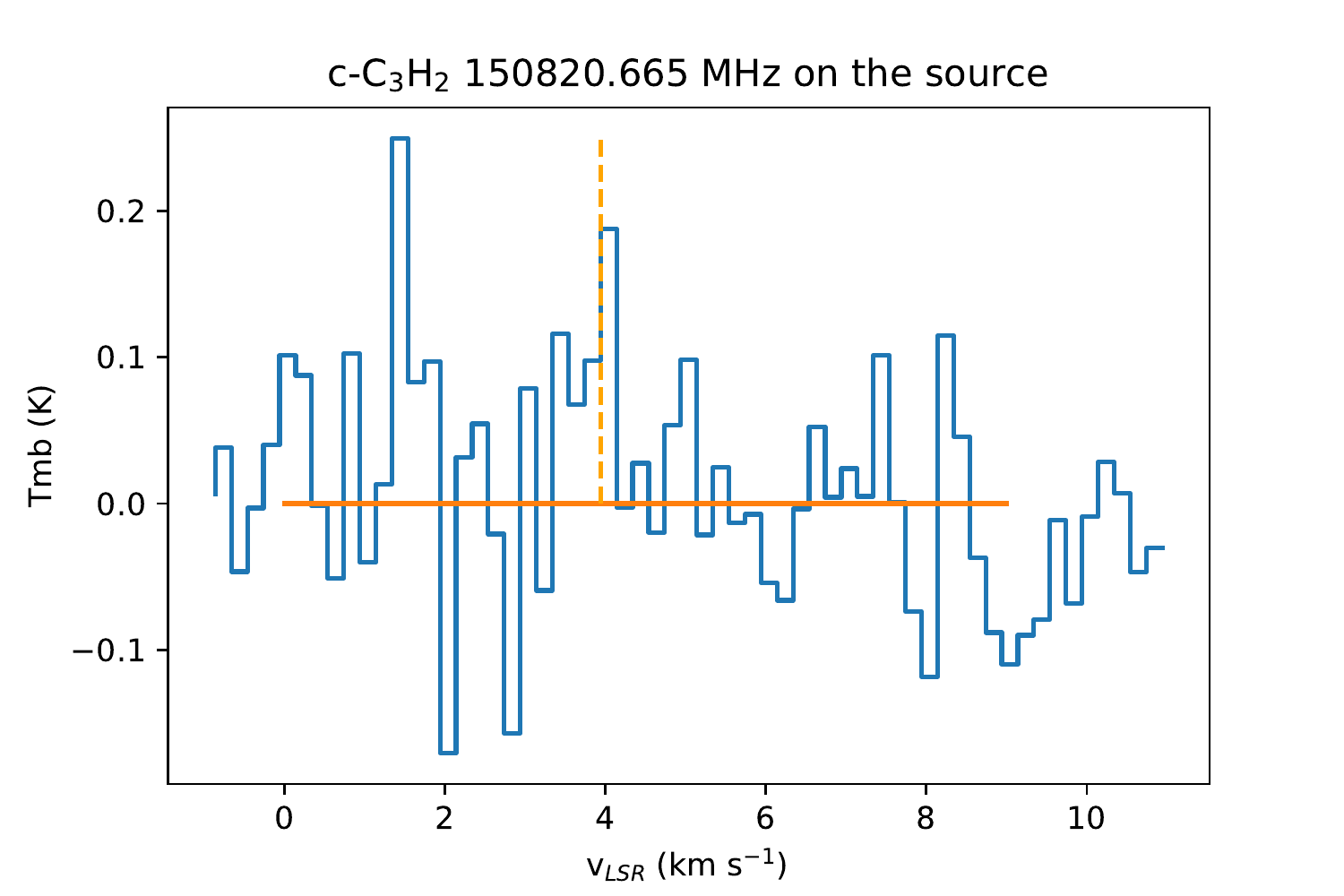}
\caption{Spectra of detected lines on the source position.\label{spectra_onsource1} }
\end{figure}

\begin{figure}[h!]
\includegraphics[width=0.95\linewidth]{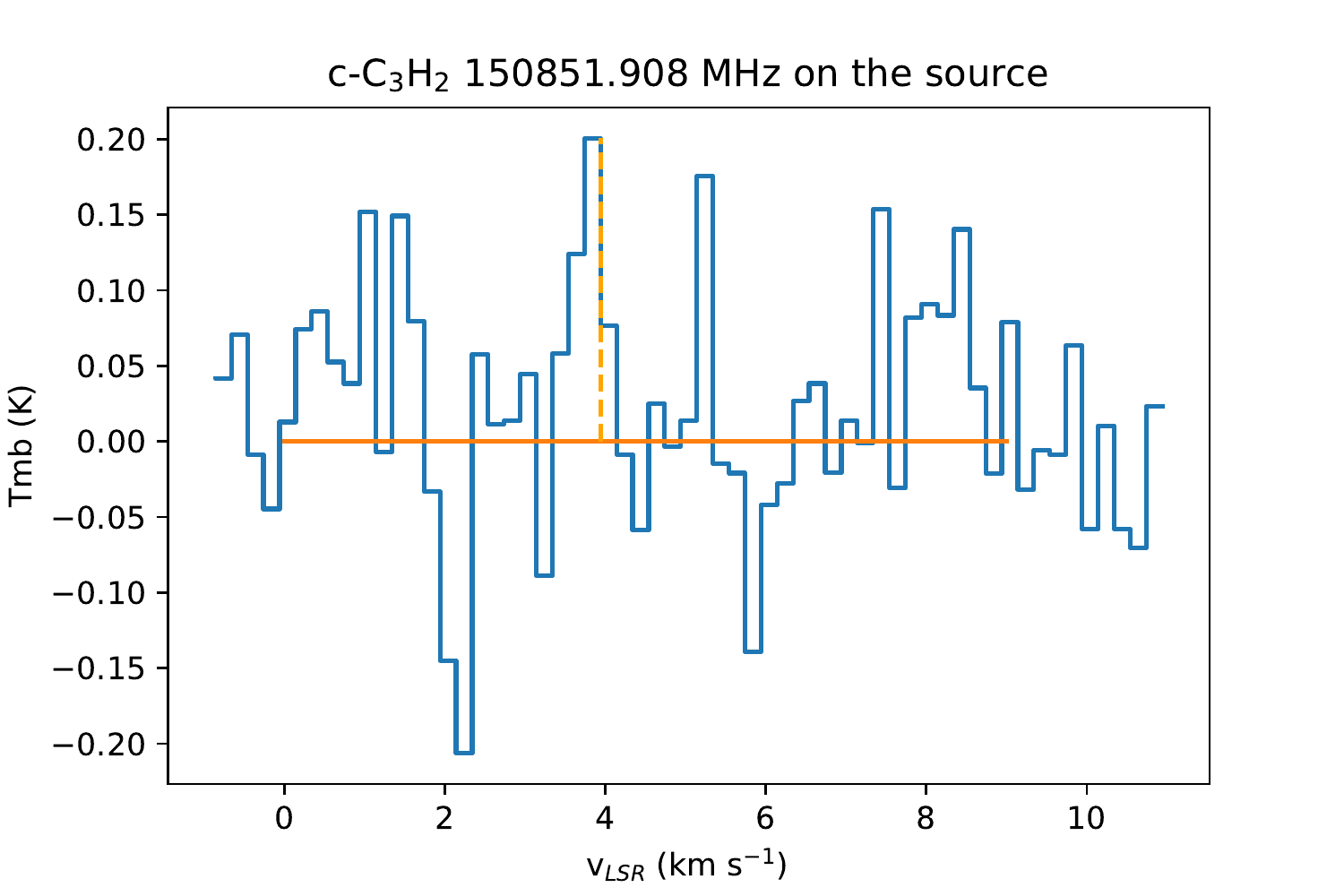}
\includegraphics[width=0.95\linewidth]{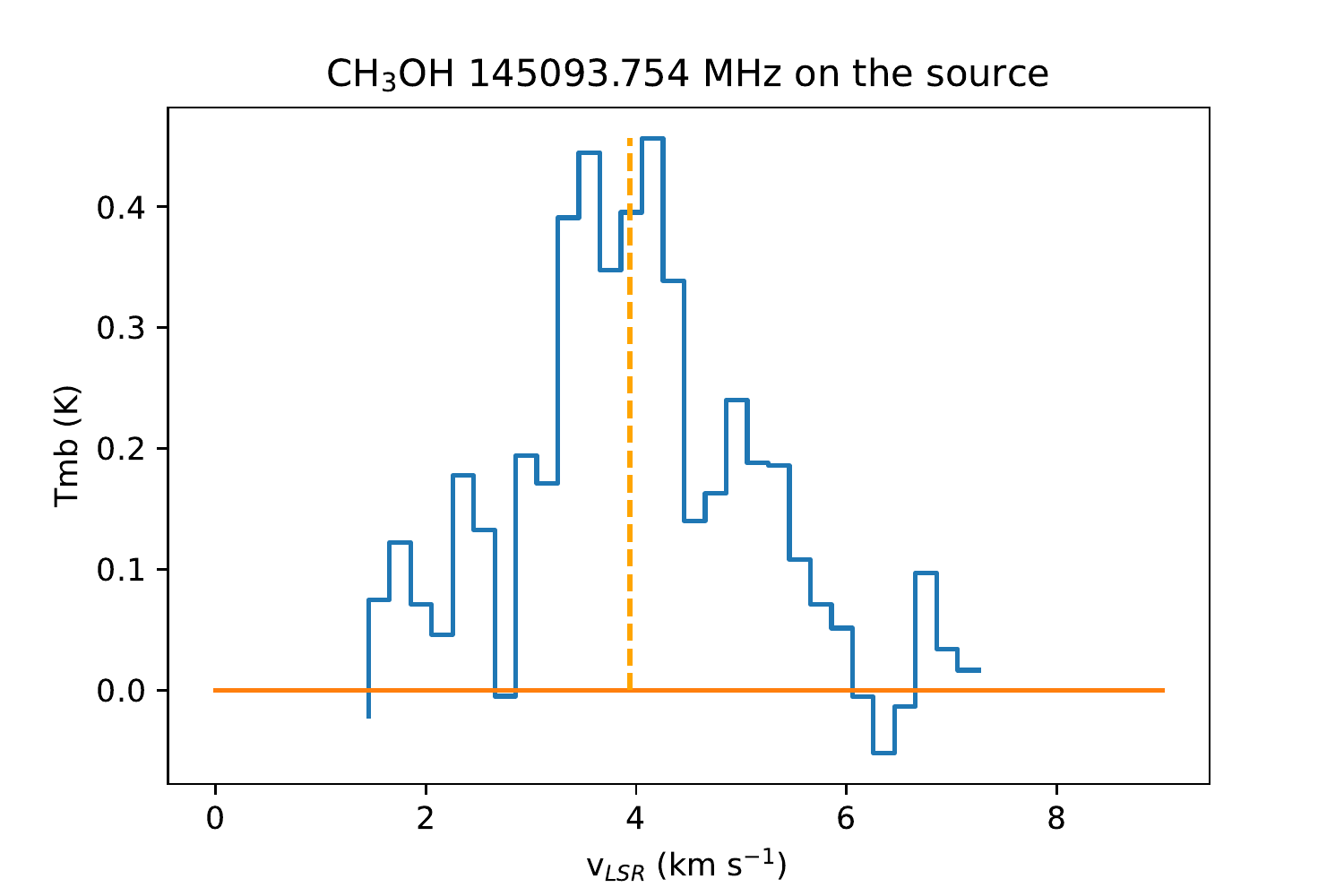}
\includegraphics[width=0.95\linewidth]{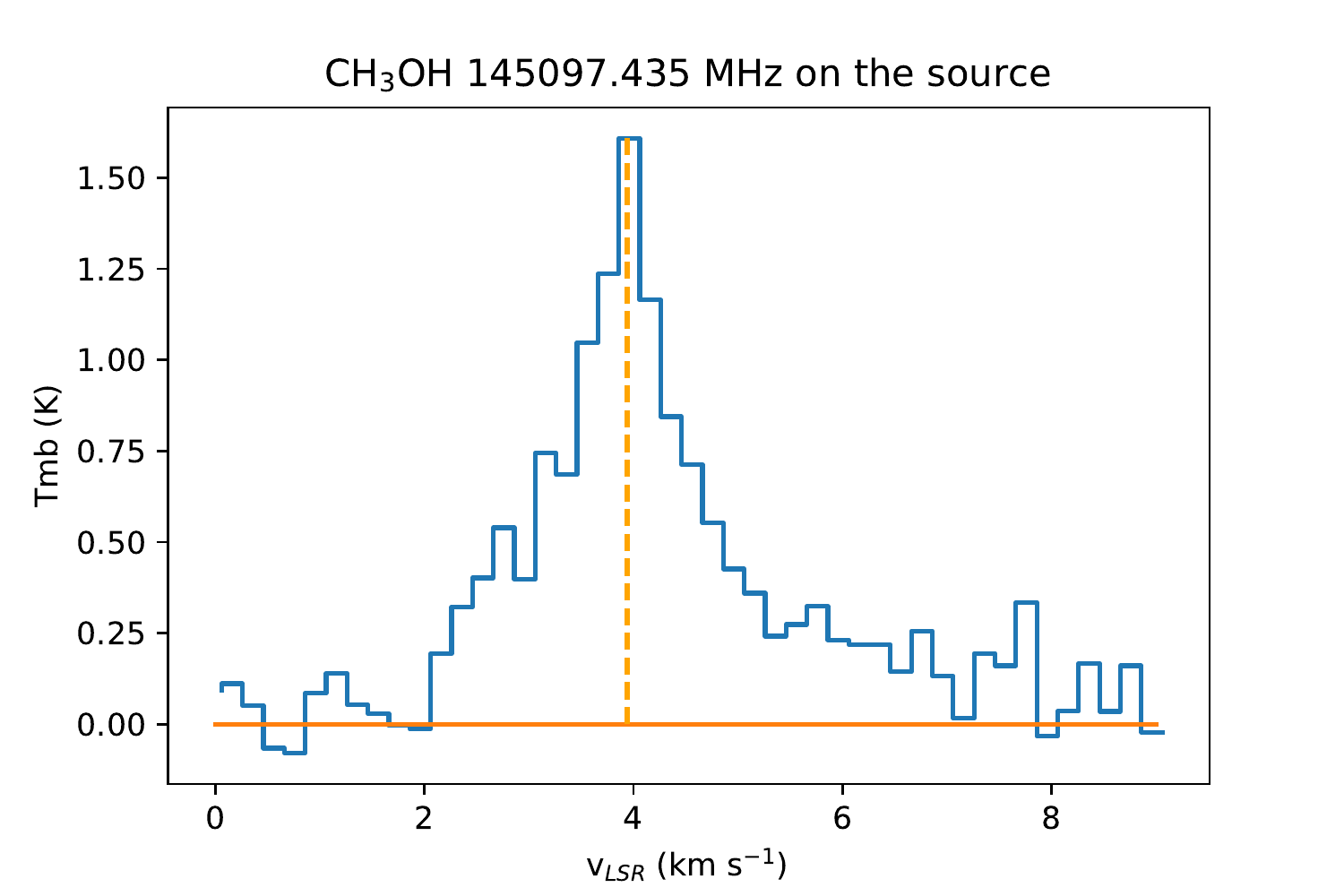}
\includegraphics[width=0.95\linewidth]{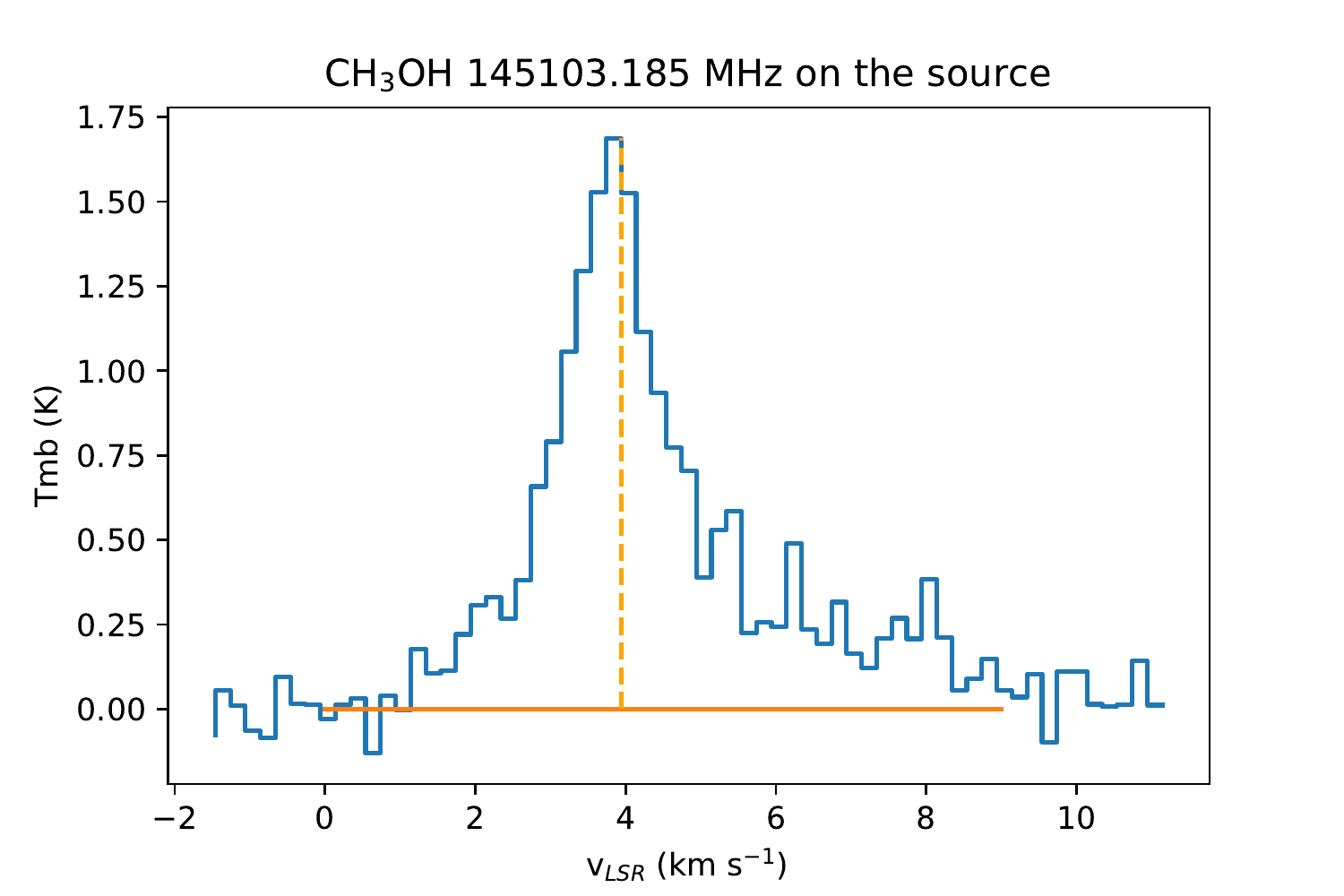}
\caption{Spectra of detected lines on the source position.\label{spectra_onsource2} }
\end{figure}

\begin{figure}[h!]
\includegraphics[width=0.95\linewidth]{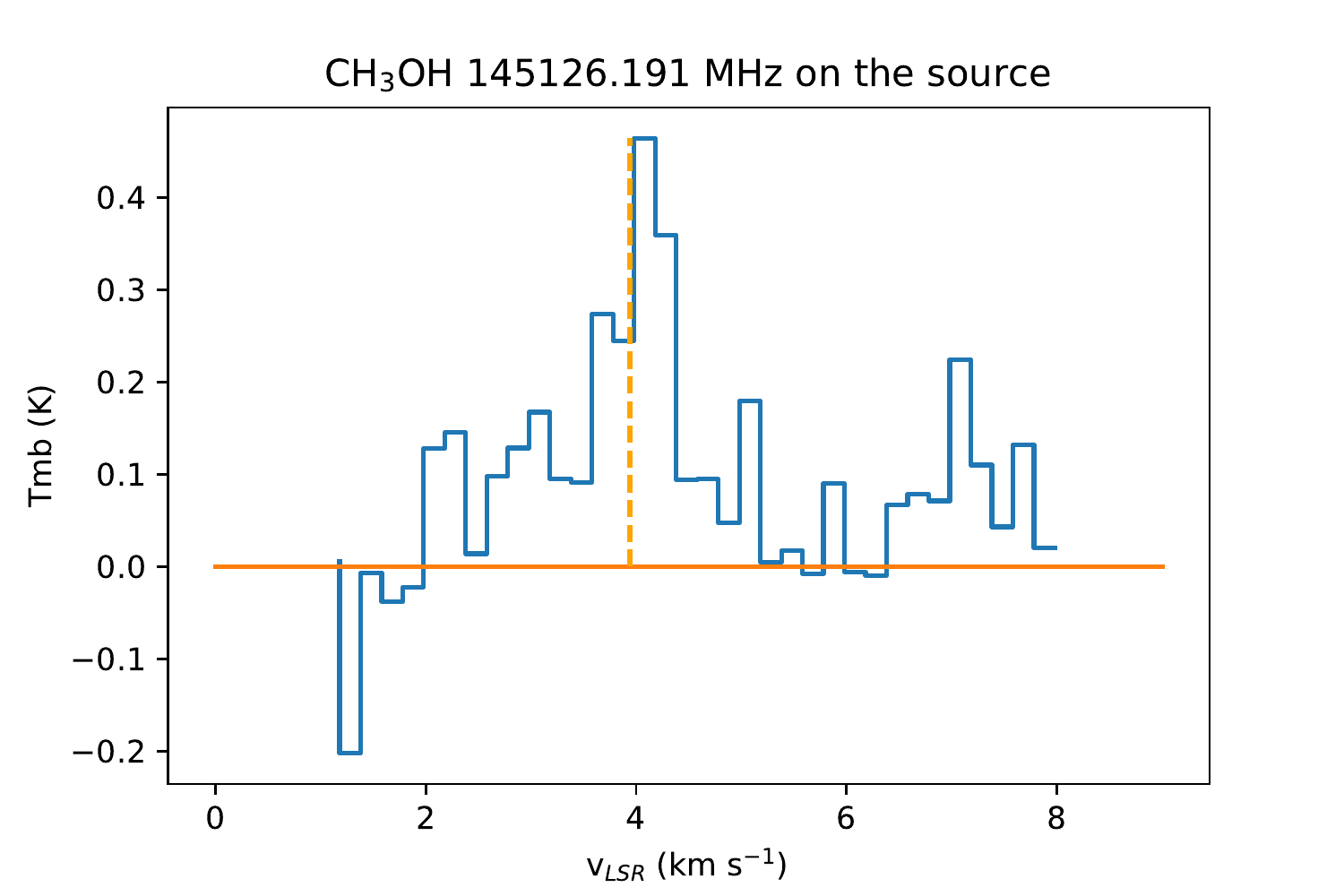}
\includegraphics[width=0.95\linewidth]{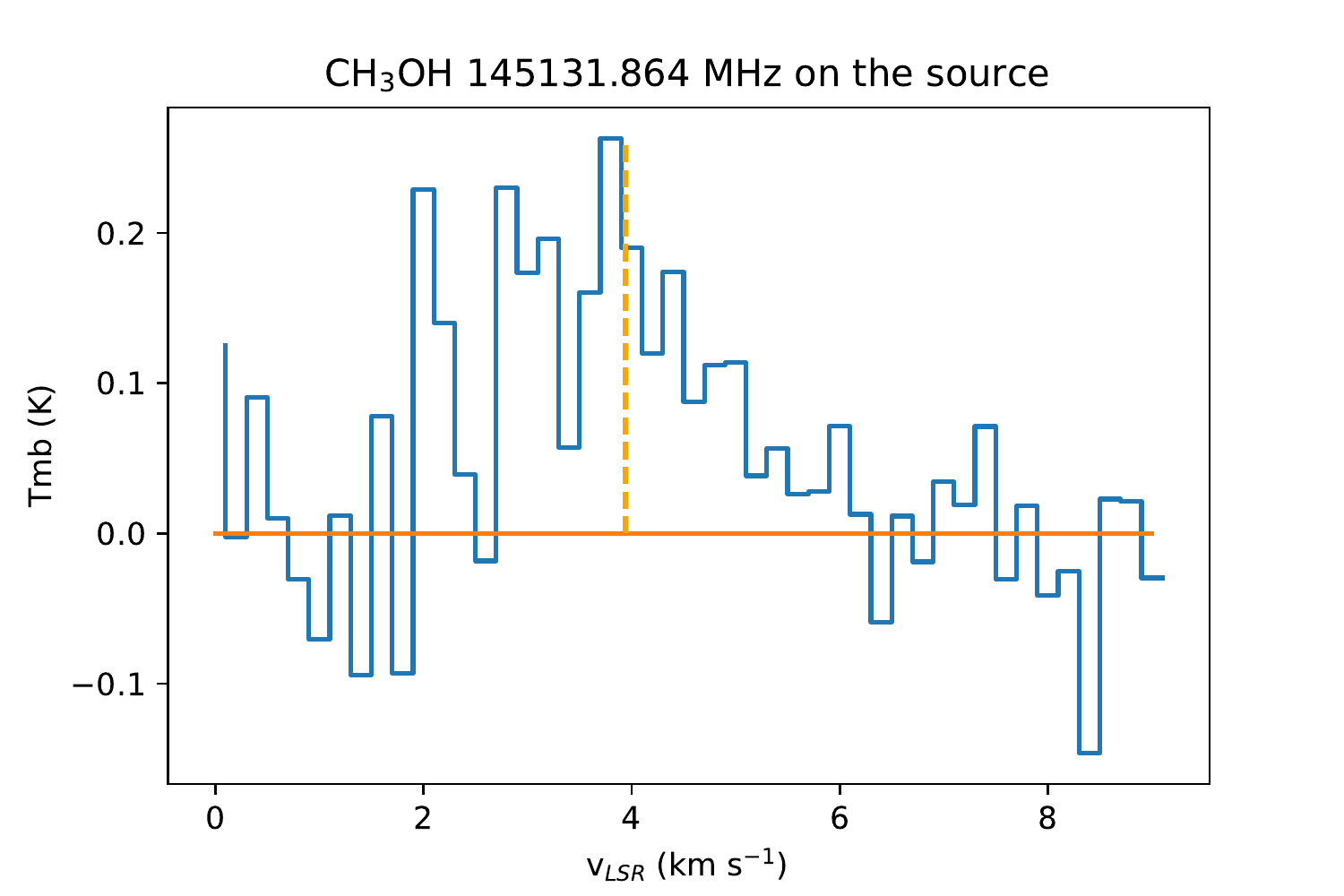}
\includegraphics[width=0.95\linewidth]{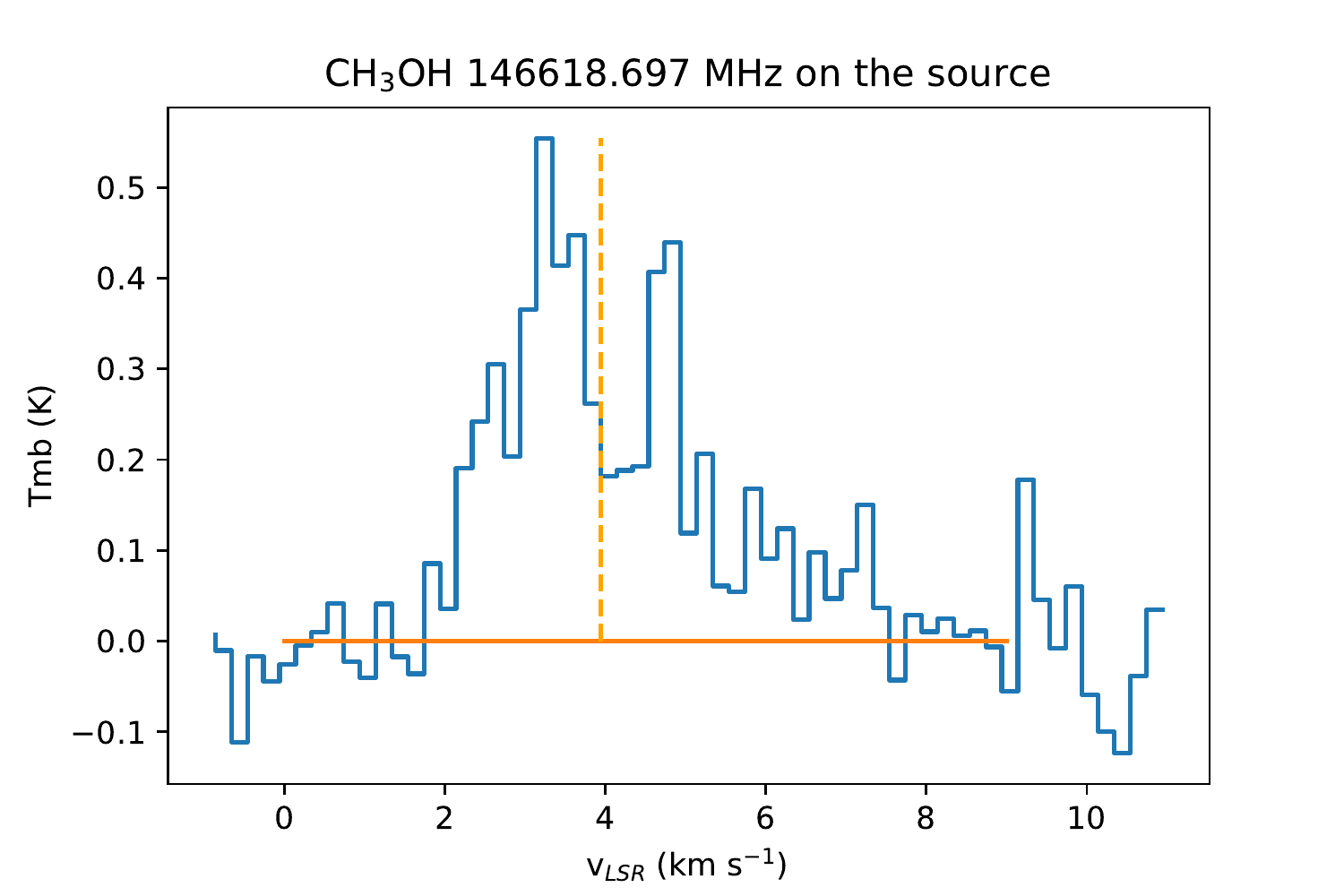}
\includegraphics[width=0.95\linewidth]{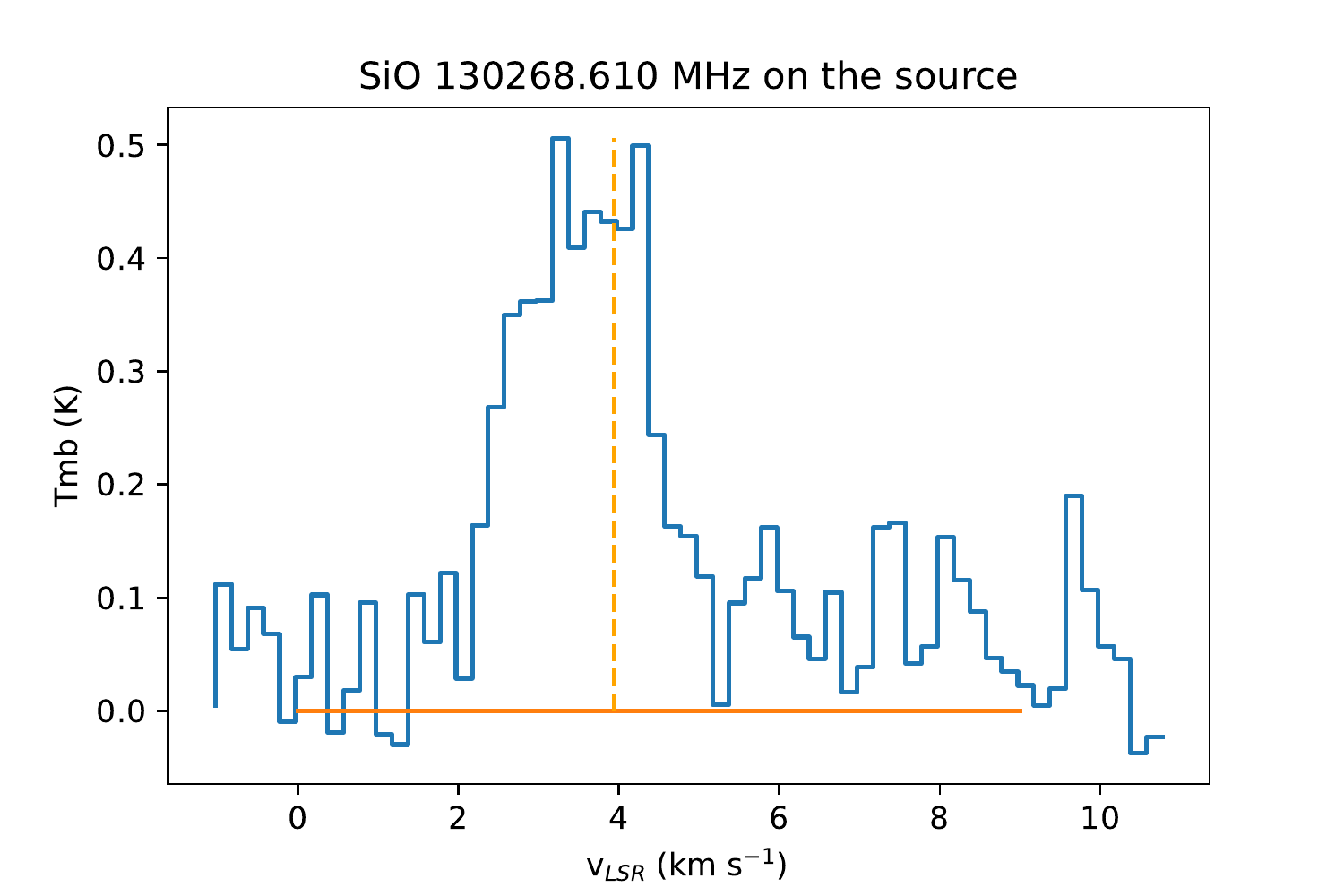}
\caption{Spectra of detected lines on the source position.\label{spectra_onsource3}}
\end{figure}

\begin{figure}[h!]
\includegraphics[width=0.95\linewidth]{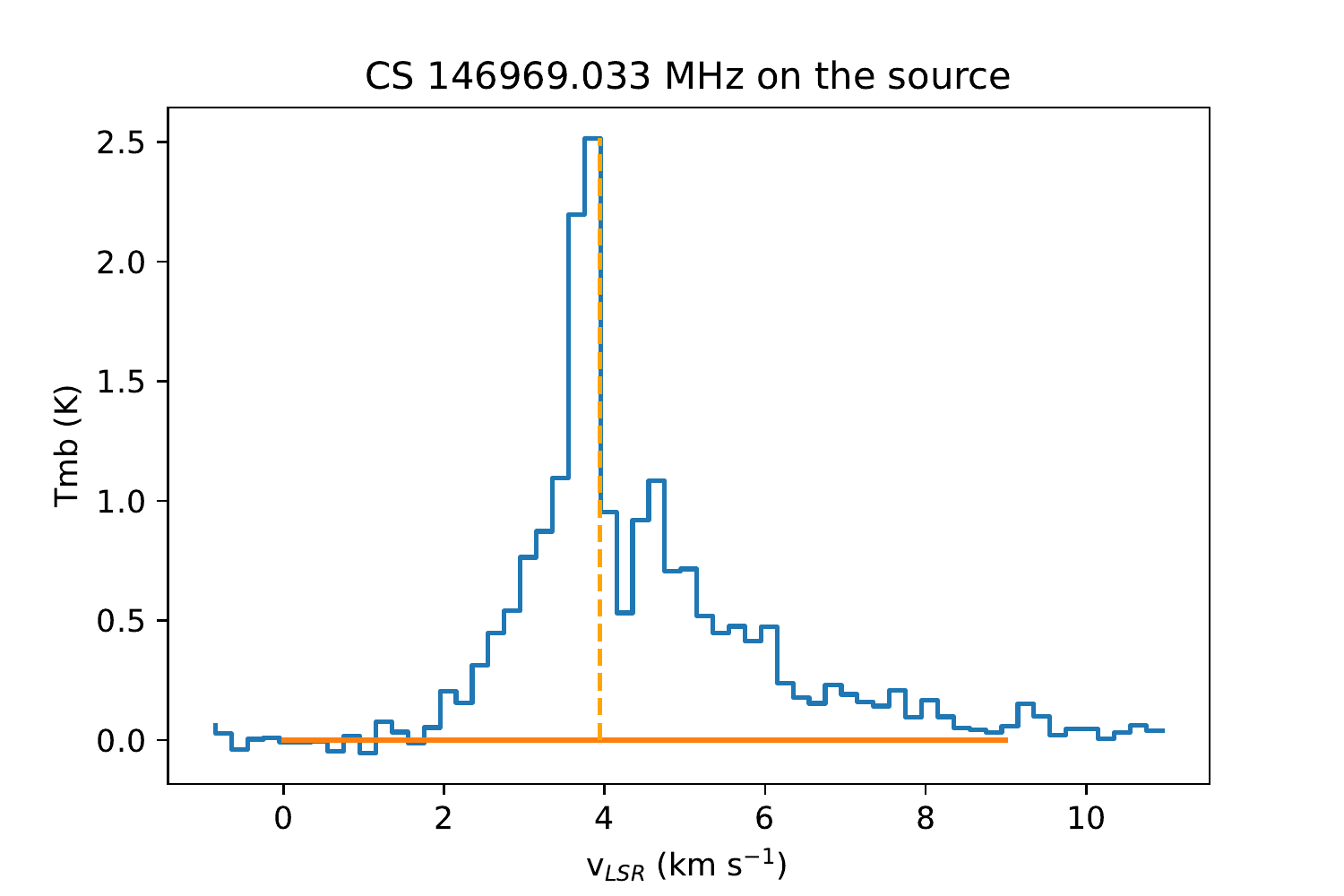}
\includegraphics[width=0.95\linewidth]{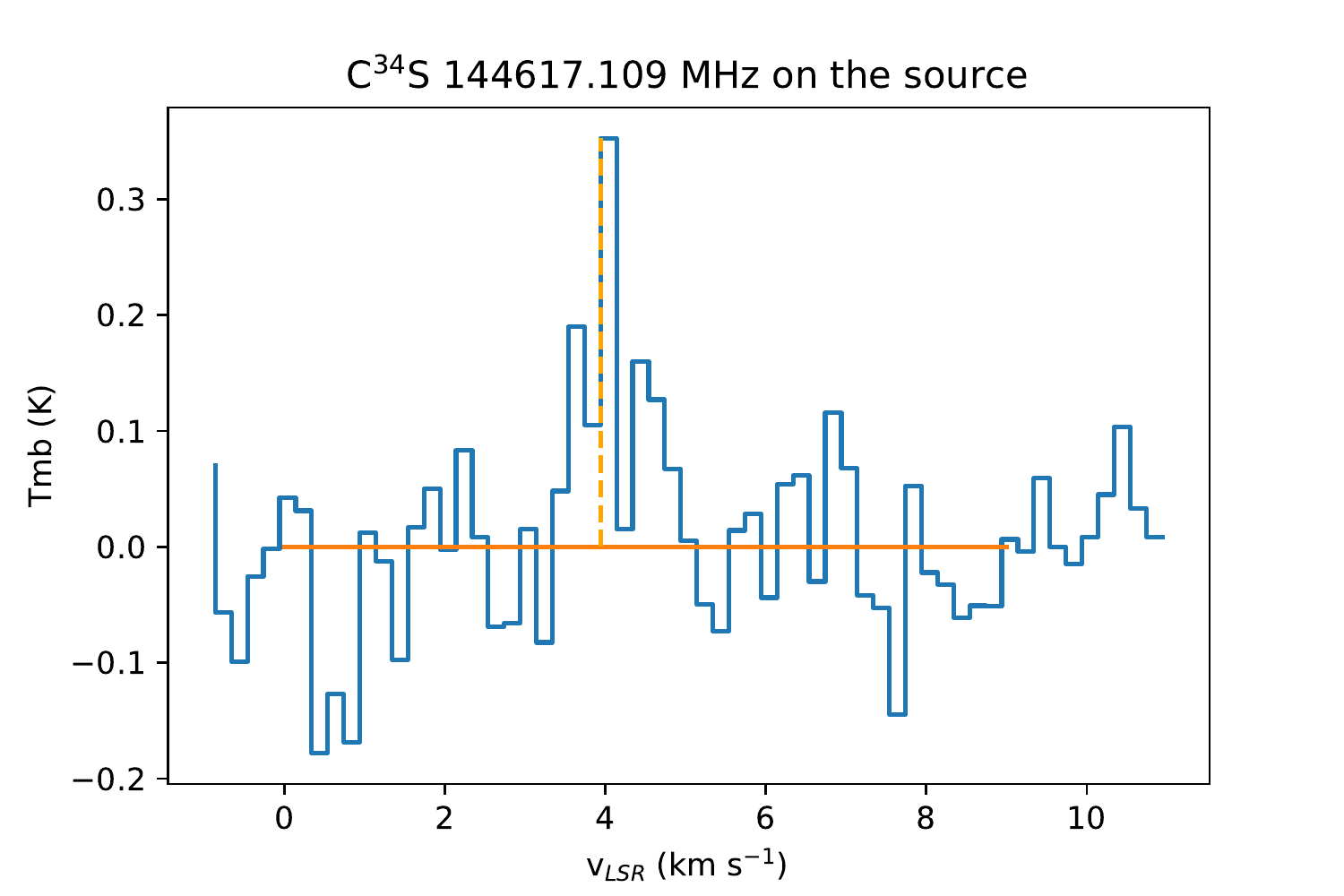}
\includegraphics[width=0.95\linewidth]{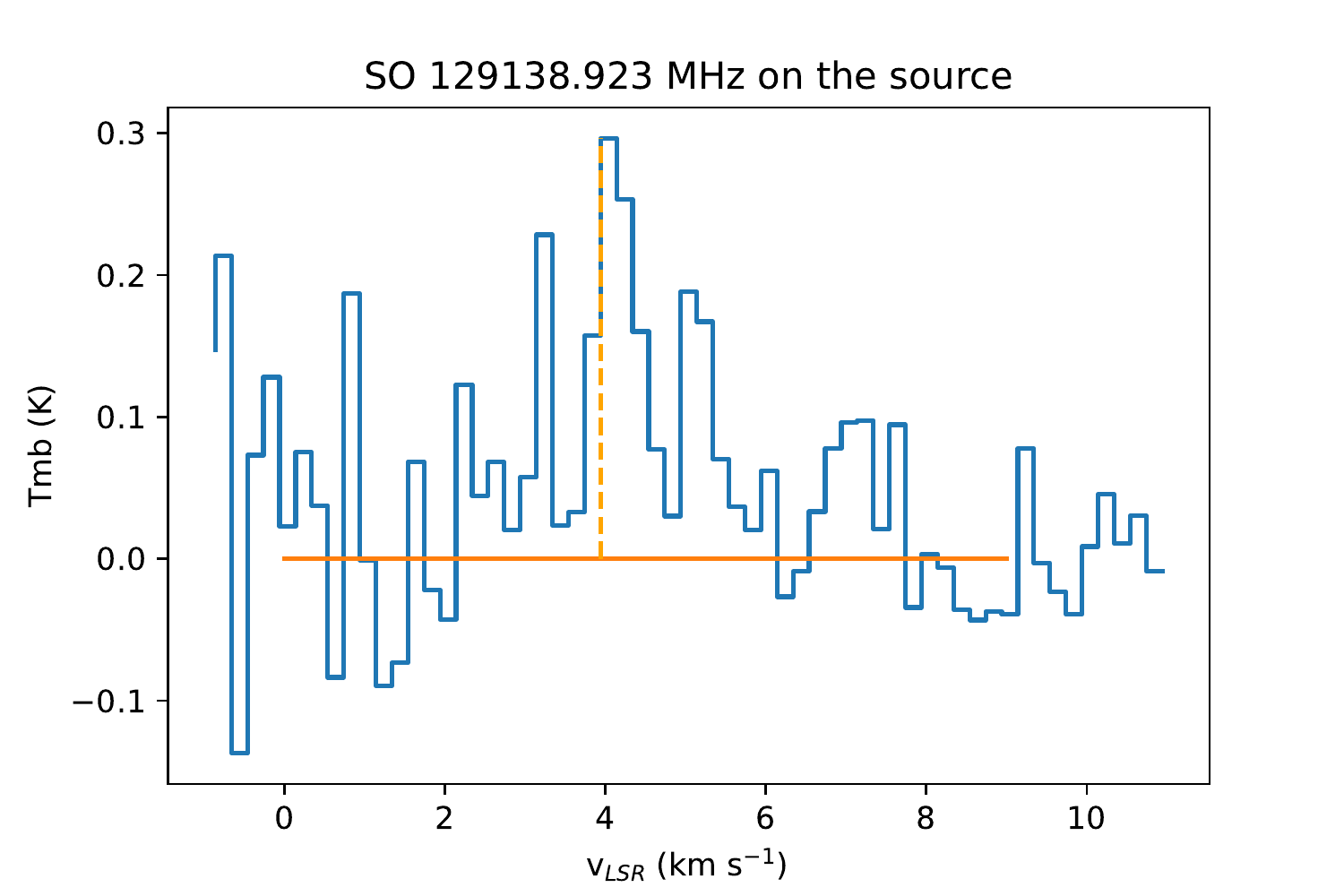}
\includegraphics[width=0.95\linewidth]{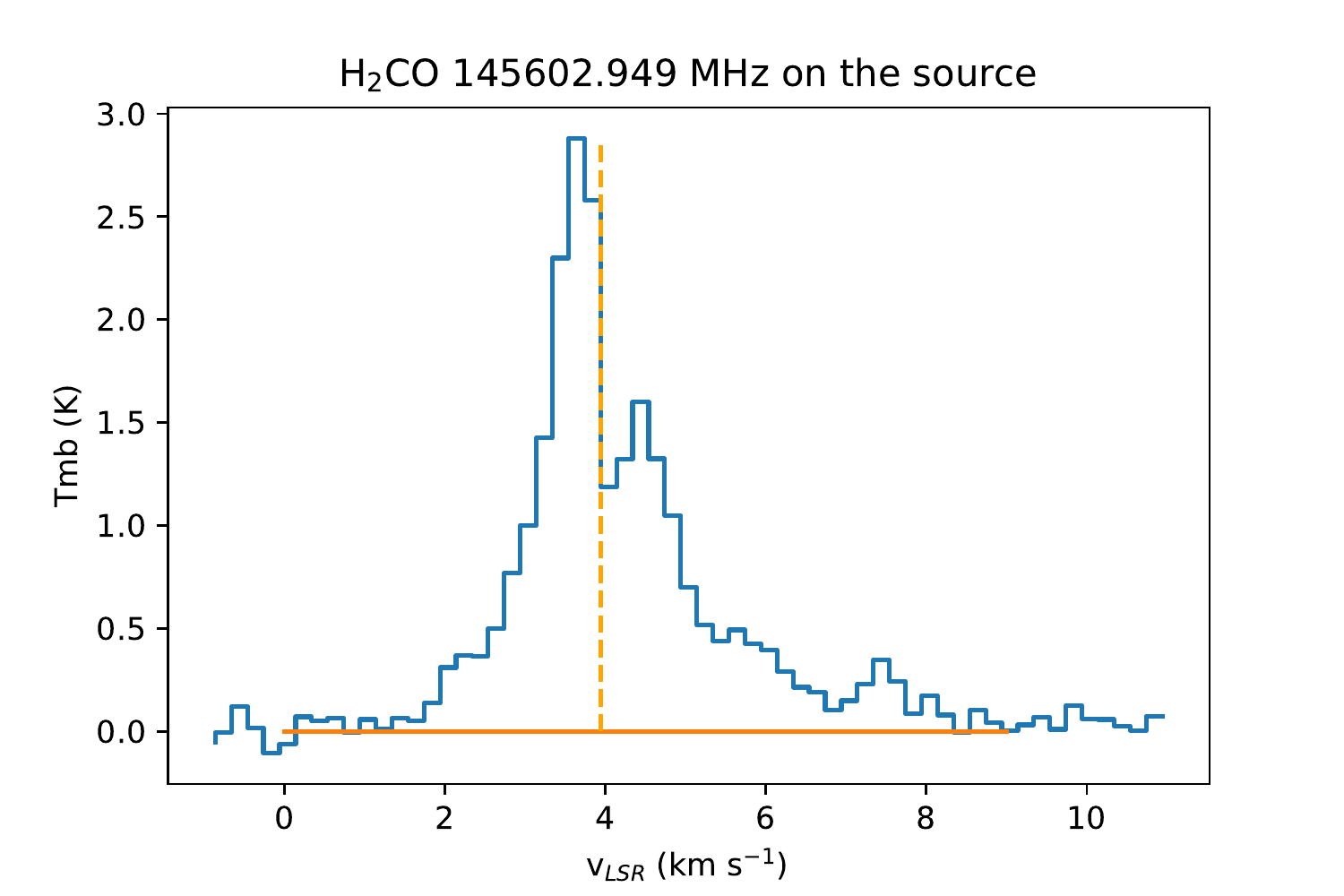}
\caption{Spectra of detected lines on the source position. \label{spectra_onsource4} }
\end{figure}

\begin{figure}[h!]
\includegraphics[width=0.95\linewidth]{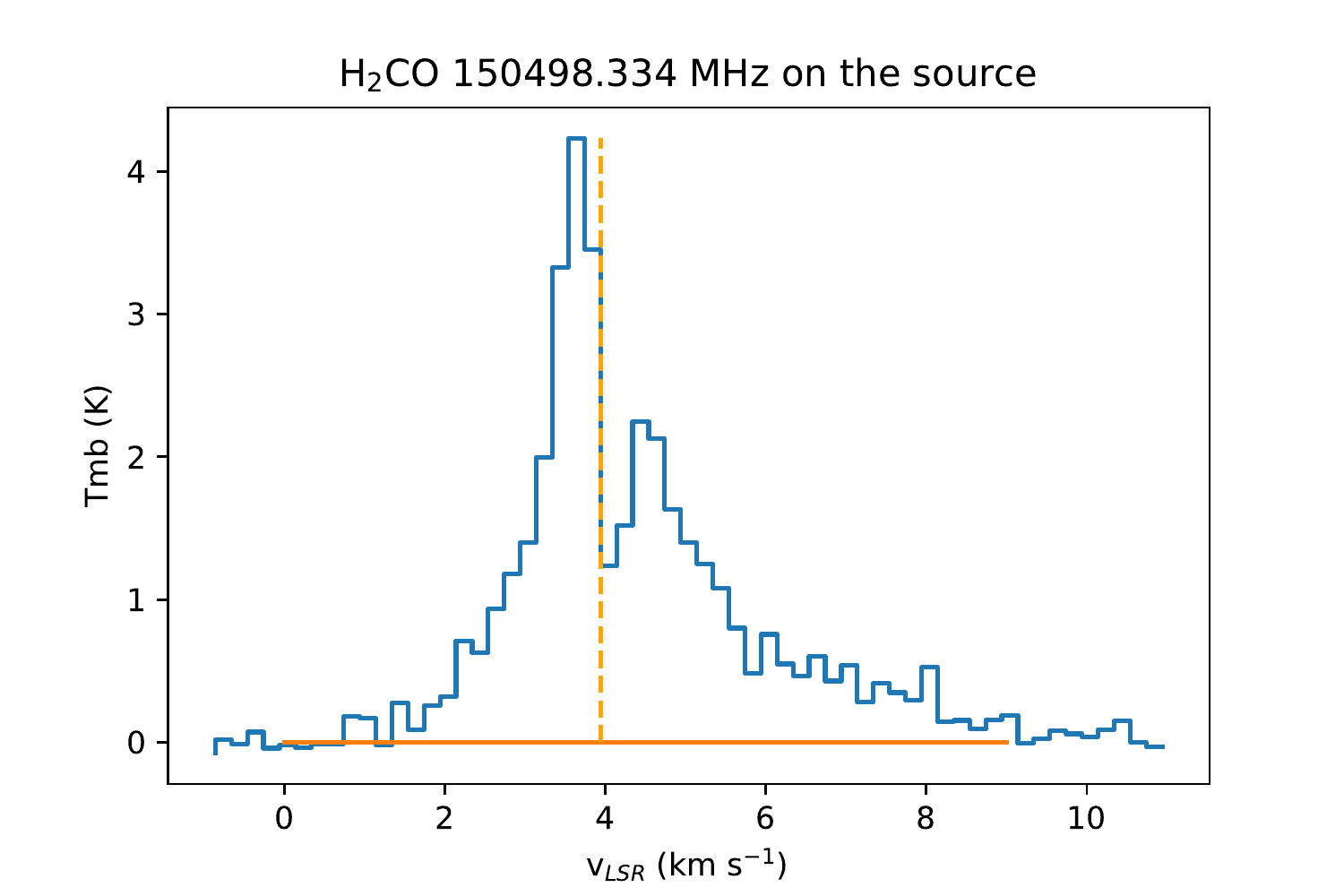}
\includegraphics[width=0.95\linewidth]{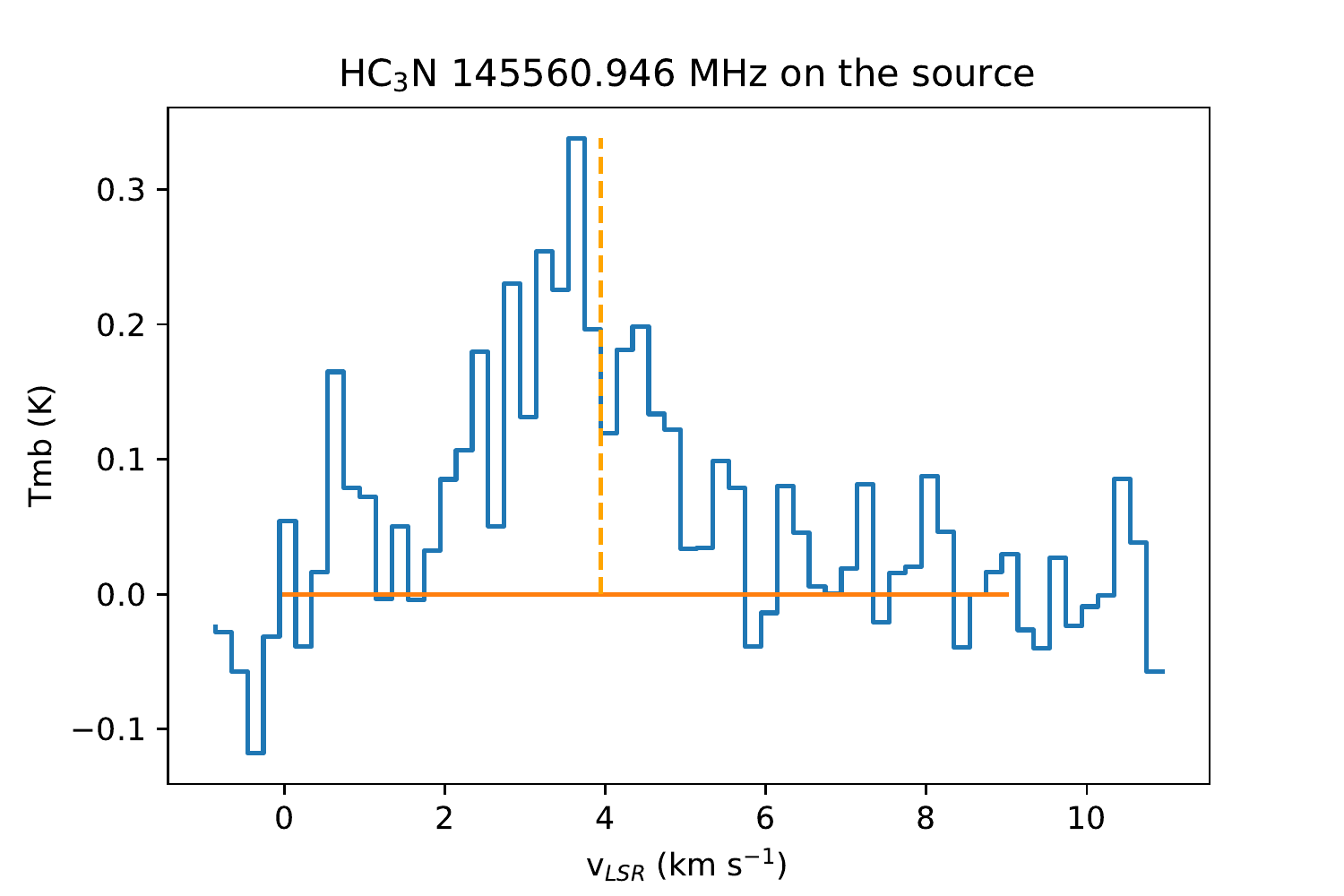}
\includegraphics[width=0.95\linewidth]{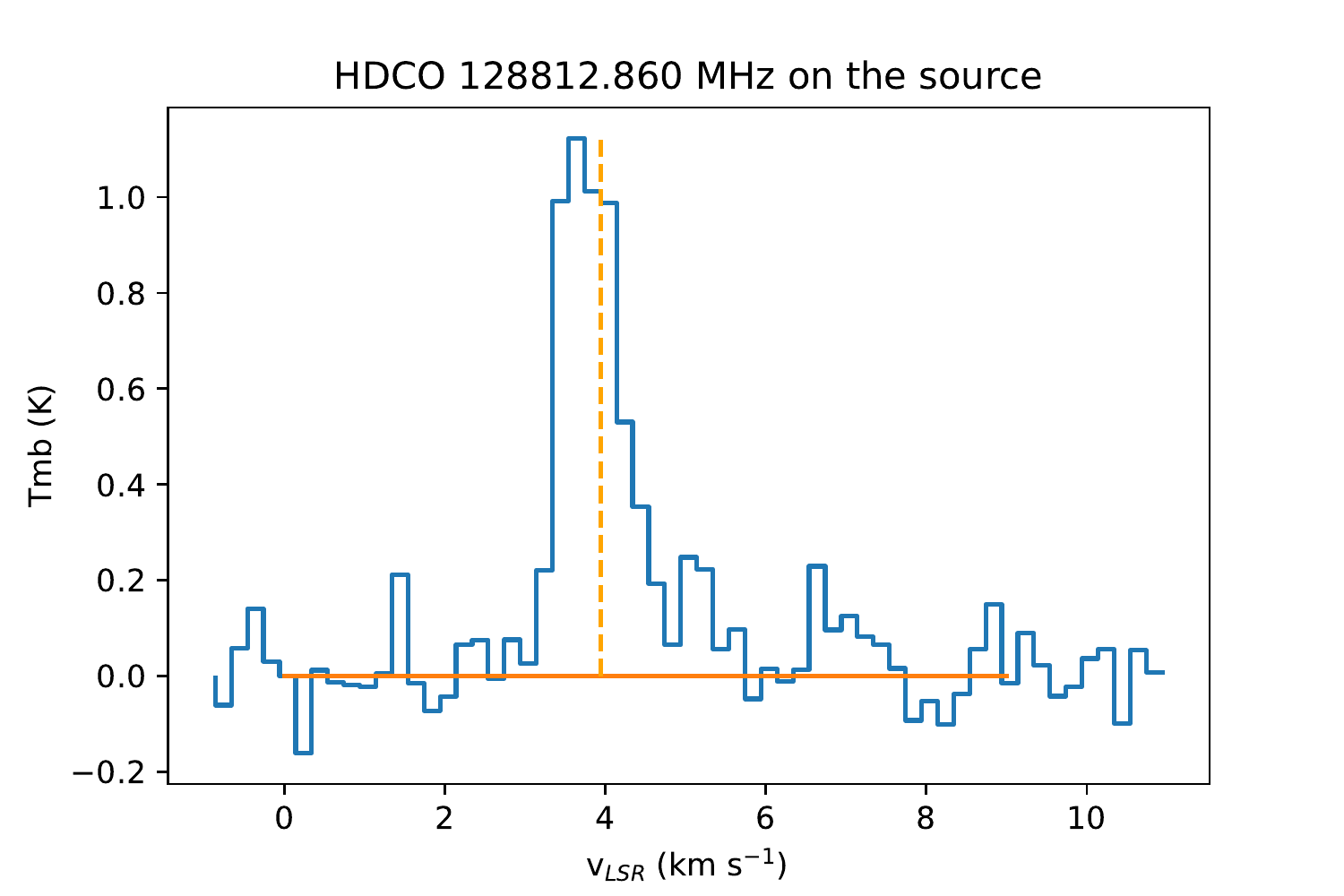}
\includegraphics[width=0.95\linewidth]{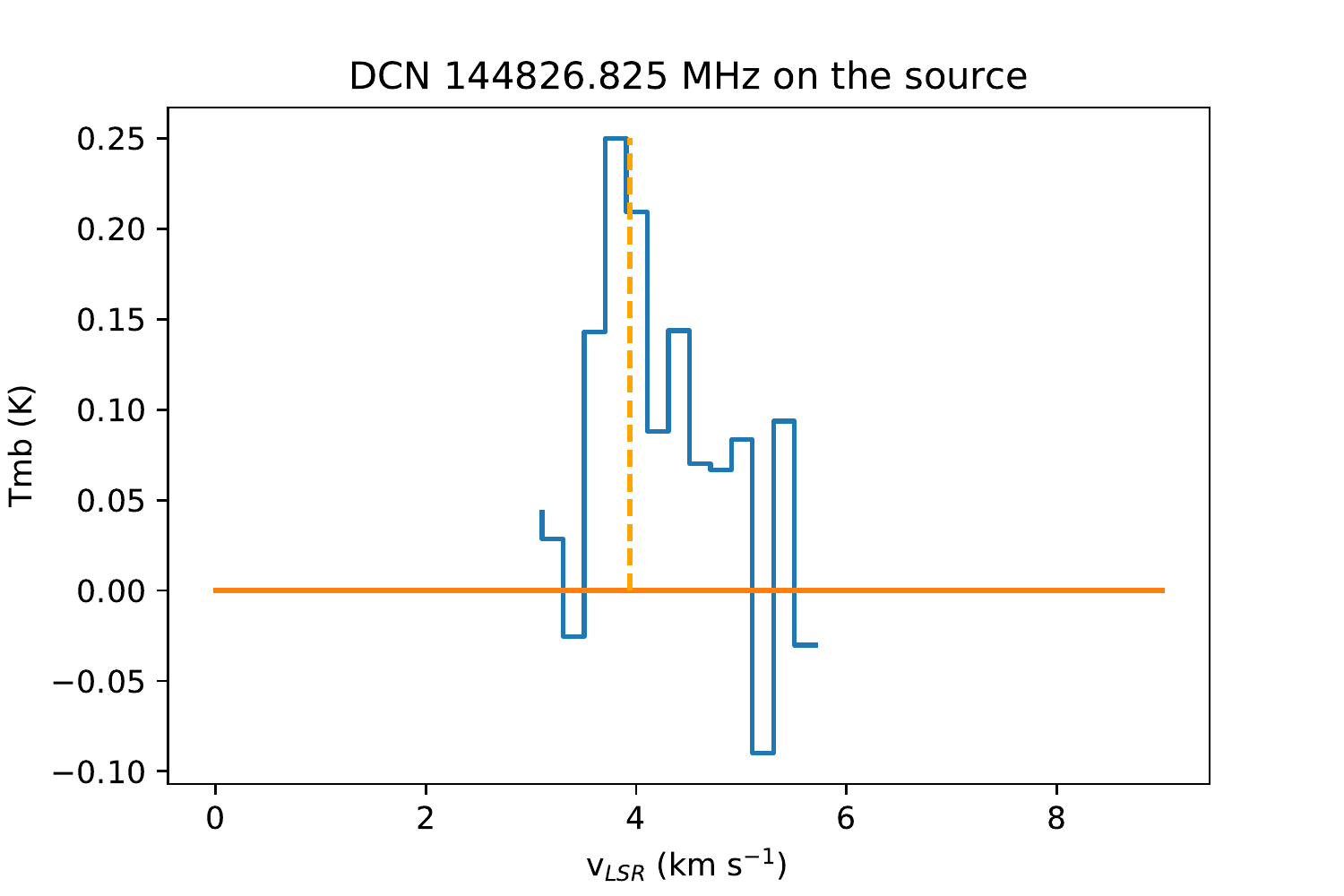}
\caption{Spectra of detected lines on the source position. \label{spectra_onsource5} }
\end{figure}

\begin{figure}[h!]
\includegraphics[width=0.95\linewidth]{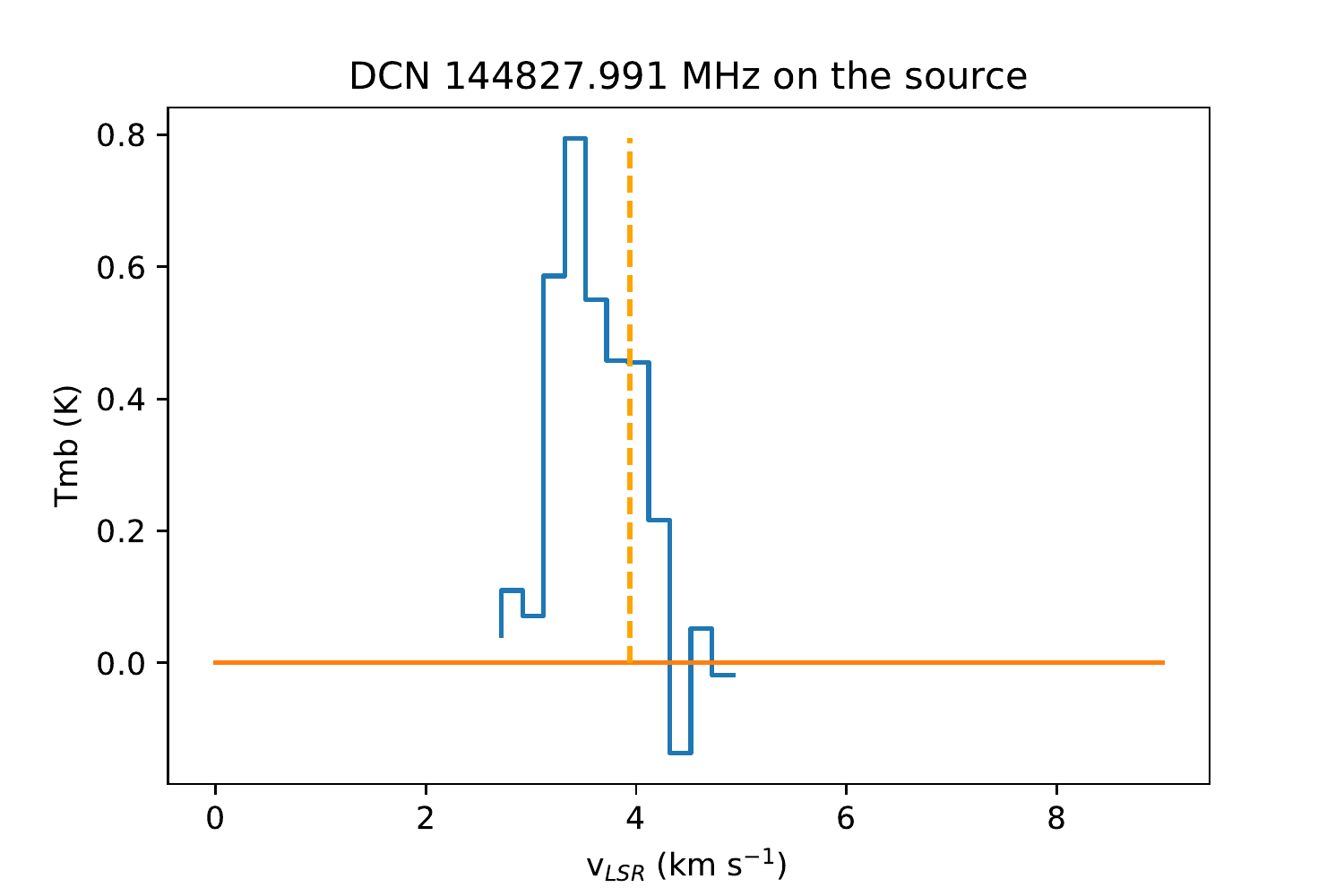}
\includegraphics[width=0.95\linewidth]{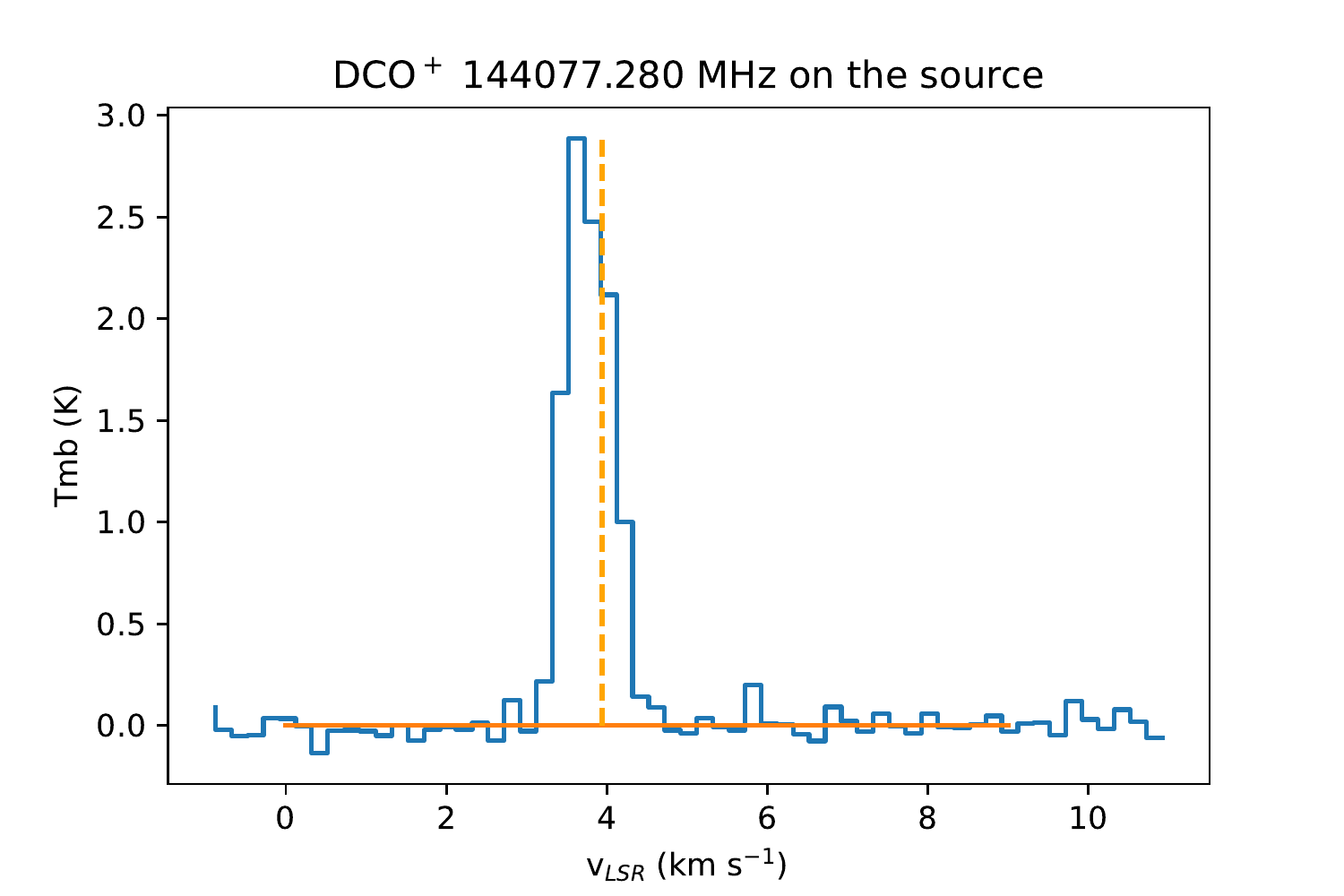}
\caption{Spectra of detected lines on the source position. \label{spectra_onsource6} }
\end{figure}

\clearpage
\section{Molecular column densities in the outflow}\label{ndradex}

\begin{figure*}
\includegraphics[width=1\linewidth]{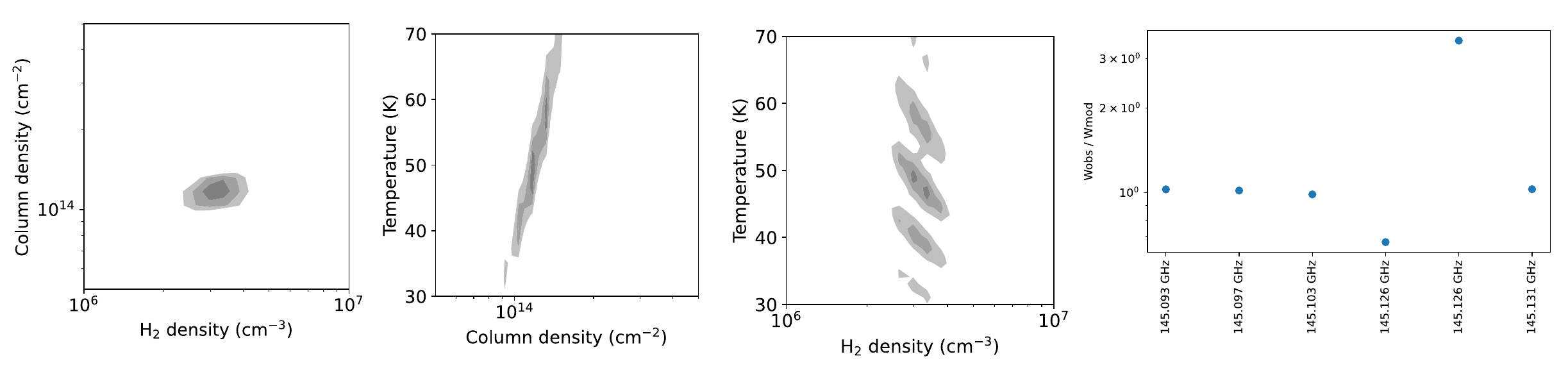}
\caption{Results of the $\chi^2$ minimization using the six observed lines of CH$_3$OH. For each of the three plots on the left, the $\chi^2$ is projected over the third axis, which is not shown in the figure. Gray contours represent  the 1$\sigma$, 2$\sigma$, and 3$\sigma$ confidence intervals. The plot on the right represents the ratio of the observed and modeled integrated intensity for each of the methanol lines and the best model obtained by the $\chi^2$ minimization. \label{chi2_6lines} }
\end{figure*}

\begin{figure*}
\includegraphics[width=1\linewidth]{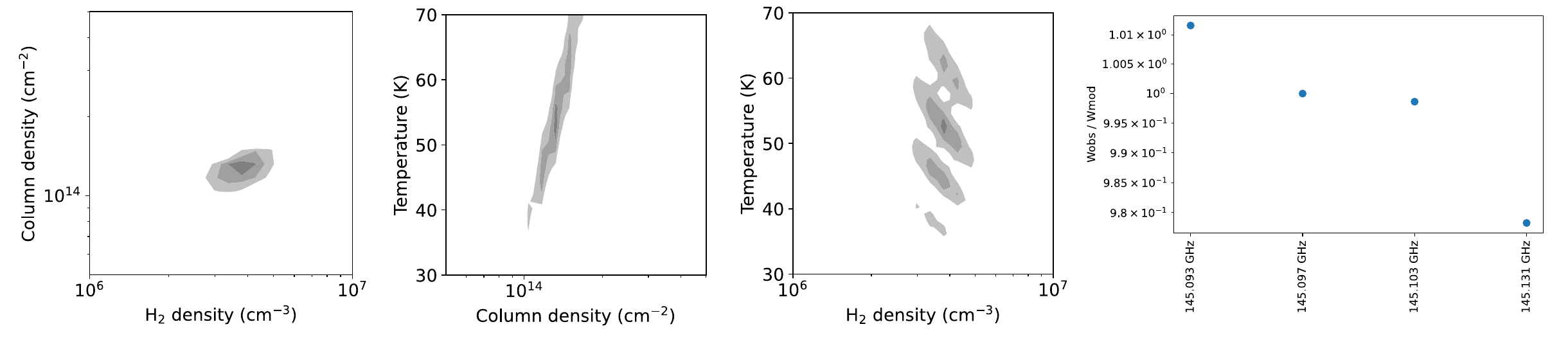}
\caption{Same as Fig.~\ref{chi2_6lines}, but for four lines of methanol only (excluding the two 145.126~GHz blended lines). \label{chi2_4lines} }
\end{figure*}

To constrain the physical conditions probed by the methanol emission on the source, we performed a $\chi^2$ minimization while comparing the observed integrated intensities of the observed lines with a grid of theoretical values. The observed integrated intensities and rms are listed in Table~\ref{Wobs}. The grid of theoretical integrated intensities was computed for the six detected lines of methanol using the ndradex\footnote{https://github.com/astropenguin/ndradex} python package, which in turn uses the non-LTE radiative transfer code RADEX \citep{2007A&A...468..627V}. The collisional coefficients are taken from \citet{2010MNRAS.406...95R}. The computation of the theoretical grid is time consuming, and therefore we first computed a large grid with low resolution, covering H$_2$ densities between $10^5$ and $10^7$~cm$^{-3}$, temperatures between 20 and 100~K, and CH$_3$OH column densities between $10^{13}$ and $10^{15}$~cm$^{-2}$. We then refined around the identified zone with the following grid: 40 values for the temperature in linear space, 20 values for the H$_2$ density between $10^6$ and $10^7$~cm$^{-3}$ in log space, and 20 values for the CH$_3$OH column density between $5\times 10^{13}$ and $5\times 10^{14}$~cm$^{-3}$ in log space. We assumed a statistical ratio of 1 of the A and E forms and a line width of 2~km~s$^{-1}$ for all the lines (representative of the mean value of the observed line widths).  The results of the $\chi^2$ minimization using the six observed lines of CH$_3$OH are shown in Fig.~\ref{chi2_6lines}.  The best model (with a 3$\sigma$ confidence level) gives a CH$_3$OH column density of $(1.2\pm 0.2)\times 10^{14}$~cm$^{-2}$, a temperature of 46$^{+26}_{-11}$~K, and a H$_2$ density of $(3.3\pm 1)\times 10^6$~cm$^{-3}$. Constraints on the methanol column density and the H$_2$ density are rather good, but the constraint on the temperature is weaker. 
The excitation temperature and optical depth of individual lines as predicted by RADEX for the best model are given in the last two columns of Table~\ref{Wobs}. Under these conditions, the 146.618~GHz maser methanol line is predicted with a negative excitation temperature (-5~K) and a peak intensity of 0.3~K (i.e., not very far from our observations). 
The observed line-integrated intensities are rather well reproduced, except for the two blended lines at 145.126~GHz (left panel of Fig.~\ref{chi2_6lines}). We therefore repeated the same $\chi^2$ minimization, but only with the four most robust lines (i.e., ignoring the two blended lines). The results are shown in Fig.~\ref{chi2_4lines}. The constraints on the physical parameters and the column density are similar, only the temperature is slightly higher. In this case, the best model gives a CH$_3$OH column density of $(1.3^{+0.2}_{-0.3})\times 10^{14}$~cm$^{-2}$, a temperature of 52$^{+23}_{-10}$~K, and a H$_2$ density of $(3.8\pm 1)\times 10^6$~cm$^{-3}$.

Using the same procedures, but with the H$_2$ density ($3.3\times 10^6$~cm$^{-3}$) and temperature (50~K) found for the methanol, we computed approximate column densities for CS, H$_2$CO, and SiO in the outflow. For these, we used the intensity integrated over the full velocity range of -0.2 and 10.9~km~s$^{-1}$ (rather than a Gaussian profile fit) because of the more complex line shapes. 
ndradex was used to compute a grid of 30 values of the column density for each molecule between $10^{13}$ and $10^{14}$~cm$^{-2}$ for H$_2$CO, $3\times 10^{12}$ and $3\times 10^{13}$~cm$^{-2}$ for CS, and $3\times 10^{11}$ and $3\times 10^{12}$~cm$^{-2}$ for SiO. We detected two lines of H$_2$CO, one for the para form and one for the ortho form. We computed their column densities separately and then summed them. 
For the line widths, we used the value of an approximate Gaussian fit of each line: 2.6~km~s$^{-1}$ for H$_2$CO, 3.4~km~s$^{-1}$ for CS, and 2.2~km~s$^{-1}$ for SiO. The collisional coefficients are taken from \citet{2013MNRAS.432.2573W} for H$_2$CO, from \citet{2006A&A...451.1125L} for CS, and from \citet{2006A&A...459..297D} for SiO. The $\chi^2$ minimization  (with a 3$\sigma$ confidence level) gives a column density of $(8.0\pm 0.3)\times 10^{13}$~cm$^{-2}$ for H$_2$CO, $(1.2\pm 0.1)\times 10^{13}$~cm$^{-2}$ for CS, and 
$(1.6 \pm 0.3) \times 10^{12}$~cm$^{-2}$ for SiO. Only the NOEMA data were used to focus on the small scales at which CH$_3$OH is seen. For SiO and CH$_3$OH, the line-integrated intensities are the same for NOEMA and NOEMA+30m data. The extended emission on the CS and H$_2$CO lines appears as a narrow intense peak at the rest velocity of the cloud. Removing this component from the lines of the combined data results in column densities similar to what we obtain with the NOEMA data only.

\begin{table*}
\caption{Fitted integrated intensity and rms of the lines used to compute the species column densities. The two last columns are the excitation temperature and optical depth predicted by RADEX (see text). }
\begin{center}
\begin{tabular}{lcccccc}
\hline
\hline
Molecule & Frequency (MHz) &  Form & W (K~km~s$^{-1}$) & rms (mK) & Tex (K) & $\tau$ \\
\hline
CH$_3$OH & 145093.754 &  E & 0.84 & 35 &41 &    0.01\\
CH$_3$OH &145097.435 & E & 2.55 & 35 &45 &0.03\\
CH$_3$OH &145103.185 &  A & 2.90 & 35 & 36  & 0.04 \\
CH$_3$OH &145126.191 &  E & 0.30 & 34 &40       &0.006\\
CH$_3$OH &145126.386 &  E & 0.20 & 34 &23       &0.001  \\
CH$_3$OH &145131.864 &  E & 0.50 & 34 & 37 &    0.007\\
H$_2$CO & 145602.949 & p & 4.61 & 34 & 73  & 0.02\\
H$_2$CO & 150498.334 & o & 7.38 & 66 & 64 & 0.04 \\
CS & 146969.033 & & 2.93 & 35 & 56 & 0.01 \\
SiO & 130268.610 & & 0.93 & 43 & 56 & 0.002\\
\hline
\end{tabular}
\end{center}
\label{Wobs}
\end{table*}%

\end{appendix}

\label{lastpage}
\end{document}